\documentclass[graybox]{svmult}

\usepackage{mathptmx}       % selects Times Roman as basic font
\usepackage{helvet}         % selects Helvetica as sans-serif font
\usepackage{courier}        % selects Courier as typewriter font
\usepackage{type1cm}        % activate if the above 3 fonts are
\usepackage{makeidx}         % allows index generation
\usepackage{graphicx}        % standard LaTeX graphics tool
\usepackage{multicol}        % used for the two-column index
\usepackage[bottom]{footmisc}% places footnotes at page bottom

\makeindex             % used for the subject index
\begin{document}

\title*{Observing the first galaxies}
\author{James S. Dunlop}
\institute{James S. Dunlop \at Institute for Astronomy, 
University of Edinburgh, Royal Observatory, Edinburgh, EH9 3HJ, UK, 
\email{jsd@roe.ac.uk}}

\maketitle

\abstract{I endeavour to provide a thorough overview of our current knowledge of galaxies and their evolution during 
the first billion years of cosmic time, corresponding to redshifts $z > 5$. After first summarizing progress 
with the seven different techniques which have been used to date in the discovery of objects at $z > 5$, I focus thereafter 
on the two selection methods which have yielded substantial samples of galaxies at early times, namely Lyman-break
and Lyman-$\alpha$ selection. I discuss a decade of progress in galaxy sample selection at $z \simeq 5 - 8$, 
including issues of completeness and contamination, and address some of the confusion which has been created 
by erroneous reports of extreme-redshift objects. Next I provide an overview of our current knowledge of the 
evolving ultraviolet continuum and Lyman-$\alpha$ galaxy luminosity functions at $z \simeq 5 - 8$, and 
discuss what can be learned from exploring the relationship between the Lyman-break and Lyman-$\alpha$ 
selected populations. I then summarize what is known about the physical properties of these galaxies 
in the young universe, before considering the wider implications of this work for
the cosmic history of star formation, and for the reionization of the universe. I conclude with a brief summary of 
the exciting prospects for further progress in this field in the next 5-10 years. Throughout, key concepts such as selection
techniques and luminosity functions are explained 
assuming essentially no prior knowledge. The intention is that 
this chapter can be used as an introduction to the observational study of high-redshift galaxies, as well as providing a 
review of the latest results in this fast-moving research field up to the end of 2011.}

\section{Introduction}
\label{sec:1}

One conclusion of this chapter will be that the very ``first'' galaxies have 
almost certainly not yet been observed. But in 
recent years we have undoubtedly witnessed an  
observational revolution in the study of early galaxies in the young Universe
which, for reasons outlined briefly below, I have chosen to define as 
corresponding to redshifts $z > 5$ (a good, up-to-date 
overview of the physical properties of galaxies at 
$z = 2-4$ is provided by Shapley 2011).

The discovery and study of galaxies at redshifts $z > 5$ is really the 
preserve of the 21st century, and has been one of the most spectacular
achievements of astronomy over the last decade. From the ages of stellar populations
in galaxies at lower redshifts it was known 
that galaxies must exist at $z > 5$  (e.g. Dunlop et al. 1996), but observationally 
the $z = 5$ ``barrier'' wasn't breached until 1998, and then only 
by accident (Dey et al. 1998). 
Although this discovery of a Lyman-$\alpha$ emitting galaxy at $z = 5.34$ was 
serendipitous, it in effect represented the first successful application at $z > 5$ 
of the long-proposed (e.g. Patridge \& Peebles 1976a,b) and oft-attempted 
(e.g. Koo \& Kron 1980; Djorgovski et al. 1985; Pritchet \& Hartwick 1990; Pritchet 1994) 
technique of searching for ``primeval'' galaxies
in the young universe on the basis of bright Lyman-$\alpha$ emission.
This discovery was important not just for chalking up the next integer 
value in redshift, but also because this was the first time that the 
redshift/distance record for any extra-galactic object was held by a ``normal'' 
galaxy which had {\it not} been discovered on the basis of powerful radio or 
optical emission from an active galactic nucleus (AGN). Later the same year, 
two more galaxies selected at $z > 5$ on the basis of their starlight 
(Fernandez-Soto et al. 1999) were spectroscopically confirmed at $z = 5.34$ 
by Spinrad et al. (1998), and the Lyman-$\alpha$ selection record was advanced 
to $z = 5.64$ (Hu et al. 1998).

In this chapter I will explain how these breakthroughs heralded a new era 
in the study of the high-redshift Universe, in which conceptually simple 
but technologically challenging techniques have now been
successfully applied to discover thousands of galaxies at $z > 5$, and to extend
the redshift record out to $z \simeq 9$. The key instrumental/observational
advances which have facilitated this work are the last two successful
refurbishments of the {\it Hubble Space Telescope} ({\it HST};
first with the ACS optical camera, and most 
recently with the near-infrared WFC3/IR imager), the
provision of wide-field optical and near-infrared imaging on $4-8$-m class
ground-based telescopes (Suprime-Cam on 8.2-m Subaru telescope, 
WFCAM on the 3.8-m UK InfraRed Telescope (UKIRT), and 
ISAAC/Hawk-I on the 8.2-m Very Large Telescope (VLT)), the remarkable performance of the 85-cm 
{\it Spitzer Space Telescope} at mid-infrared wavelengths, and finally 
the advent of deep red-sensitive optical spectroscopy on the 10-m Keck telescope 
(with LRIS \& DEIMOS), the VLT (with FORS2), and 
on Subaru (with FOCAS).

I will also endevour to summarize what we have learned about the {\it properties} 
of these early galaxies from this multi-frequency, multi-facility investigation 
and, as a result, what new information we have gleaned about
the evolution of the universe during the first $\simeq 1$\,Gyr of cosmic
time. I conclude with a very brief discussion of the prospects for further progress 
over the next decade; 
a more detailed description of future facilities is included elsewhere in this volume.  

The cosmological 
parameters of relevance to this work are summarized briefly in the 
next section. Where required, all magnitudes are reported in the AB system, 
where $m_{AB} = 31.4 - 2.5\log(f_{\nu}/1\,{\rm nJy})$ (Oke \& Gunn 1983).
 
\section{Why redshift ${\bf z > 5}$}
\label{sec:2}

It is perhaps useful to first pause briefly to review what a redshift 
of $z = 5$
actually means, and why it matters.

Redshift, $z$, is, of course, simply a straightforward
way to quantify the ratio of the observed wavelength ($\lambda_o$)
to the emitted wavelength ($\lambda_e$) of
light:

\begin{equation}
1 + z = \frac{\lambda_o}{\lambda_e}.
\end{equation}

\noindent
The longitudinal relativistic Doppler effect is:

\begin{equation}
1 + z = \frac{\lambda_o}{\lambda_e} = \sqrt\frac{1+v/c}{1-v/c}
\end{equation}

\noindent
and so $z = 5$ corresponds to a recession velocity of $v = 0.946 c$ (where 
$c$ is the speed of light in vacuum).

However, in a Universe with matter, at least some of any observed redshift
should be attributed to gravitational effects, and in any case the precise
recessional velocity of a galaxy several billion
years ago is of little real interest. What is more helpful
is to recognise that the stretching 
of the wavelength of light simply reflects the overall expansion of the
Universe, i.e.

\begin{equation}
1 + z   = \frac{\lambda_o}{\lambda_e} = \frac{R(t_{now})}{R(t_{e})}
\end{equation}

\noindent
where $R(t)$ is simply the scale factor which describes the time evolution of
our apparently isotropic, homogeneous Universe.

Thus, when we observe a galaxy at $z=5$ we are observing light which was
emitted from that galaxy when the Universe was 1/6th of its present size (and
at the highest redshifts currently probed, $z \simeq 9$, the
Universe was 1/10th of its present size).

The precise age at which the Universe was 1/6th of its present size of
course depends on the dynamics of the expansion. With our current
``best-bet'' concordance cosmology of a flat Universe with a matter
density parameter of $\Omega_m = 0.27$, a vacuum energy (or dark energy) density parameter
of $\Omega_{\Lambda} = 0.73$, and a Hubble Constant $H_0 = 71\,{\rm km s^{-1}
Mpc^{-1}}$ (WMAP7; Komatsu et al. 2011; Larson et al. 2011), $z = 5$ corresponds to an age of 1.2\,Gyr, equivalent to
$\simeq 9$\% of current cosmic time. Thus, to a very reasonable approximation,
the study of the universe at $z > 5$ can be thought of as a direct window
into the first Gyr, or first $\simeq 10$\% of the growth and evolution of
cosmic structure.

Finally, at the risk of stating the obvious, it must always be remembered that
different redshifts correspond not only to different times, but also to
different places. Thus, when we presume to connect observations of galaxies
at different redshifts to derive an overall picture of cosmic evolution,
we are implicitly assuming homogeneity; i.e. that ``back-then, over there'' is
basically the same as ``back-then, over here''. For this to be true it is 
crucial that surveys for high-redshift galaxies contain sufficient cosmological
volume to be ``representative'' of the Universe at the epoch in question.
As we shall see, at $z > 5$ this remains a key challenge with current
observational facilities.

\begin{figure}

\includegraphics[scale=0.565]{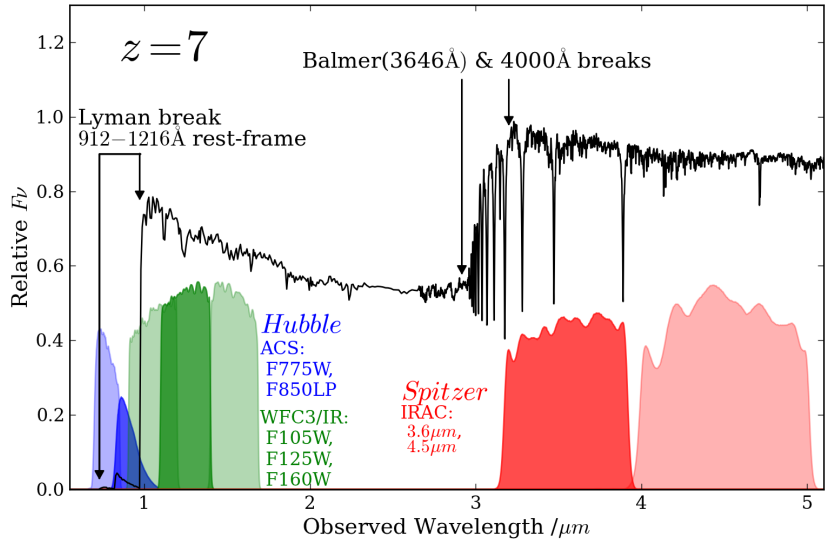}
\caption{An illustration of the redshifted 
form of the rest-frame ultraviolet spectral energy 
distribution (SED) anticipated from a young galaxy at $z \simeq 7$, showing 
how the ultraviolet light is sampled by the key 
red optical ($i_{775}$, $z_{850}$) and near-infrared 
($Y_{105}$, $J_{125}$, $H_{160}$) filters on-board {\it HST} (in the ACS and 
WFC3/IR cameras respectively), while the longer-wavelength 
rest-frame optical light is probed by the 3.6\,$\mu$m and 4.5\,$\mu$m IRAC channels 
on-board {\it Spitzer}. Wavelength is plotted in the observed frame, 
with flux-density plotted as relative $f_{\nu}$ (i.e. per unit frequency).
The spectrum shows the sharp drop at $\lambda_{rest} = 1216$\,\AA\ due to the 
strong ``Gunn-Peterson'' absorption by intervening neutral hydrogen
anticipated at this redshift (here predicted following Madau (1995); 
see also the observed spectrum of the
most distant quasar shown in Fig. 2). Longward of this ``Lyman-break''
the spectrum shown is simply that of the intrinsic integrated 
galaxy starlight as predicted for a 
0.5\,Gyr-old galaxy by the evolutionary spectral synthesis models 
of Bruzual \& Charlot (2003) (using Padova-1994 tracks, assuming constant star formation, 
zero dust extinction, and 1/5th solar metallicity). The characteristic sharp step in the galaxy 
continuum at $\lambda_{rest} = 1216$\AA\ (which at $z \simeq 7$ 
is predicted to result in a very red $z_{850}-Y_{105}$ colour) holds the key to the effective selection of 
Lyman-break galaxies at $z > 5$, as discussed in detail in section 3.1. 
The theoretical spectrum shown here does not include the Lyman-$\alpha$ emission 
line which is produced by excitation/ionization of hydrogen atoms in the inter-stellar medium of the galaxy; 
this offers the main current alternative route for the 
selection of high-redshift galaxies (see Fig. 8 and section 3.2),
and the only realistic hope for spectroscopic confirmation of galaxy redshifts
at $z > 5$ with available instrumentation. 
Also not shown are other nebular emission lines at rest-frame optical 
wavelengths, which can complicate the apparent strength of the key Balmer or 4000\AA\ break 
as measured by the IRAC photometry. In the absence of 
serious line contamination, the strength 
of this break offers a key estimate of the age of the stellar population, with 
consequent implications for a meaningful measurement of galaxy stellar mass. The gap 
between the WFC3/IR and IRAC filters can be filled for brighter objects 
with ground-based $K$-band imaging, but will not be covered from space until the advent 
of {\it JWST} (courtesy S. Rogers).}
\label{fig:1}       % Give a unique label
\end{figure}

\section{Finding galaxies at ${\bf z > 5}$: selection techniques}
\label{sec:3}

There are, in principal, several different ways to attempt to 
pinpoint extreme-redshift galaxies amid the overwhelming numbers 
of lower-redshift objects on the sky. The two methods that have proved 
most effective in recent years both involve optical to near-infrared 
observations of rest-frame ultraviolet light, and both rely 
on neutral Hydrogen. The first method, the so called Lyman-break technique, 
selects Lyman-break galaxies (LBGs)
via the distinctive ``step'' introduced into their 
blue ultraviolet continuum emission by the blanketing effect of 
neutral hydrogen absorption (both within the galaxy itself, and by intervening 
clouds along the observer's line-of-sight; see Fig. 1). 
The second method selects galaxies which are 
Lyman-$\alpha$ emitters (LAEs), via their highly-redshifted 
Lyman-$\alpha$ emission lines, produced by hydrogen atoms in their 
interstellar media which have been excited by the ultraviolet light from 
young stars. Both of these techniques have now been used to discover 
large numbers of galaxies out to $z \ge 7$, and are therefore 
discussed in detail in the two subsections below. 

The only real drawback of these two techniques is that they are only capable of selecting 
galaxies which are young enough to produce copious amounts of ultraviolet light, and 
are sufficiently dust free for a fair amount of this light to leak out in our direction. 
In an attempt to find galaxies at $z > 5$ which are at least slightly 
older (remembering there is only $\simeq 1$\,Gyr available) some authors (e.g. 
Wiklind et al. 2008) have endeavoured to select 
galaxies on the basis of the Balmer break, even though, at $\lambda_{rest} = 3646$\,\AA\, 
this break is moved 
to $\lambda_{obs} > 2.4\,{\rm \mu}$m at $z > 5$. Since this lies beyond the near-infrared 
wavelength range accessible 
from the ground, this work is only 
possible due to the power of the IRAC camera on board 
{\it Spitzer}, which can be used to observe from 3 to 8\,${\rm \mu}$m. As discussed 
later in section 5.1, {\it Spitzer} has certainly proved very effective at measuring the strength 
of Balmer breaks in high-redshift galaxies which have already been discovered via their 
ultraviolet emission but, to date, Balmer-break {\it selection} has yet to uncover a galaxy at $z > 5$ 
which could not have been discovered via other techniques (i.e. the only spectroscopically-confirmed
Balmer-break selected galaxy in the sample compiled by Wiklind et al. is also a Lyman-break galaxy).
This lack of success may of course simply be telling us that there are not many (or indeed any) 
galaxies at these early epochs of the correct age and star-formation history 
($\ge 0.5$\,Gyr-old and no longer forming stars) to be {\it better} selected 
via their Balmer break than their ultraviolet emission; in an era of essentially limitless
gas fuel, and almost universal star-formation activity this would not be altogether surprising. 
However, it may also be the case that Balmer-break selection is simply premature 
with current facilities; the spectral feature itself is not nearly as strong 
(a drop in flux density of a factor of $\simeq 2$ at most) as the Lyman break at these 
redshifts, and at $z > 5$ its detection currently relies on combining 
{\it Spitzer} 3.6\,${\rm \mu}$m photometry with ground-based $K$-band (2.2\,${\rm \mu}$m) photometry (Fig. 1). 
Thus, while Balmer-break selection is undoubtedly important for ensuring we have a 
complete census of the galaxy population at the highest redshifts, its successful 
application may have to await the advent of the {\it James Webb Space Telescope} ({\it JWST}), 
and so it is not discussed further here.

A fourth technique, which is only now coming of age at $z > 5$, 
involves the selection of extreme-redshift galaxies via sub-mm/mm observations of their 
redshifted thermal dust emission. By definition this approach is incapable of detecting the 
{\it very} first primeval galaxies, devoid of any of the elements required for dust, but 
chemical enrichment appears to be a very rapid process, 
and both dust and molecular emission have certainly been detected in objects at 
$z > 6$ (Walter et al. 2003; Robson et al. 2004; Wang et al. 2010). Although this dust emission is 
powered by ultraviolet emission from young stars, it is already clear that at least some of the 
galaxies successfully discovered via sub-mm observations at more modest redshifts display such strong 
dust extinction that they could not have been selected by rest-frame ultraviolet observations. Thus, 
while we might expect dust to become less prevalent at extreme redshifts (and there is evidence 
in support of this presumption - e.g. Bouwens et al. 2009; Zafar et al. 2010; Bouwens et al. 2010a; 
Dunlop et al. 2012; Finkelstein et al. 2012) 
it will be important to pursue sub-mm/mm selection
at $z > 5$ over the coming decade to ensure a complete picture of early galaxy evolution.
As with Balmer-break selection, this is a technique which (for technical reasons) is still in its infancy,
although excitingly the first spectroscopic 
redshift at $z > 5$ for a mm-selected galaxy has recently been measured 
purely on the basis of redshifted CO line emission ($z \simeq 5.3$; Riechers et al. 2010). Over the next 
few years this whole field should be revolutionized by the advent of the Atacama Large Millimetre Array (ALMA).

\begin{figure}

\includegraphics[scale=0.4]{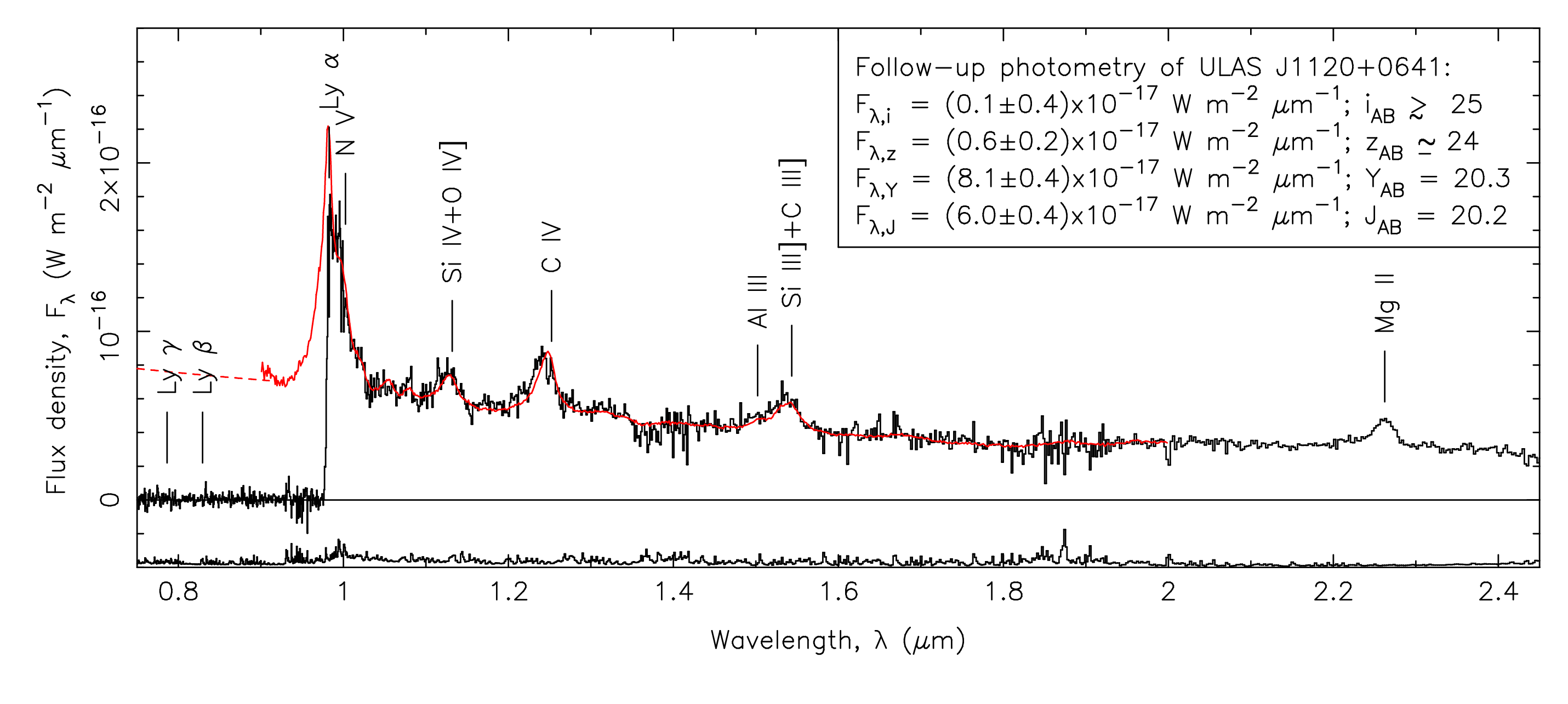}
\caption{The near-infrared spectrum of the most distant known 
quasar ULASJ112001.48+064124.3, the first (and, to date, only) 
quasar discovered at redshifts $z > 7$ ($z = 7.085$; Mortlock et al. 2011). The data 
are shown in black, with the 1-$\sigma$ error spectrum shown at the base of the plot.
Despite being observed only 0.77 billion years after the Big Bang,
this quasar has an intrinsic spectrum essentially identical to that displayed 
by lower-redshift quasars with, for example, strong Carbon lines indicating
approximately solar metallicity (the red curve shows the 
average spectrum of 169 quasars in the redshift range $2.3 < z < 2.6$). 
However, shortward of 
$\lambda_{rest} \simeq 1216$\,\AA\, the spectrum provides an excellent 
demonstration of the Gunn-Peterson effect, whereby the increased 
fraction of neutral hydrogen 
along the line-of-sight has completely obliterated the UV continuum 
emission from the quasar. This sudden drop in flux-density shortward of 
Lyman-$\alpha$ is the key spectral feature 
which facilitates not just the selection of rare extreme-redshift 
quasars such as this, 
but also the selection of fainter but much more numerous
``Lyman-break galaxies'' (LBGs) at 
redshifts $z > 5$ (see Fig. 1) (courtesy D. Mortlock).}
\label{fig:2}       % Give a unique label
\end{figure}

A fifth approach, which it is important not to forget, is that high-redshift ``galaxies'' 
continue to be located on the basis of both optical and radio emission powered 
by accretion onto their central super-massive black holes. 
Indeed, the quasar redshift record has recently crossed the $z = 7$ 
threshold ($z = 7.085$; Mortlock et al. 2011), and significant numbers of quasars are 
now known at $z > 6$ (e.g. Fan et al. 2003, 2006; Willott et al. 2010). 
High-redshift quasars are rare but, because of their brightness, have the potential to 
provide much useful information on the state of the inter-galactic medium (IGM) at early times 
(e.g. Carilli et al. 2010), as well as providing signposts 
towards regions of enhanced density in the young universe. 
However, the very strong active nuclear emission
which facilitates the discovery of high-redshift quasars 
also makes it extremely difficult to detect, never mind study, the stellar 
populations in their host galaxies (e.g. Targett, Dunlop \& McLure 
2012), and so they are inevitably 
of limited use for the detailed investigation 
of early galaxy evolution. 
By contrast it is perfectly possible to study the stellar populations 
in high-redshift radio galaxies (e.g. McCarthy 1993; Dunlop et al. 1996; Seymour et al. 2007), 
and indeed for many years essentially all of our knowledge of galaxies 
at $z > 3$ was derived from the optical--infrared--sub-mm 
study of objects which 
were originally selected on the basis of radio-frequency synchrotron emission powered 
by super-massive black holes (e.g. Lilly 1988; Dunlop et al. 1994; Rawlings et al. 1996). 
However, in recent years the search for increasingly high-redshift radio galaxies 
has rather run out of 
steam; the $z = 5$ threshold was passed in 1999 ($z = 5.197$; van Breugel et al. 1999), 
but 12 years later the radio-galaxy redshift record remains unchanged.
This difficulty in further progress is perhaps not unexpected, given the now 
well-established decline in the number density of powerful radio sources beyond $z \simeq 
3$, and the unhelpfully strong k-correction
provided by steeply-falling power-law synchrotron emission (Dunlop \& Peacock 1990; Rigby et al.
2011). Nevertheless, searches for higher-redshift radio galaxies will continue, 
motivated at least in part by the desire to find even a few strong radio 
beacons against which to measure the 21-cm analogue of the Lyman-$\alpha$ 
forest as we approach the epoch of reionization (the ``21-cm forest''; 
Carilli, Gnedin \& Owen 2002; Furlanetto \& Loeb 2002; Mack \& Wyithe 2012). 
However, at least for now, radio-continuum selected objects offer little 
direct insight into galaxy evolution at the very highest redshifts. A thorough 
review of what is currently known about distant radio galaxies and their environments is 
provided by Miley \& De Breuck (2008), who also include a compendium of known 
high-redshift radio galaxies. 

A relatively new sixth, and remarkably effective route to pinpointing high-redshift 
objects has recently arrived with the discovery of Gamma-Ray Bursts (GRBs). These are 
now regularly detected via monitoring with gamma-ray satellites such as {\it Swift} 
(Gehrels et al. 2004), and then rapidly followed up 
with a range of ground-based observations (e.g. Fynbo et al. 2009). 
Long-duration GRBs are thought to arise from the death 
of very massive, possibly metal-poor stars (Woosley \& Bloom 2006), and observationally 
have been associated with Type 1c supernovae (e.g. Hjorth et al. 2003). Regardless of their 
precise physical origin, they have proved to be very luminous events which 
are visible out to the highest redshifts, $z > 8$ (the gamma-ray positions are 
poor, but rapid follow-up can pinpoint the fading optical/near-infrared 
afterglow unambiguously and, if quick enough, can also yield robust redshift information).   
GRBs broke the $z = 5$ ``barrier'' very quickly after their discovery, with 
a redshift of $z = 6.295$ measured for GRB\,050904 by Haislip et al. (2006) and  
Kawaii et al. (2006). Another GRB at $z > 5$ followed the next year with
the discovery of GRB\,060927 at $z = 5.467$ (Ruiz-Vesco et al. 2007). Two 
years later, GRBs wrested the redshift record from quasars and LAEs, 
with Greiner et al. (2009) reporting a redshift of $z=6.7$ for GRB\,080913. Then,
most spectacularly, a GRB became the first spectroscopically confirmed object
at $z > 8$, with GRB\,090423 being convincingly shown to lie at $z=8.23$ 
(Salvaterra et al. 2009; Tanvir et al. 2009). Most recently, it has been argued
that GRB\,090429B lies at $z \sim 9.4$ (Cucchiara et al. 2011) but the robustness of 
this (photometric) redshift is currently a matter of debate. Given this impressive 
success in redshift record breaking, the reader may be surprised to learn 
that I have chosen not to consider GRBs further in this chapter. The reason is 
that, to date, while the hosts of many lower-redshift GRBs have been uncovered
(e.g. Perley et al. 2009) follow-up observations targetted on the positions of faded 
GRB remnants at $z > 5$ have yet to yield useful information on their host galaxies. 
This is, of course, an interesting result in its own right. It indicates 
that, as arguably expected, GRBs largely occur in  
faint dwarf galaxies which lie below the sensitivity limits of even our 
very best current instrumentation. Specifically, 
the follow-up {\it HST} WFC3/IR imaging of the $z=8.23$ GRB\,090423
reaching $J_{125} \simeq 28.5$ has failed to detect the host galaxy
(Tanvir et al. in prep), while the host of GRB\,090429B is apparently undetected to 
$Y_{105} \simeq 28$ (Cucchiara et al. 2011). Thus, while as 
discussed by Robertson \& Ellis (2012), high-redshift GRBs 
can already provide important insights into global cosmic star-formation history, their 
usefulness as transient signposts towards extreme-redshift galaxies is 
unlikely to be properly exploited until the advent of {\it JWST}.

Finally, over the next decade we are likely to see the emergence of a seventh technique 
for finding extreme-redshift galaxies via radio-wavelength spectroscopy. Specifically, 
following the first successful mm-to-radio CO-line redshift determinations, 
in addition to the above-mentioned targetted CO line follow-up of 
pre-selected mm/sub-mm sources with ALMA, we can expect to see ``blind'' spectroscopic surveys 
for CO and for highly-redshifted 21-cm atomic Hydrogen emission 
with the new generation of radio facilities (e.g. Carilli 2011). 

\subsection{Lyman-break selection}
\label{subsec:2}

In the absence dust obscuration, young star-forming galaxies are expected to 
be copious emitters of UV continuum light, with a star-formation rate $SFR~=~1\,{\rm~M_{\odot} 
\,yr^{-1}}$ predicted to produce a UV luminosity 
at $\lambda_{rest}~\simeq~1500$\,\AA\ of 
$f_{\nu}~\simeq~8~\times~10^{27}\,{\rm erg\,s^{-1}\,Hz^{-1}}$ 
for a Salpeter (1955) initial mass 
function (Madau et al. 1998). 
For reference, this corresponds to an absolute magnitude
of $M_{1500} \simeq -18$ which, at $z \simeq 7$, translates 
to an observed 
near-infrared $J$-band magnitude of $J \simeq 28.5$. 
As we shall see, this is very
comparable to the detection limit of 
the deepest {\it HST} WFC3/IR imaging currently available. 

The basic idea of selecting distant objects (galaxies or quasars) via the 
signature introduced by hydrogen absorption of this ultraviolet 
light goes back several 
decades
(e.g. Meier 1976a,b). As first successfully implemented in the modern 
era by Guhathakurta et al. (1990) and Steidel \& Hamilton (1992), the aim
was to select galaxies at $z \sim 3$ by searching for sources in which 
the Lyman-limit at $\lambda_{rest} = 912$\,\AA\ had been redshifted to lie  
between the $U$ and $B_{j}$  filters at $\lambda_{obs} \simeq 3600$\,\AA. All 
ultraviolet-bright astrophysical objects display an intrinsic drop in their spectra at 
$\lambda_{rest} = 912$\,\AA\ (which corresponds to the ionization energy of 
the hydrogen atom in the ground state), and the expectation was that, 
in young galaxies, this drop would be very strong (roughly an order-of-magnitude in flux density)
due to a combination of the hydrogen edge in stellar photospheres, 
and photo-electric absorption by the interstellar neutral hydrogen gas (expected 
to be abundant in young galaxies). At the highest redshifts, the ever denser intervening 
neutral hydrogen clouds also produce increasing Lyman-$\alpha$ absorption
(between energy levels 1 and 2 in the hydrogen atom) resulting in an 
ever-thickening Lyman-$\alpha$ forest which impacts on 
the continuum of the target galaxy between $\lambda_{rest} = 1216$\,\AA\ and 
$\lambda_{rest} = 912$\,\AA. At moderate redshifts the average blanketing effect of this
forest simply produces an additional (and useful) signature in the galaxy spectrum 
in the form of an apparent step in the continuum below Lyman-$\alpha$ (a factor of $\sim 2$ 
drop in flux density at $\lambda_{rest} = 1216$\,\AA\ at $z \sim 3$; Madau 1995). However, 
as discussed further below (and illustrated in Figs. 1 \& 2) 
ultimately the forest becomes so optically thick that it kills virtually 
all of the galaxy light at $\lambda_{rest} < 1216$\,\AA, rendering the original 912\,\AA\ 
break irrelevant, and Lyman-break selection in effect becomes the selection of objects
with a sharp break at $\lambda_{rest} = 1216$\,\AA.

The beauty of the Lyman-break selection technique is that it can be applied 
using imaging with broad-band filters, allowing potentially large samples 
of high-redshift galaxies to be selected 
for spectroscopic follow-up and confirmation. When selecting galaxies in this way,
what one is looking for are objects which are repeatedly visible (and fairly blue)
in the longer wavelength images, but then effectively disappear in the bluest 
image under consideration. For this reason such objects are often called ``dropout''
galaxies. Thus,``$U$-dropouts'' (or simply ``$U$-drops'') are galaxies which disappear 
in the $U$-band filter, and are therefore expected to have their Lyman limit moved 
to $\lambda_{obs} \simeq 3500$\,\AA\ implying a redshift $z \sim 3$ (in practice 
$2.5 \leq z \leq 3.5$). Similarly, ``$B$-drops'' (or ``$G$-drops'')
are expected to be galaxies at $z \sim 4$, while ``$V$-drops'' should have 
$z \sim 5$. Thus, deep broad-band optical imaging can be used to select 
samples of galaxies in bands of increasing redshift. 

The simple act of colour 
selection yields redshifts accurate to $\delta z \simeq 0.1-0.2$. Consequently, with 
the aid of simulations to estimate the effective redshift distribution and cosmological
volume probed by each specific drop-out criterion, luminosity functions (LFs) can 
be derived in broad redshift bands without recourse to optical spectroscopy. 
However, for proper assessment of completeness/contamination, and the determination 
of redshifts with sufficient accuracy to allow robust clustering measurements, 
spectroscopic follow-up is essential.

The huge break-throughs enabled by the successful application of the ``dropout'' 
technique in the 1990s are perhaps best exemplified by the work of Steidel
and collaborators (who were able to use the 10-m 
Keck telescope to spectroscopically 
confirm large samples of LBGs, enabling LF and clustering 
measurements - e.g. Steidel et al. 1996, 2000) and by the study of Madau et al. (1996) 
who applied the technique to the deep {\it HST} WFPC2 $U_{300}$,\,$B_{450}$,\,$V_{606}$,\,$I_{814}$ 
imaging in the Hubble Deep Field (HDF; Williams et al. 1996, Ferguson et al. 2000) 
to produce the first measurement of the average cosmic star-formation density 
out to $z \sim 4$. A full overview of this ``low-redshift'' work is beyond 
the scope of this Chapter, but a thorough review of the 
success of the Lyman-break technique in enabling the discovery 
and study of galaxies in the redshift range $2 < z < 5$ can be found in  
Giavalisco (2002). 

\begin{figure}

\includegraphics[scale=0.95]{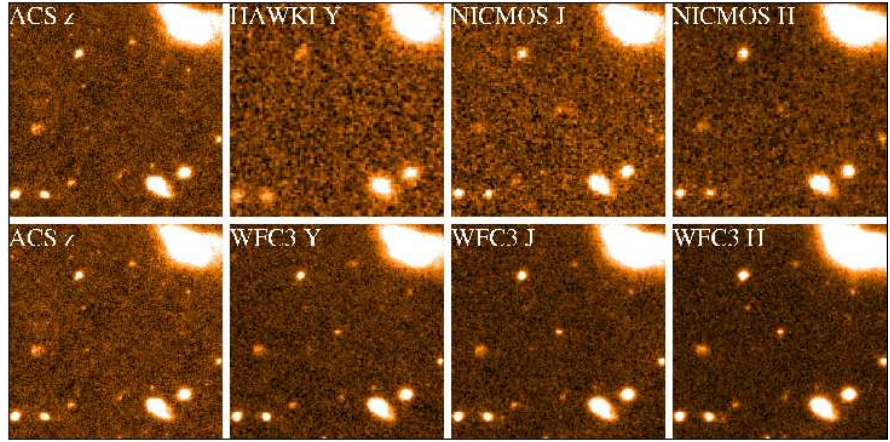}
\caption{The Lyman-break selection of a $z \simeq 7$ 
galaxy uncovered in the Hubble Ultra-Deep Field (HUDF). The upper row of 
plots shows postage stamps of the available data at $z_{850}$, $Y$, $J_{110}$,
$H_{160}$ prior to the advent of the new WFC3/IR near-infrared camera on 
{\it HST} in 2009. The lower row of plots shows the hugely-improved 
near-infrared imaging provided by WFC3/IR for the same object; 
it can be clearly seen that this galaxy is strongly detected in the three 
longest-wavelength passbands ($H_{160}$, $J_{125}$ and $Y_{105}$) but 
drops out of the $z _{850}$ image altogether, due to the presence of the 
Lyman-break redshifted to $\lambda_{obs} \simeq 1$\,$\mu$m, as was illustrated
in Fig. 1 (courtesy R. McLure).}
\label{fig:3}       % Give a unique label
\end{figure}

The ensuing decade has seen rapid progress from $z \simeq 5$ to $z \simeq 8$,
in part because this selection technique is, in
principle, even more straightforward at $z \ge 5$ than at lower redshifts. This is
because by $z = 5$ the Lyman-$\alpha$ forest produced by intervening clouds of
neutral Hydrogen is expected to be so dense that the anticipated  break in the
continuum level at $\lambda_e \simeq 1216$\,\AA\ is $\simeq 1.8$ mag.,
or a factor $\simeq 5$ in average flux density (Madau 2005). 
This is more than twice as
strong as any of the other intrinsically strong breaks
displayed by the starlight from galaxies
(e.g. the $\lambda = 4000$\,\AA\ break in an old stellar population,
produced by an acculmulation 
of absorption lines from ionized metals, 
(especially Ca II H and K lines at 3933 and 3968\,\AA), or the $\lambda = 3646$\,\AA\ 
Balmer break in a $\sim 0.5$\,Gyr-old post-starburst galaxy,
most prominent in A stars, with $T \sim 10000$\,K). By 
$z > 6.5$, observations of the highest-redshift quasars indicate 
that essentially all flux shortward of Lyman-$\alpha$ is extinguished 
(Fig. 2), and LBG selection effectively 
becomes the selection of galaxies with a complete 
``Gunn-Peterson Trough'' (Gunn \& Peterson 1965).

Thus, given sufficiently good signal:noise, and appropriate
broad-band filters, the selection of Lyman-break galaxies at $z > 5$ should be easy and 
reasonably clean,
and indeed has proved to be so once detector and telescope developments were
successfully combined to deliver the necessary deep, red-sensitive imaging. 

\subsubsection{Lyman-break galaxies at ${\bf z > 5}$}

The main reason for a delay in progress in LBG selection beyond $z \simeq 5$ was the 
need for sufficiently deep imaging in at least two wavebands longer than the putative 
Lyman break; as illustrated in Figs. 1, 3, 4 and 5, 
at least two colours (hence three wavebands) are needed 
to confirm both the existence of a strong spectral break, and a blue colour longward 
of the break (as anticipated for a young, ultraviolet-bright galaxy; see subsection 3.1.3 
on potential contaminants). This need was finally met with the refurbishment of the {\it HST} in March 
2002 with a new red-sensitive optical camera, the Advanced Camera for Surveys (ACS), and a new cooling 
system for the Near Infrared Camera and Multi-Object Spectrometer (NICMOS). 
Crucially, the ACS was 
quickly used to produce and release the 
deepest ever optical image of the sky, the 4-band ($B_{435}$, $V_{606}$, $i_{775}$, $z_{850}$) 
Hubble Ultra Deep Field (HUDF; 
Beckwith et al. 2006), 
covering an area of $\simeq 11$\,arcmin$^2$ to typical depths of $m_{AB} \simeq 29$ 
for point sources. This field (or at least 5.7\,arcmin$^2$ of it) 
was also imaged with NICMOS, in the $J_{110}$ and $H_{160}$  bands 
by Thompson et al. (2005, 2006) to depths of $m_{AB} \simeq 27.5$. 
Around the same time the ACS was also used as part of the Great Observatories
Deep Survey (GOODS) program to image two 150\,arcmin$^2$ fields (again 
in $B_{435}$, $V_{606}$, $i_{775}$, $z_{850}$) to more moderate depths, 
$m_{AB} \simeq 27.5 - 26.5$
(GOODS-North, containing the HDF, and GOODS-South, containing the HUDF; Giavalisco et al. 2004).
Deep {\it Spitzer} IRAC imaging (at 3.6, 4.5, 5.6, 8\,${\rm \mu}$m) 
was also obtained over both GOODS fields, and a co-ordinated effort was made to obtain deep $K_s$-band imaging 
for GOODS-South 
from the ground with ISAAC on the 8.2-m VLT (Retzlaff et al. 2010).

The result was a flood of papers reporting the discovery of ``$i$-drop'' galaxies 
at $z \simeq 6$ (Bouwens et al. 2003, 2004a, 2006; Bunker et a. 2003, 2004; Dickinson et al. 2004; Stanway et al. 2003, 2004, 2005; Yan \& Windhorst 2004; Malhotra et al. 2005; Beckwith et al. 2006; Grazian et al. 2006), 
and even an (arguably premature, but partially successful) attempt to uncover 
``$z_{850}$-drop'' galaxies at $z \simeq 7$ (Bouwens et al. 2004c) 
and set limits at even higher redshifts (Bouwens et al. 2005). 

Spectroscopic follow-up was rapidly achieved for several of the brighter ``i-drops'' 
yielding the first spectroscopically-confirmed LBGs at $z \simeq 6$ (Bunker et al. 2003; Lehnert 
\& Bremer 2003; Vanzella et al. 2006; Stanway et al. 2007), and some of these 
were even successfully detected with {\it Spitzer} at 3.6\,${\rm \mu}$m and 4.5\,${\rm \mu}$m, 
yielding some first estimates of their stellar masses 
and star-formation histories (e.g. Labb\'{e} et al. 2006; Yan et al. 2006; Eyles et al. 2007)

Further spectroscopic follow-up of $z \ge 5$ LBGs in the GOODS fields has 
been steadily pursued with Keck and the VLT over the last few years, (e.g. Stark et al. 2009, 2010, 2011; 
Vanzella et al. 2009) yielding interesting results on mass density, evolution, and Lyman-$\alpha$ 
emission from LBGs which are discussed further in sections 4 and 5. 

From 2005, progress in wide-area red optical and near-infrared imaging 
with Suprime-Cam (Miyasaki et al. 2002) on the Subaru telescope, 
and WFCAM (Casali et al. 2007) on UKIRT (via the UKIDSS survey; Lawrence et al. 2007) 
led to the first significant samples of brighter $z \simeq 6$ galaxies
being selected from ground-based surveys covering areas approaching 
$\simeq 1$ deg$^2$ (Kashikawa et al. 2004; Shimasaku et al. 2005; Ota et al. 2005; 
McLure et al. 2006, 2009; Poznanski et al. 2007; Richmond et al. 2009). 
As discussed further below in section 4.1, this work complements the deeper but much smaller-area 
{\it HST} surveys by providing better sampling of the bright end of the LF.

Motivated by the availability of 12-band 
CFHT+Subaru+UKIRT+{\it Spitzer}-IRAC photometry in the UKIDSS Ultra Deep Survey (UDS) field
(coincident with the Subaru/XMM-Newton Deep Survey (SXDS); Furusawa et al. 2008), 
McLure et al. (2006) also introduced a new approach to 
selecting galaxies at $z > 5$, replacing simple two-colour ``dropout''
criteria with multi-band redshift estimation via model spectral energy distribution 
(SED) fitting 
(a technique commonly adopted at lower redshifts -- e.g. Mobasher et al. 2004, 2007;
Cirasuolo et al. 2007, 2010). 
This approach has the advantage of using all of the data in a consistent way 
(including multiple non-detections) and captures the uncertainty in
redshift (and resulting uncertainty in stellar masses etc) for each individual object (Fig. 4).
In addition, it provides better access
to redshift ranges where the simple two-colour dropout technique is sub-optimal 
(due, for example, to the Lyman-break lying within rather than at the edge of a
filter bandpass; see Fig. 5). It can also yield a redshift probability distribution
for each source (e.g. Finkelstein et al. 2010; 
although there is a debate to be had about appropriate priors), 
and explicitly exposes alternative acceptable
redshift solutions (e.g. Dunlop et al. 2007), enabling targetted spectroscopic 
follow-up to reject these if desired. Nevertheless, careful simulation work is 
still required to estimate incompleteness and contamination, and the SED-fitting 
approach can arguably be harder for others to replicate than simple
two-colour selection.

One disadvantage of ground-based imaging is the potential for $z > 5$ LBG sample contamination 
by cool dwarf stars (see section 3.1.3). On the other hand, because the LBG candidates are relatively 
bright, spectroscopic follow-up has proved very productive, and has now yielded Lyman-$\alpha$ redshifts 
in the range $z \simeq 6 - 6.5$ for
$\simeq 30$ LBGs selected from ground-based surveys
(Nagao et al. 2004, 2005, 2007; Jiang et al. 2011; Curtis-Lake et al. 2012). These spectroscopic 
programs provide not only more accurate redshifts, but also enable measurement of the prevalence 
and strength of Lyman-$\alpha$ emission from LBGs as a function of redshift and 
continuum luminosity. Such measurements have the potential to shed light on the connection 
between LBGs and LAEs, the cosmic evolution of dust, and the process of reionization
(see sections 4.3 \& 6.2)

At $z > 6.5$ ground-based selection of LBGs becomes extremely difficult due to the 
difficulty in reaching the necessary near-infrared depths, although quite how difficult 
depends of course on the shape of the LF
at $z \simeq 7$. Such progress as has been made with Subaru and the VLT between $z \simeq 
6.5$ and $z \simeq 7.3$ is discussed in the next subsection.

\begin{figure}

\includegraphics[scale=0.7, angle=270]{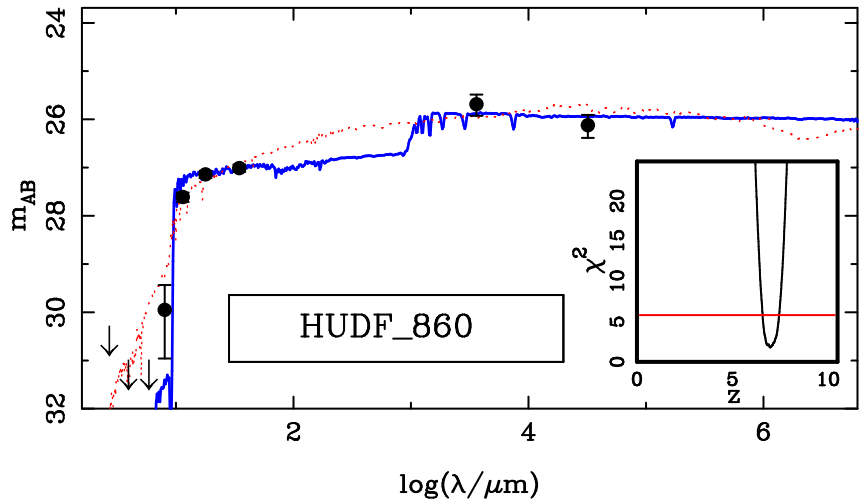}

\caption{An example of the galaxy-template 
SED-fitting analysis employed by McLure et al.
 (2009, 2010, 2011) for high-redshift galaxy selection, which makes 
optimum use of the available multi-wavelength photometry (including de-confused
{\it Spitzer} IRAC fluxes; McLure et al. 2011). 
Based on the evolutionary synthesis models of 
Bruzual \& Charlot (2003), not only redshift, but also 
age, star-formation history, dust extinction/reddening, mass 
and metallicity are 
all varied in search of the best-fitting solution. This also enables robust 
errors to be placed on the range of acceptable photometric redshifts, after 
marginalising over all other parameters. In this case the photometry provides
more than adequate accuracy and dynamic range to exclude all redshift
solutions other than that indicated by the blue line, which 
yields $z \simeq 6.96 \pm 0.25$. 
The thin dotted red line shows the best-fitting alternative 
solution at low redshift, albeit in this case this 
alternative is completely unacceptable.}
\label{fig:4}       % Give a unique label
\end{figure}

\subsubsection{Lyman-break galaxies at ${\bf z = 7 - 10}$}

By redshifts $z \simeq 7$, the Lyman break has moved to $\lambda_{obs} \simeq 1$\,${\rm \mu}$m, 
beyond the sensitivity regime of even red-sensitive CCD detectors. As a result,
efforts to uncover LBGs at $z > 6.5$ were largely hamstrung by the lack of 
sufficiently-deep near-infrared imaging, until the installation of the long-awaited new camera,  
WFC3, in the {\it HST} in May 2009. Due to its exquisite sensitivity and (by space standards)
wide field-of-view (4.8\,arcmin$^2$), the 
infrared channel of this camera, WFC3/IR, offered 
a $\sim40$-fold improvement in mapping speed over NICMOS for deep near-infrared surveys.
This, coupled with the availability of an improved near-infrared filter set, immediately rendered 
obsolete the few heroic early attempts to uncover LBGs at $z > 6.5$ with NICMOS (e.g. Bouwens
et al. 2004, 2010c),
even those assisted by gravitational lensing (e.g. Richard et al. 2008; Bradley et al;. 2008; Bouwens et al. 2009a; Zheng et al. 2009). 

\begin{figure}

\includegraphics[scale=0.6]{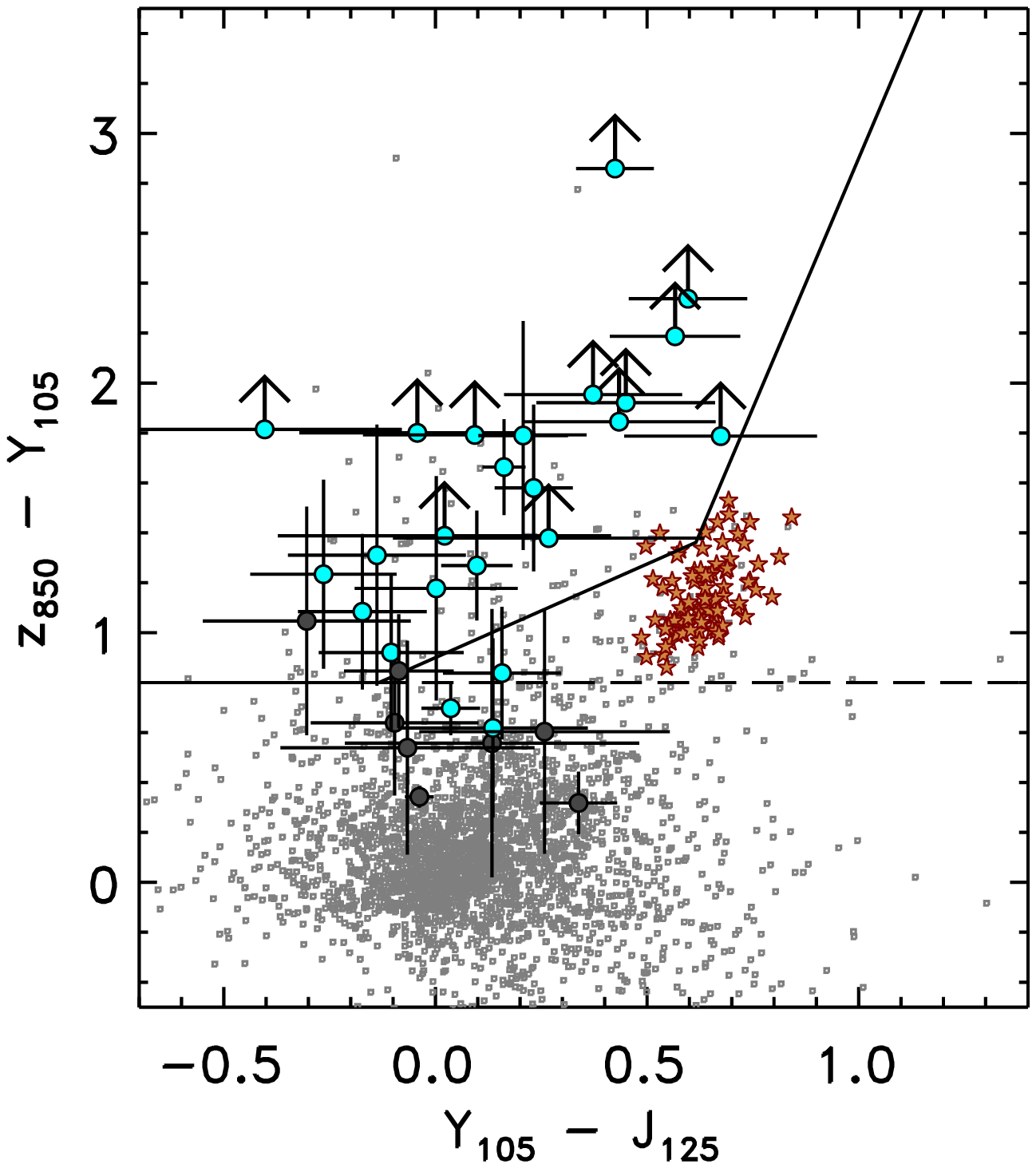}
\caption{An illustration provided by Finkelstein et al. (2010)
of some of the limitations of using a simple, 
strict, colour-colour criterion to select high-redshift LBGs. 
Filled circles 
indicate high-redshift galaxies selected by SED 
fitting, with the dark-grey circles indicating galaxies with $6.0~<~z_{phot}~<~6.3$ 
and the blue circles highlighting galaxies at $6.3~<~z_{phot}~<~7.5$. Arrows are
1-$\sigma$ limits.
The solid lines show the selection cuts adopted by Oesch et al. (2010), and the dashed line is from Yan et al. (2010) (both designed to select LBGs at $z \simeq 7$). The small grey squares are low-redshift 
galaxy contaminants with $z _{phot} < 6.0$, and the red stars indicate 
the colours of galactic brown-dwarf stars. The colour cuts result
in the inclusion of many contaminants as well as 
the exclusion of genuine high-redshift candidates
that are identified via full SED fitting (which makes more optimal use
of all the available data, including marginal detections at optical
wavelengths) (courtesy S. Finkelstein).}
\label{fig:5}       % Give a unique label
\end{figure}

The remarkable improvement offered by WFC3/IR at near-infrared wavelengths 
is illustrated in Fig. 3, 
which shows the imaging data 
available before and after Sept 2009 for (arguably) the only moderately-convincing ``$z_{850}$-drop'' $z \simeq 7$ galaxy uncovered with NICMOS+ACS (Bouwens et al. 2004; Oesch
et al. 2009; McLure et al. 2010). These images are extracted from the first 
(Sept 2009) release of the WFC3/IR $Y_{105}$,$J_{125}$,$H_{160}$ imaging of the HUDF, taken  
as part of the HUDF-09 treasury program (PI: Illingworth). This 
reached previously unheard-of depths $m_{AB} \simeq 28.5$, and immediately  
transformed our knowledge of galaxies at $z > 6.5$,  
with four independent groups reporting the first substantial samples of galaxies 
with $6.5 < z < 8.5$
(Oesch et al. 2010a; Bouwens et al. 2010b; McLure et al. 2010; Bunker et al. 2010; Finkelstein et al.
2010). Both the above-mentioned alternative approaches to LBG selection were applied to these new data,
with McLure et al. (2010) and Finkelstein et al. (2010) undertaking full SED fitting (Fig. 4), 
while the other groups applied standard two-colour ``drop-out'' criteria (Fig. 5). Three independent 
reductions of the raw data were also undertaken prior to LBG selection. Given this, the level of 
agreement between the $6.5~<~z~<~8.5$ source lists was (and remains) undeniably impressive. The era 
of galaxy study at $z~>~7$ has now truly arrived.

The initial HUDF WFC3/IR data release was rapidly followed by the release of the WFC3 
Early Release Science (ERS) data (Windhorst et al. 2011). The infrared component
of this dataset comprised 2-orbit depth WFC3/IR imaging
in $Y_{098}$,$J_{125}$,$H_{160}$ over 10 pointings in the northern part of the GOODS-South 
field, and thus complemented the HUDF imaging by delivering imaging of $\simeq 40$\,arcmin$^2$
to $m_{AB} \simeq 27.5$. The intervening two years have seen the completion of the HUDF-09 program,
involving deeper imaging of the HUDF itself to $m_{AB} \simeq 29$, 
and $Y_{105}$,$J_{125}$,$H_{160}$ imaging of two parallel fields to $m_{AB} \simeq 28.5$. 
The combined HUDF-09 and ERS dataset has now been analysed in detail for LBGs at $z > 6.5$ 
(again by several independent groups; Wilkins et al. 2010, 2011a; Bouwens et al. 2011b; 
Lorenzoni et al. 2011; McLure et al. 2011),
and has yielded samples of $\sim 70$ candidate LBGs at $z \sim 7$, 
$\sim 50$ at $z \sim 8$, and possibly one galaxy at $z \sim 10$ (Bouwens et al. 2011a;
Oesch et al. 2012). As with the original data release, despite disagreement over certain
individual sources (see, for example, the careful cross-checking 
performed by McLure et al. 2011) there is generally good agreement over the $z \sim 7$ and $z 
\sim 8$ galaxy samples, especially if attention is 
restricted to the brighter objects. However, 
where the data have been pushed to the limit, potential 
contamination by low-redshift interlopers becomes more of an issue (see below), and in 
particular there is some debate over the robustness of the $z \sim 10$ galaxy. This 
discovery relies on detection in a single band ($H_{160}$) because the proposed Lyman-break
lies at the long-wavelength edge of the $J_{125}$ filter. Therefore, while there is little 
doubt that this is a real object, there is currently no direct observational evidence that 
it displays a blue slope longward of the break. As illustrated in Fig. 6, to push LBG 
selection beyond $z \simeq 8$ really 
requires still deeper imaging, and the additional use of the $J_{140}$ filter
(to provide two detections of the galaxy ultraviolet continuum above the Lyman-break 
at $z \sim 9$). Such a program has now been approved in the HUDF, and is planned with 
{\it HST} WFC3/IR in summer 2012.

In addition to this ground-breaking ultra-deep near-infrared imaging, 
wider field surveys with WFC3/IR are now underway. In particular Trenti et al. (2011, 2012) and 
Yan et al. (2011) have recently used parallel WFC3/IR imaging to search for ``brighter'' 
$Y$-drop $z \sim 8$ LBGs, yielding several candidates which are potentially bright enough
to be amenable to spectroscopic folow-up with ground-based near-infrared spectrographs. 
The 3-year, 902-orbit, Cosmic Assembly Near-infrared Deep Extragalactic Survey 
(CANDELS) Treasury Program has also recently commenced (Grogin et al. 2011; 
Koekemoer et al. 2011)\footnote{http://candels.ucolick.org}. 
This will ultimately deliver WFC3/IR imaging (with parallel 
ACS optical imaging) to $m_{AB} \sim 27$ over $\simeq 0.25$\,deg$^2$
spread over 5 different well-studied fields, including deeper survey regions 
reaching $m_{AB} \sim 28$ over 
$\simeq 0.04$\,deg$^2$ (split between GOODS-North and GOODS-South). This survey 
is expected to provide the area and depth required to enormously clarify our understanding
of the prevalence and properties of moderate luminosity ($L^*$) galaxies at $z \simeq 6.5 - 8.5$.

Finally, progress is also expected from WFC3/IR imaging of lensing clusters. Imaging 
of the Bullet Cluster has already yielded several $z \simeq 7$ LBG candidates (Hall et al. 2012) 
and a second major (524 orbit) {\it HST} multi-cycle Treasury Program, the Cluster Lensing and Supernova Survey with Hubble 
(CLASH)\footnote{http://www-int.stsci.edu/~postman/CLASH}, will deliver multi-band
WFC3 imaging of 25 clusters over the next 3 years (Postman et al. 2012).

These rapidly-growing 
samples of WFC3/IR-selected LBGs at $z > 6.5$ are providing a wealth of new 
information on galaxies and their evolution in the first billion years, not least 
because many of the the brighter ones have also proved to be detectable
at $3.6$\,${\rm \mu}$m with {\it Spitzer} IRAC. 
As a result, even without spectroscopic redshifts, 
it has already been possible not only 
to obtain the first meaningful measurements 
of the galaxy luminosity function at $z \sim 7$ and $z \sim 8$
(see section 4) but also to explore the physical properties of these 
young galaxies (i.e. masses, stellar populations, sizes; see section 5).

\begin{figure}

\includegraphics[scale=0.35, angle=0]{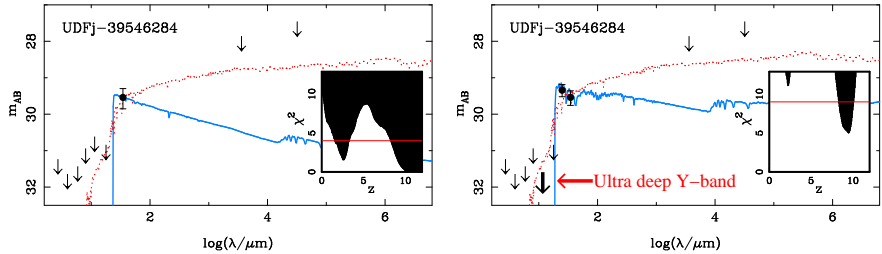}

\caption{Recovering a reliable galaxy population at $z\simeq 9-10$ requires {\it ultra-deep} exposures across the 
Lyman-break and a strategically-chosen deployment of at least {\it two} WFC3/IR filters for source detection. 
(Left): The marginal nature of the $z\simeq 10$ candidate claimed by Bouwens et al. (2011a) based on a sole 
$H_{160}$ detection. In addition to a possible high $z$ SED (blue line), an acceptable solution also 
exists at $z\simeq 2$ (red dotted line). The inset shows $\chi^2$ as a function of redshift. 
Possible `flux-boosting' at $H_{160}$ is an additional concern. (Right): Deeper $Y_{105}$ imaging (as planned in 2012), coupled with 
the security of two detections ($J_{140}$ \& $H_{160}$) above the Lyman break, should allow secure identification of this 
source and the elimination the low-$z$ solution if 
this galaxy really does lie at extreme redshifts $z > 9$ (simulated $J_{140}$ photometry was here inserted assuming $z_{true} \simeq 9.5$).  
The planned ultra-deep WFC3/IR UDF12 imaging of the HUDF in {\it HST} Cycle 19 
may detect up to $\simeq 20$ sources beyond $z \simeq 8.5$ to $H_{160} = 29.5$.}
\label{fig:6}   
\end{figure}

Nevertheless, spectroscopic follow-up is being vigorously pursued (e.g. Schenker et al.
2012). It is to be hoped that the new, wider area WFC3/IR surveys yield more bright
$z \sim 7 - 8$ LBGs which are amenable to spectroscopic follow-up, as effective 
ground-based near-infrared spectroscopy of the most distant galaxies revealed 
via the HUDF09 imaging at $m_{AB} \simeq 28.5$ has, unsurprisingly, 
proved extremely challenging. In particular, while Lehnert et al. (2010) reported
a spectroscopic redshift $z \simeq 8.55$ for the most distant credible 
HUDF LBG discovered by McLure et al. (2010) and Bouwens et al. (2010b), this observation 
took 15 hours of integration
with the near-infrared spectrograph SINFONI on the VLT, and the claimed marginal detection of Lyman-$\alpha$ 
has {\it not} been confirmed by independent follow-up spectroscopy (Bunker et al., in prep). Indeed, 
as discussed
further below, follow-up spectroscopy of even the brighter $z \simeq 7$ LBG candidates 
selected from ground-based surveys has, to date, not been particularly productive, 
for reasons that are still a matter of some debate (see section 4.3). 
But it must be noted that the 
current lack of spectroscopic redshifts should {\it not} be 
taken as implying that most of the 
$z \simeq 7$ and $z \simeq 8$ are not robust, as given sufficiently deep 
photometry all potential contaminants can be excluded, and a redshift estimated
accurate to $\delta z \simeq \pm 0.1$. In fact, it may well 
be the case that, by $z \simeq 7$, many galaxies do not produce measurable Lyman-$\alpha$
emission, and much of the current ongoing spectroscopic effort is really directed  
at trying to better quantify the evolution of Lyman-$\alpha$ emission from 
LBGs, a measurement which has the potential to shed light on the physics of reionization
(see sections 4.3 and 6.2).

From the ground, LBG selection has now been pursued with some success right up to (but not significantly 
beyond) $z \simeq 7$, due to the advent of deep $Y$-band imaging on both Subaru/Suprime-Cam (Ouchi et al. 2009b) and 
Hawk-I on the VLT (Castellano et al. 2010a,b). From the deep $Y$-band and $z'$-band imaging of both the Subaru Deep Field (SDF) and 
GOODS-North, Ouchi et al. (2009) reported 22 $z'$-drops to a depth of 
$y = 25.5-26$ over a combined area of $\simeq 0.4$\,deg$^2$, 
but the lack of 
comparably-deep near-infrared data at longer wavelengths forced them to make major corrections (by about 
a factor $\simeq 2$) for contamination. 
Nevertheless, three of these LBGs now have spectroscopically-confirmed redshifts
at $z \sim 7$ based on Lyman-$\alpha$ emission-line detections with the DEIMOS 
spectrograph on the Keck telescope (Ono et al. 2012). The VLT Hawk-I imaging undertaken by Castellano et 
al. (2010a,b) covered a smaller area ($\simeq 200$\,arcmin$^2$), but 
to somewhat deeper depths, and has yielded
$\simeq~20$ $z$-drops to $Y~\simeq~26.5$. 
Spectroscopic follow-up of this sample with FORS2 on the VLT has now provided
five Lyman-$\alpha$ spectroscopic redshifts in the range $6.7~<~z~<~7.1$
(Fontana et al. 2010; Vanzella et al. 2011; Pentericci et al. 2011).

Given the current concerns over the validity of the Lehnert et al. (2010) 
redshift, at the time of writing the robust 
{\it spectroscopic} redshift record for
an LBG (or indeed any galaxy or quasar) stands at 
$z = 7.213$ (Ono et al. 2012). 
Further spectroscopic follow-up
at $z \simeq 7$ is, of course, in progress, but 
wide-area ground-based exploration of the bright end of the 
LBG luminosity function at even higher redshifts must await 
deeper $Y$, $J$, $H$, $K$-band imaging (now underway 
with UltraVISTA; see section 7).

\subsubsection{Contaminants and controversies}

Spectroscopic follow-up (or improved multi-frequency photometry) of 
LBG samples has revealed, not unexpectedly, that three different types of interloper
can contaminate samples of LBGs at $z > 5$.

The first class of contaminant comprises very red dusty galaxies, or AGN, at lower redshifts. Such objects 
can produce a rapid drop in flux density over a relatively short wavelength range, which can 
sometimes be so severe as to be mistaken for a Lyman-break, especially if the two 
filters designed to straddle the break are actually not immediately adjacent in 
wavelength (e.g. $z'$ and $J$). Because such red dusty objects do not rapidly turn over to produce 
very blue colours at longer wavelengths, LBG sample contamination by such objects 
is not too serious an issue provided {\bf i)} a sufficently strong Lyman-break 
criterion is enforced (unfortunately not always the case), {\bf ii)} sufficiently deep multi-band imaging is available
at longer wavelengths to properly 
establish the longer wavelength SED slope, and {\bf 
iii)} LBG selection is confined to young, unreddened, reasonably-blue 
galaxies. However, if, as 
attempted by Mobasher et al. (2005), one seeks to select more evolved objects without very 
blue slopes longward of the proposed Lyman-break, then things can become difficult. As shown 
by Dunlop et al. (2007), very dusty objects at $z \simeq 1.5 - 2.5$ can easily be mistaken for evolved, 
high-mass LBGs at $z \simeq 5-6$, and templates with reddening as extreme as $A_V > 6$ 
sometimes need to be considered to reveal the alternative low-redshift solution. Such reddening is extreme,
but the point is that ``dropout'' selection specifically
designed to find LBGs at $z > 5$ 
transpires to also be an excellent 
method for selecting the rare, most extremely-reddened objects in the field at the appropriate lower 
redshifts. As discussed by Dunlop et al. (2007), often {\it Spitzer} MIPS 24\,$\mu$m detections can 
help to reveal 
low-redshift dust-enshrouded interlopers, but this experience illustrates how difficult 
it will be to robustly uncover any significantly-evolved or reddened galaxies at $z > 5$. 

\begin{figure}
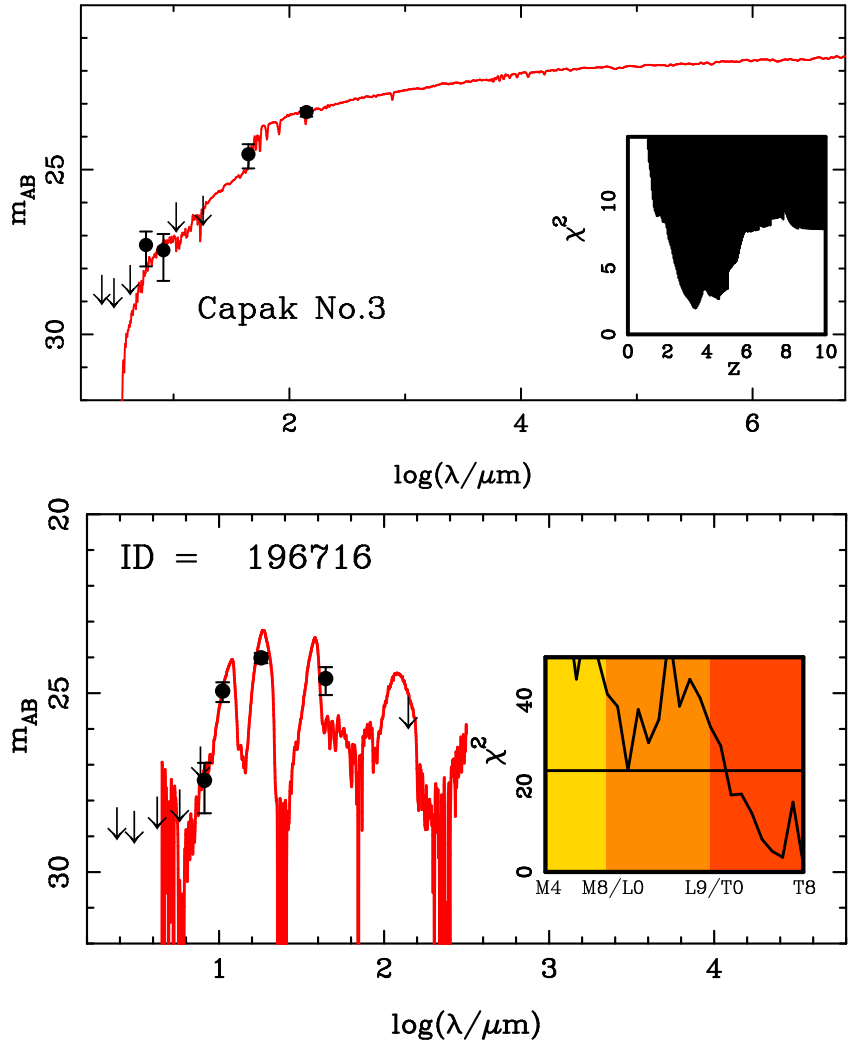


\includegraphics[scale=0.69, angle=270]{fig7a.eps}
\includegraphics[scale=0.75, angle=270]{fig7b.eps}
\caption{Examples of two different types of interlopers which can 
contaminate LBG samples, especially those selected at brighter magnitudes 
from ground-based imaging. The upper panel shows the best 
SED fit and $\chi^2$ versus $z$ for a COSMOS 
galaxy claimed by Capak et al. (2011b) 
to meet the standard LBG selection criterion at $z \simeq 7$, and to be tentatively 
confirmed by near-infrared spectroscopy at $z = 7.69$. In fact, with the 
improved near-infrared photometry provided by the 
UltraVISTA survey it is clear that no acceptable 
redshift solution for this galaxy exists at $z > 5$ and that the original 
object selection was based on inadequate photometric error analysis
(Bowler et al. 2012). The 
true best-fitting model solution corresponds to a moderately dusty galaxy at
$z \simeq 3.5$. 
The lower panel shows another object selected from the UltraVISTA 
imaging in the COSMOS field which really does meet the standard $z'-J$:$J-H$ 
colour criterion for a $z = 7$ galaxy, but which as shown here is in fact 
a T-dwarf galactic star. In this case the photometry 
(especially in the crucial $Y$-band) is of high enough quality
that no acceptable solution could be found with a galaxy SED at any redshift, 
but this is not always the case. Fortunately, both these types of contaminant 
become (at least statistically) less of a problem for $z \simeq 7$ LBG surveys 
at fainter magnitudes, because the most dusty galaxies at $z \simeq 2-4$ 
tend to be high-mass objects, and because the number counts of cool dwarf
stars fall (or at least certainly plateau) beyond $J=24$ (Ryan et al.
2011) due to the scale-height of the galactic plane.}
\label{fig:7}       % Give a unique label
\end{figure}

This confusion lies behind several dubious/erroneous claims of extreme-redshift
galaxies in the literature. Examples include not only the supposed $z \simeq 6$ ultra-massive galaxy uncovered
by Mobasher et al. (2005) in the HUDF, but also the claimed discovery of a bright $z \simeq 9$ 
galaxy reported by Henry et al. (2008) (subsequently retracted when deeper optical imaging revealed 
a significant detection in the $i'$-band;  Henry et al. 2009).
It is also the likely reason that most of the $z > 6.5$ 
galaxies tentatively uncovered by Hickey et al. (2010) from the 
VLT Hawk-I $Y$-band imaging of GOODS-South have proved to be false
(in the light of the subsequent ERS+CANDELS WFC3/IR imaging of the field) and,
as illustrated in Fig. 7, is
part of the explanation (in combination with inadequate photometric 
error analysis) for recent claims of very bright $z \simeq 7$ galaxies in the COSMOS field 
(despite supposed ``tentative'' spectroscopic confirmation at $z \simeq 7$ for two objects; Capak et al. 2011b).
Fortunately, continuity arguments indicate that this may become less of a problem when attempting to select LBGs 
at the highest redshifts and faintest magnitude limits, as the reddened lower-redshift interloper population
seems to become (relatively) much less prevalent in this region of parameter space.

The second class of contaminant comprises cool galactic stars, specifically M, L and 
(in the case of $z \simeq 7$ LBG selection) T dwarfs.
This is a long-established problem in the colour-selection of high-redshift 
radio-quiet quasars which are unresolved in all but the very deepest images 
(Hewett et al. 2006). However, as discussed by many authors (e.g. 
McLure et al. 2006; Stanway et al. 2008a,b; Vanzella et al. 2009; Hickey et al. 2010) 
the compactness of high-redshift galaxies 
(see section 5.4) means that contamination by cool dwarf stars has also become an 
important issue 
in the search for high-redshift LBGs. The problem is most acute for ground-based surveys both because 
most $z > 5$ LBGs are unresolved with even good ground-based seeing, and because the brighter LBGs
are so much rarer on the sky (McLure et al. 2009; Capak et al. 2011b) than the fainter more numerous population
revealed by the deeper {\it HST} imaging. 

The particular problem of T-dwarf contamination of $z \simeq 7$ LBG searches has arguably 
been under-estimated until very recently, in part because our knowledge of T dwarfs has 
evolved in tandem with LBG searches over the last decade (Knapp et al. 2004; Chiu et al. 
2008; Burningham et al. 2010). Specifically, early $z \simeq 7$ LBG ``dropout'' criteria 
appear to have assumed that T dwarfs did not display colours redder than $z-J \simeq 1.8$ 
(e.g. Bouwens et al. 2004), but cooler dwarfs have since 
been found with $z - J > 2.5$ (e.g. Burningham et al. 2008, 2010; 
Delorme et al. 2008; Leggett et al. 2009; Lucas et al. 2010;
Liu et al. 2011). 
For ground-based $z \simeq 7$ LBG searches, the key to excluding 
dwarf-star contamination 
lies in having sufficiently-accurate multi-band 
infrared photometry since, for example, 
T-dwarfs have redder $Y-J$ colours (by $\simeq 1$\,mag) than 
genuine $z \simeq 7$ LBGs (and different IRAC colours; Stanway et al. 2008a).
This is a further argument in favour of multi-band SED fitting which, given $Y,J,H,K$ 
and IRAC photometry can often reveal a stellar contaminant on the basis of failure 
to achieve an acceptable fit with any galaxy template (as 
shown in the lower panel of Fig. 7). Given the above-mentioned 
high level of spectroscopic completeness achieved by Curtis-Lake et al. (2012) and Jiang et al. 
(2011) (and the results of stacking analyses; McLure et al. 2006, 2009) 
it seems unlikely that the published ground-based $z \simeq 6$ LBG samples 
are seriously contaminated by dwarf stars, but the situation remains 
more confused for bright surveys at $z \simeq 7$.

Fortunately, due to the combination of image depth, small field-of-view, 
and high angular resolution, T-dwarf contamination of the $z \simeq 7$ 
LBG samples revealed by the new deep WFC3/IR imaging is expected to be extremely small.
This is confirmed by considering that the typical
absolute $J-$band (AB) magnitude of T-dwarf stars is $J\simeq 19$
(Leggett et al. 2009). At the depths probed by the WFC3/IR imaging of the 
HUDF, a T-dwarf contaminant would
thus have to be located at a distance of $0.5-1.0$\,kpc. Given this
distance is $2 \rightarrow 3$ times the estimated galaxy thin disk
scale-height of $\simeq 300$\,pc (e.g. Reid \& Majewski 1993; Pirzkal
et al. 2009), it is clear that significant contamination is unlikely.
This is not to suggest that dwarf stars cannot be found at such
distances as, for example, Stanway et al. (2008a) report the discovery
of M dwarfs out to distances of $\simeq 10$\,kpc. However, 
the surface density is low, with the integrated
surface density over all M-dwarf types contained within $\simeq
1$\,kpc amounting to $\simeq 0.07$\,arcmin$^{-2}$.  Extrapolating
these results to T dwarfs is somewhat uncertain, but a comparable
surface density for L and T-dwarf stars is supported by the search for
such stars in deep fields undertaken by Ryan et al. (2005, 2011). The
results of this work suggest that the 4.5-arcmin$^2$ field-of-view
of WFC3/IR data should contain $\leq 0.5$ T-dwarf
stars down to a magnitude limit of $z_{850} = 29$.

The final class of contaminant, as revealed by the lower-redshift secondary 
solutions in SED-based redshift estimation (McLure et al. 2010; Finkelstein et al. 2010) 
consists of fairly blue, $\simeq 0.5$\,Gyr-old post-starburst galaxies which display 
a strong Balmer break. Given sufficient signal-to-noise there is really no room for confusion, as 
the Balmer break can never approach the strength of the anticipated Lyman-break at $z > 5$ 
(e.g. before it faded the $z = 8.2$ GRB displayed $Y-J > 4$). 
However, the SED-fits shown by McLure et al. (2010) 
demonstrate that, with inadequate photometric dynamic range, 
a Balmer break at $z \simeq 2$ can be mistaken for a Lyman-break 
at $z \simeq 8$. Fortunately the potential contaminants occupy a rather 
specific regime of parameter space (i.e. they must lie in a narrow redshift 
range, a narrow age range, be virtually dust-free, and have very low stellar masses
to be confused with $z \simeq 7-8$ LBGs selected at the faintest magnitudes) and continuity
arguments can be advanced that they are likely rare (e.g. Bouwens et al. 2011b), 
but the real lesson here is the importance of ensuring that any 
imaging shortward of any putative Lyman-break is sufficiently deep to 
exclude lower-redshift interlopers (not necessarily easy with the 
deepest WFC3/IR imaging, given the depth of the available complementary ACS optical imaging).

\subsection{Lyman-${\bf \alpha}$ selection}

The intrinsic Lyman-$\alpha$ emission from young galaxies is expected to be strong, reaching large 
rest-frame equivalent widths $EW_{rest} \simeq 200$\,\AA\ if driven by star formation (Charlot \& Fall 1993).
A star-formation rate of $SFR = 1\,{\rm M_{\odot}\,yr^{-1}}$ 
corresponds to a Lyman-$\alpha$ luminosity of $\simeq 1 
\times 10^{42}\,{\rm erg\,s^{-1}}$ (Kennicutt 1998).
 
However, for many years, blank-field searches for Lyman-$\alpha$ emitters (LAEs) at even moderate redshifts 
were disappointingly unsuccessful (e.g. Koo \& Kron 1980; Pritchet \& Hartwick 1990), raising fears that 
observable Lyman-$\alpha$ in high-redshift galaxies might be severely compromised by dust, because of the potentially 
long path lengths traversed by Lyman-$\alpha$ photons
through the interstellar medium due to resonant scattering (Charlot \& Fall 1991; 1993). 
However, as mentioned at the beginning of this Chapter, 
by the end of the 20th century a few $z > 5$ LAEs had been uncovered through the complementary 
techniques of long-slit spectroscopy (covering small areas but a broad redshift range) and narrow-band 
imaging (covering larger areas but a narrower redshift range). For a while these two techniques 
were competitive (Stern et al. 2000) but, with the advent of genuinely 
wide-field CCD imaging cameras on 8-m class telescopes, 
narrow-band searches for LAEs have surged ahead, and have proved spectacularly successful in uncovering
large samples of galaxies at $z > 5$ (e.g. Ouchi et al. 2005, 2008).

Modern narrow-band imaging searches are sensitive to Lyman-$\alpha$ 
rest-frame equivalent widths down to $EW_{rest} \simeq 15$\,\AA\ (helped at high redshift by the 
fact that $EW_{obs} = (1+z) EW_{rest}$) and limiting line flux-densities
$f \simeq 5 \times 10^{-18}\,{\rm erg\,s^{-1}\,cm^{-2}}$. At $z \simeq 7$ this corresponds to 
a Lyman-$\alpha$ luminosity $L \simeq 2.5 \times 10^{42}\,{\rm erg\,s^{-1}}$ which, in the absence of obscuration,
is equivalent to a star-formation rate $SFR \simeq 2\,{\rm M_{\odot}\,yr^{-1}}$. Thus, like the most sensitive 
LBG surveys at high redshifts, LAE selection can now 
detect galaxies at $z \simeq 7$ with a star-formation rate comparable to that 
of the Milky Way (e.g. Chomiuk \& Povich 2011).

The basic technique involves comparing
images taken through a narrow-band ($100-200$\,\AA\ wide) filter with a broad-band 
(or nearby narrow-band) image 
at comparable wavelengths. At very high redshifts, the efficiency of this approach is sensibly optimized 
by designing filters to image in low-background regions between the OH atmospheric emission lines which begin 
to plague substantial wavelength ranges beyond 
$\lambda_{obs} \simeq 7000$\,\AA. For this reason, 
samples of $z > 5$ LAEs are generally confined to pragmatically-selected redshift bands (Fig. 8).

Narrow-band searches for LAEs complement broad-band surveys for LBGs by probing a largely distinct region of 
parameter space. The weaknesses of narrow-band searches are that they probe 
smaller redshift ranges and hence smaller cosmological volumes
(for a given survey area), 
and obviously can only uncover that fraction of the galaxy population which actually 
displays relatively bright Lyman-$\alpha$ emission. They are also subject to severe contamination by 
emission-line galaxies at lower-redshifts, which can only be sorted out via follow-up spectroscopy, or 
additional broad-band (or further tuned narrow-band) imaging (see subsection 3.2.2). 
On the other hand, narrow-band imaging 
is sensitive to objects with much fainter continua than can be detected in LBG surveys, delivers targets 
for follow-up spectroscopy which are at least already known to contain an emission line, 
and is extremely effective at uncovering large-scale structures where many objects 
lie within a relatively narrow redshift band (e.g. Capak et al. 2011a).

As with LBG selection, a detailed overview of LAE studies at $z < 5$ is beyond the scope of this Chapter, 
but a helpful overview of this ``lower-redshift'' work is provided by Ouchi et al. (2003), who first used 
narrow-band imaging through the NB711 filter on Subaru to uncover substantial numbers of LAEs at 
$z \simeq 4.8$. The successful 
use of Lyman-$\alpha$ selection at $z > 5$ is now described in detail below.

 \begin{figure}

\includegraphics[scale=0.27, angle=0]{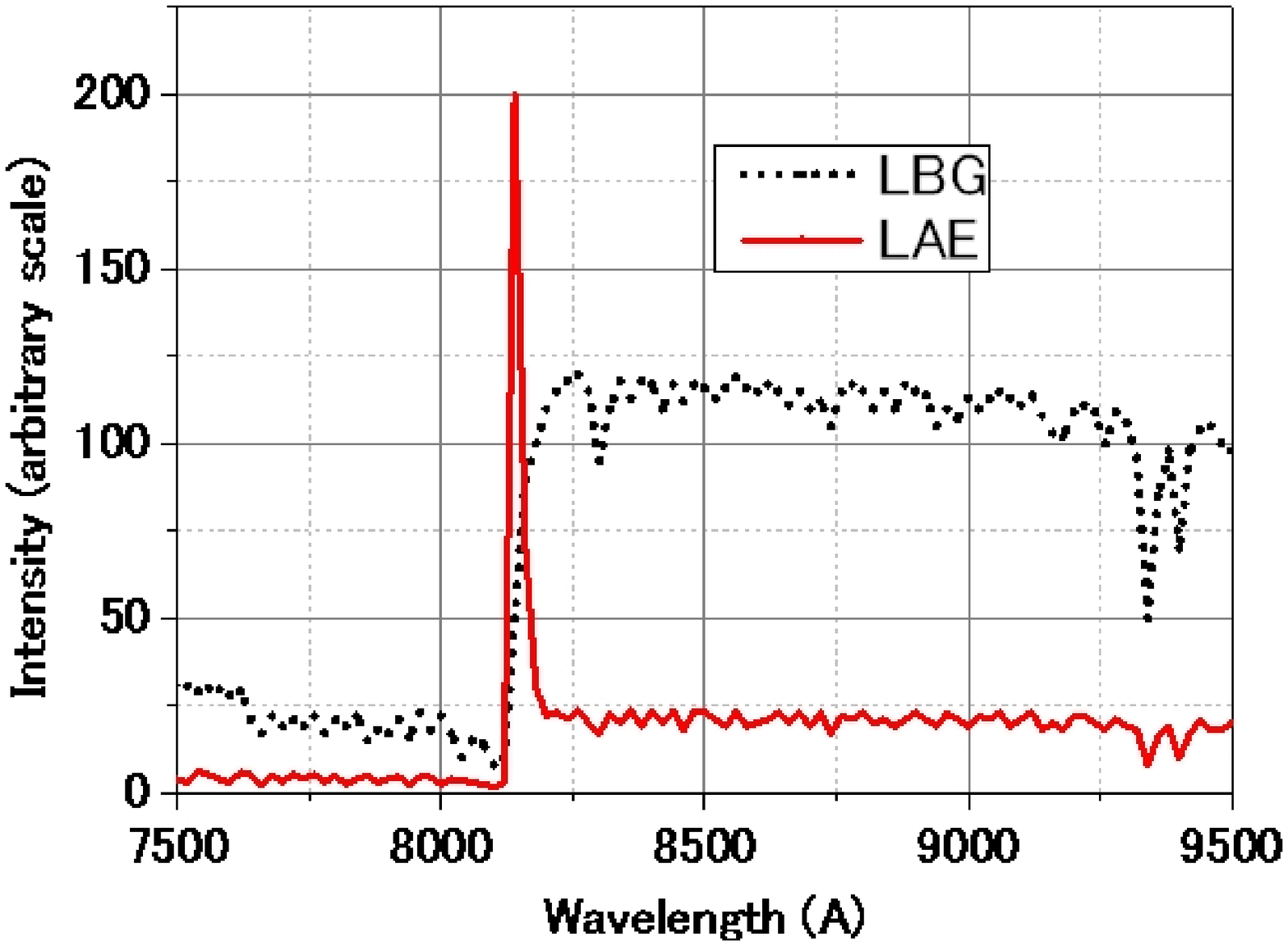}
\includegraphics[scale=0.31, angle=0]{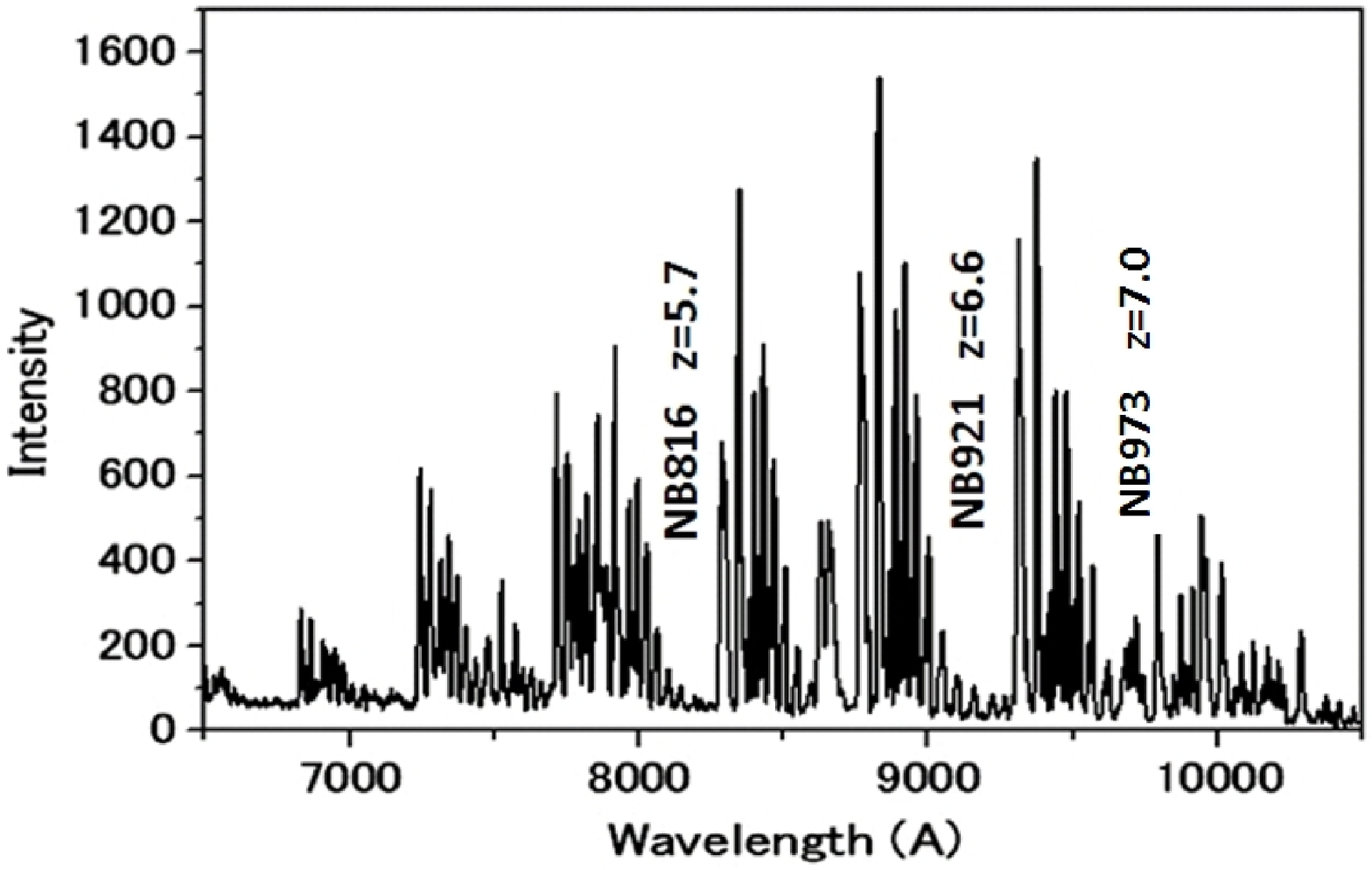}

\caption{The selection of high-redshift galaxies via Lyman-$\alpha$ emission.
The left-hand panel illustrates the typical spectrum of a Lyman-$\alpha$ emitter (solid line) 
compared with a Lyman-break galaxy (broken line) at an assumed redshift $z = 5.7$, 
showing the Lyman-$\alpha$ emission-line redshifted from $\lambda_e = 1216$\AA\ to $\lambda_{obs} = 8150$\AA\, 
and the stellar continuum long-ward of the Lyman-$\alpha$ emission line. The right-hand plot 
shows the OH night sky emission bands, highlighting the few gaps within which narrow-band filters can be most effectively 
targetted. The Subaru narrow-band filters whose transmission profiles are matched to these dark windows are used to 
detect LAEs at $z =5.7$ (NB816), $z=6.6$ (NB921) and $z=7.0$ (NB973), as discussed in section 3.2.1 (courtesy M. Iye).}
\label{fig:8}       % Give a unique label                                                                                                                             
\end{figure}

\subsubsection{Lyman-$\alpha$ galaxies at ${\bf z > 5}$}

After passing the $z = 5$ threshold in 1998, the redshift record 
for LAEs rapidly advanced beyond z = 6.5 (Hu et al. 2002; Rhoads et al. 2003, 2004), and 
indeed LAEs were to provide the most distant known objects for the rest of the decade.

Since 2004, the discovery of LAEs at $z > 5$ has been largely driven by 
narrow-band imaging with the wide-field optical camera Suprime-Cam on the 
Subaru telescope, coupled with follow-up spectroscopy with the 
FOCAS spectrograph on Subaru, and the LRIS and DEIMOS spectrographs
on Keck. A consortium of Subaru astronomers developed the required 
series of narrow-band filters at ever increasing wavelengths. As shown in Fig. 8, the band-passes of these filters
are designed to fit within the most prominent dark gaps between the bands of strong telluric OH emission which 
come to increasingly-dominate the night-sky spectrum at $\lambda_{obs} > 7000$\,\AA. 

A filter at 8160\AA\ (NB816) is able to target Lyman-$\alpha$ emission at $z \simeq 5.7$. 
This was used by Ouchi et al. (2005) to produce a very large sample of $\simeq 500$ $z \simeq 5.7$ LAEs
from imaging of the Subaru XMM-Newton Deep Survey field (SXDS; Furusawa et al. 2008) 
and by Shimasaku et al. (2006) to produce another large and independent sample of 
$z \simeq 5.7$ LAEs from imaging of the Subaru Deep Field (SDF; Kashikawa et al. 2004). 
The NB816 filter was also used by Ajiki et al. (2006) to image both GOODS fields,
and a fourth sample of NB816-selected LAEs was uncovered in the COSMOS field by Murayama et al. (2007).

Imaging of these survey fields through another, redder filter (NB921) 
led to the first substantial samples of potential LAEs at $z \simeq 6.6$ 
(Taniguchi et al. 2005; Kashikawa et al. 2006; Ouchi et al. 2010), 
and imaging of the SDF through the even redder NB973 filter yielded what 
remains to this day the most distant narrow-band selected galaxy. 
This LAE, IOK-1, was spectroscopically confirmed at $z = 6.96$ by Iye et al. (2006) 
and was, for four years, the most distant object known. The discovery image and 
spectrum of IOK-1 is shown in Fig. 9; the spectrum clearly shows the asymmetric 
emission-line profile which is characteristic of Lyman-$\alpha$ emission at 
extreme redshift (produced by neutral Hydrogen absorption of the blue wing of the 
emission line; Hu et al. 2010) and helps to enable single-line spectroscopic 
confirmation of narrow-band selected LAE candidates at these high redshifts (see 
below for potential contaminants). 

\begin{figure}

\includegraphics[scale=0.28, angle=0]{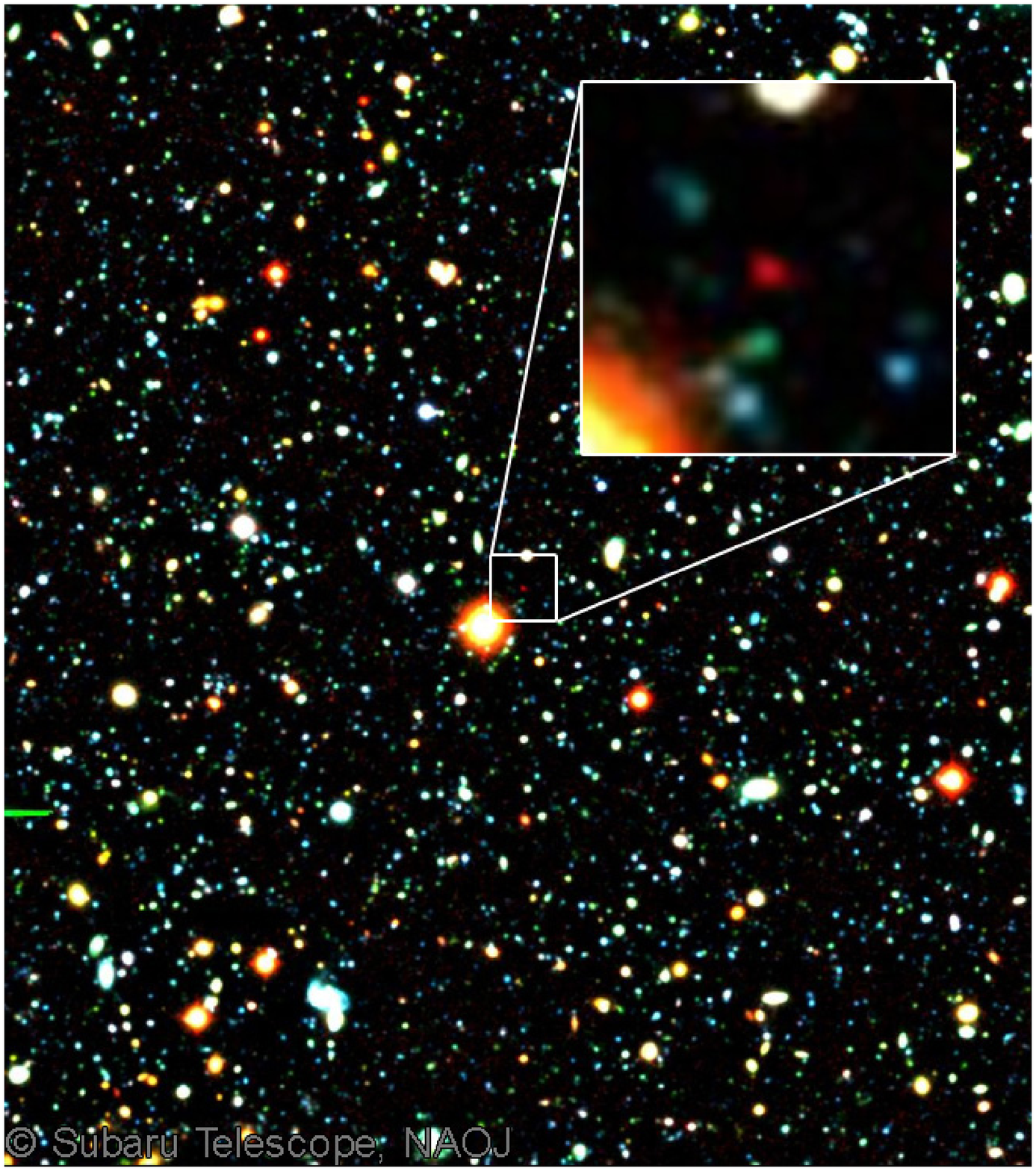}
\hspace*{0.1in}
\includegraphics[scale=0.297, angle=0]{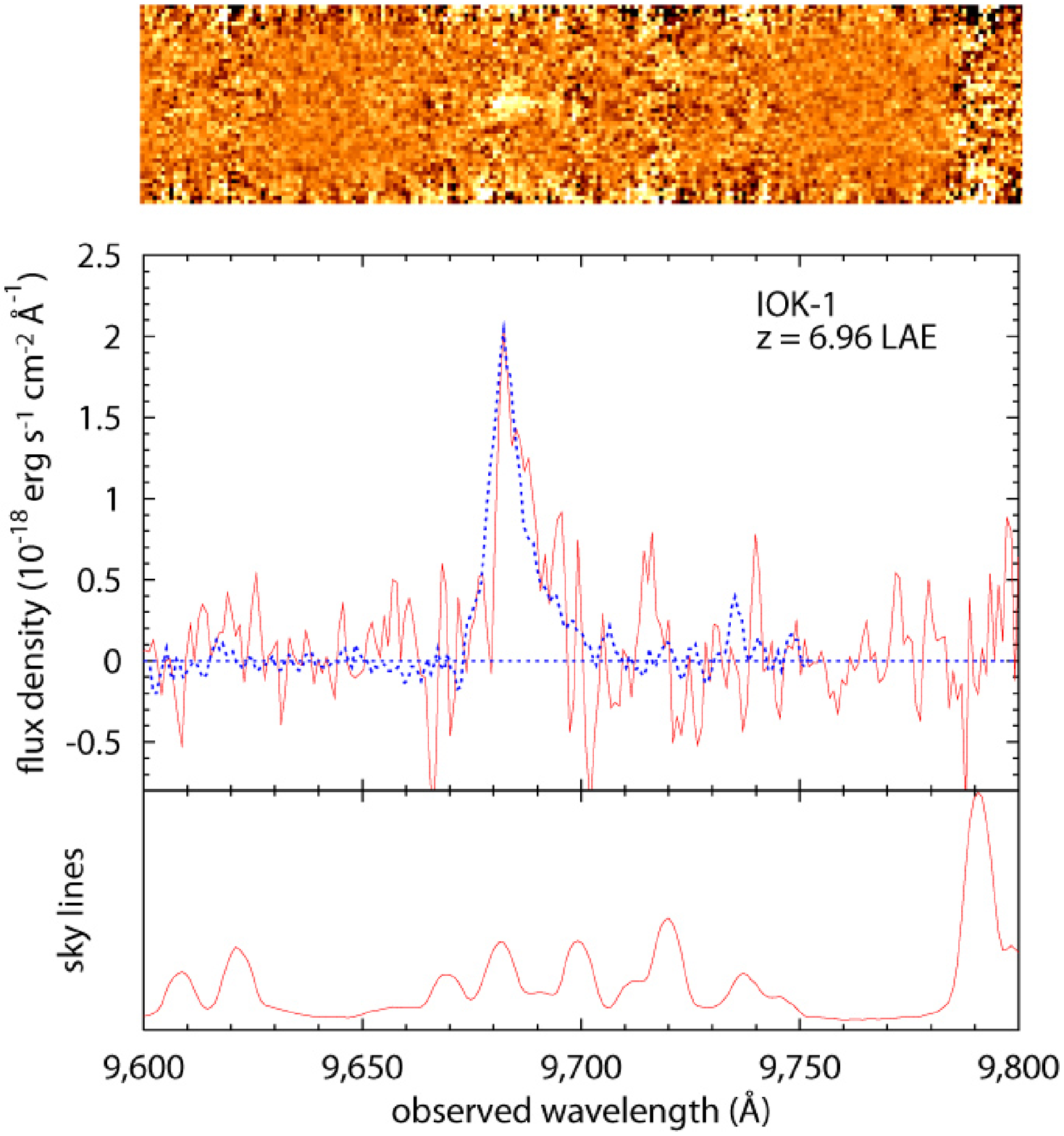}

\caption{The most distant spectroscopically-confirmed 
LAE selected via narrow-band imaging, the galaxy IOK-1,
is shown as a red blob in the colour postage-stamp insert image which covers $8\times8$ arcsec.  
The entire field of view shown in the larger image 
covers $254\times284$ arcsec (North is up and East to the left). The 2-dimensional and 1-dimensional
Subaru FOCAS spectrum of IOK-1 is shown in the right-hand panels (Iye et al. 2006). The spectrum clearly shows 
an asymmetric Lyman-$\alpha$ emission line at a wavelength corresponding to a redshift $z \simeq 6.96$
(courtesy M. Iye).}
\label{fig:9}       % Give a unique label
\end{figure}

The NB973 filter has now been used on Subaru 
to provide a few more candidate LAEs 
at $z \simeq 7$ (Ota et al. 2008, 2010a). Most recently, 
following refurbishment of Suprime-Cam with new red-sensitive CCDs, the 
NB1006 filter has been installed to allow searches for LAEs at $z \simeq 7.3$ (Iye 2008).

Complementary deeper (but smaller-area) narrow-band searches for LAEs at $z > 7$ have recently 
been conducted on the VLT, but have not yet yielded any spectroscopically-confirmed candidates
(Cuby et al., 2007; Cl\'{e}ment et al. 2012). 
As discussed above in the context of the spectroscopic follow-up 
of the highest-redshift LBGs, there of course 
exists the interesting possibility that Lyman-$\alpha$ 
emission may not be so easily produced by many galaxies as we enter the epoch of reionization
(see section 4.3.6). This issue may soon be clarified by further deeper narrow-band 
imaging searches in the near-infrared. Finally, it is probably fair to say that existing 
attempts to uncover extreme redshift LAEs up to $z \simeq 10$ 
via long-slit infrared spectroscopy targetted
on the critical lines in strong-lensing clusters remain controversial (Stark et al. 2007b).

\subsubsection{Potential contaminants}

It must be emphasized that narrow-band selected LAE candidates at $z > 5$ need
to be confirmed with spectroscopy because the vast majority of objects with a narrow-band 
excess will be contaminants. Many of these are genuine emission-line objects
(galaxies or AGN) at lower 
redshifts, with the narrow-band excess being produced by, for example, CIV emission
at 1549\,\AA, MgII at 2798\,\AA, [OII] at 3727\,\AA, [OIII] at 5007\,\AA, or H-$\alpha$ 
at 6563\,\AA. Isolation of genuine extreme-redshift LAEs is of course helped by 
the fact that, like LBGs, they should show essentially no emission at wavelengths 
shortward of $\lambda_{rest} = 1216$\,\AA. Thus, broad-band imaging at bluer wavelengths
can be used to reject many low-redshift objects without recourse to spectroscopy.
A second alternative to spectroscopy as a means to rule out at least some sub-samples 
of lower-redshift emission-line objects is observation through a second narrow-band filter
at a wavelength specifically designed to pick up a second emission line (e.g. Sobral et al. 2012). However,
this is rarely practical, and at least multi-object spectroscopy is reasonably efficient when 
targetting a subset of objects which are already known to likely display detectable emission lines.

Another potential source of LAE sample contamination is transient objects (e.g. variable AGN or 
supernovae) because often the narrow-band image is compared with a broad-band image which was taken 
one or two years earlier. Finally, the sheer size of the images means that rare, apparently
significant (5-$\sigma$) noise peaks can occur in a single narrow-band image, and these need 
to be excluded by either repeated imaging or spectroscopy (this is the same single-band 
statistical detection problem which can afflict searches for extreme-redshift LBGs 
in the longest-wavelength broad-band filter; Bouwens et al. 2011a).

\section{Luminosity Functions}
The evolving luminosity function is generally regarded as the best way to 
summarize the changing demographics of high-redshift galaxies. It is defined 
as the number of objects per unit comoving volume per unit
luminosity, and the data are most 
often fitted to a Schechter function (Schechter 1976):

\begin{equation}
\frac{dn}{dL} = \phi(L) = \left(\frac{\phi^*}{L^*}\right)\left(\frac{L}{L^*}\right)^{\alpha} e^{-(L/L^*)}
\end{equation}

\noindent
where $\phi^*$ is the normalization density, $L^*$ is a characteristic luminosity, 
and $\alpha$ is the power-law slope at low luminosity, $L$. The faint-end 
slope, $\alpha$, is usually negative ($\alpha \simeq -1.3$
in the local Universe; e.g. Hammer et al. 2012) 
implying large numbers of faint galaxies.

In the high-redshift galaxy literature, 
the UV continuum luminosity function is usually presented in units 
of per absolute magnitude, $M$, rather than luminosity $L$, in which case, making the 
substitutions $\phi(M) dM = \phi (L) d(-L)$ and $M-M^* = -2.5 \log{(L/L^*)}$, 
the Schechter function becomes

\begin{equation}
\phi(M) = \frac{\ln 10}{2.5} \phi^* \left(10^{0.4(M^*-M)}\right)^{(\alpha+1)} \exp \left[-10^{0.4(M^*-M)}\right]
\end{equation} 

\noindent
and this function is usually plotted in log space (i.e. $\log\left[\phi(M)\right]$ vs. $M$).

The Schechter function can be regarded as simply one way of describing the 
basic shape of any luminosity function which displays a steepening above 
a characteristic luminosity $L^*$ (or below a characteristic 
absolute magnitude $M^*$). 
Alternative functions, such as a double power-law can often 
also be fitted, and traditionally have been used 
in studies of the luminosity 
function of radio galaxies and quasars (e.g. Dunlop \& Peacock 1990). 
Given good enough data, especially extending to the very faintest 
luminosities, such simple parameterizations of the luminosity function 
are expected to fail, but the Schechter function is more than adequate 
to describe the data currently available for galaxies at $z \ge 5$.
A recent and thorough
overview of the range of approaches to determining and fitting luminosity 
functions, and the issues involved, can be found in Johnston (2011).

At the redshifts of interest here, the luminosity functions derived 
from optical to near-infrared observations are rest-frame ultraviolet 
luminosity functions. Continuum luminosity functions for LBGs 
are generally defined at $\lambda_{rest} \simeq 1500$\,\AA\,
or $\lambda_{rest} \simeq 1600$\,\AA, while the luminosity functions derived for LAEs 
involve the integrated luminosity of the Lyman-$\alpha$ line. Because 
of the sparcity of the data at the highest redshifts, and the typical 
redshift accuracy of LBG selection, the evolution of the luminosity function
is usually described in unit redshift intervals, although careful simulation
work is required to calculate the volumes actually 
sampled by the filter-dependent 
selection techniques used to select LBGs and LAEs. 
Detailed simulations (involving input luminosities and sizes) 
are also required to estimate incompleteness corrections when 
the survey data are pushed towards the detection limit, and the form of these 
simulations can have a significant effect on the shape of the derived 
luminosity functions, especially at the faint end (as discussed
by, for example, Grazian et al. 2011).

Different reported Schechter-function fits can sometimes exaggerate 
the discrepancies between the basic data gathered by different research groups. 
In particular, without good statistics and dynamic range, there can be severe 
degeneracies between $\phi^*$, $L^*$ and $\alpha$, and 
very different values can be deduced for these parameters even when the basic
statistics (e.g. integrated number of galaxies above the flux-density limit) 
are not very different (e.g. Robertson 2010).

This is an important point, because the luminosity-integral of the evolving 
luminosity function 

\begin{equation}
j(L) = \int_{L}^{\infty} L \phi(L) dL = \phi^* L^* \Gamma(2 + \alpha, L/L^*)
\end{equation}

\noindent
(where $\Gamma$ is the incomplete gamma function) is often 
used to estimate the evolution of average 
{\it luminosity density} as a function
of redshift (from which the cosmic history of star-formation density
and ionizing photons can be inferred; e.g. Robertson
et al. 2010). For this reason care must be taken not to over-interpret 
the implications of extrapolating the fitted function (e.g. Su et al. 2011), 
especially when, as
appears to be the case at very high redshift (see below), the faint-end 
slope is very steep. 
Formally, luminosity density diverges for $\alpha < -2$ if the 
luminosity function is integrated to zero, but in practice the integral needs 
to be terminated at some appropriate faint luminosity (see section 6.1). The key point 
is that, for any steep faint-end slope even approaching $\alpha \simeq -2$,
the value of the integral depends critically on $\alpha$ and 
the adopted value of the faint-end luminosity cutoff (which, 
for obvious reasons, is still a matter of debate and 
could be a function of environment; Hammer et al. 2012).

At the bright end of the luminosity 
function the problem is generally 
not completeness but small-number statistics,
and authors are often tempted to push their survey to produce a derived 
value for a brightest luminosity bin which depends on only a 
handful of objects. Given the small numbers, contamination by even rare 
populations (such as the brown dwarf stars discussed in section 3.1.3) 
can often 
be a problem at the bright end. An additional issue for a steeply-falling 
luminosity function is correcting for ``Eddington bias'', which 
tends to boost apparent average luminosity
at the bright end. This again requires careful simulation
to achieve a consistent solution.
Finally it must be remembered that all luminosity functions are afflicted 
to some extent by cosmic variance (Sommerville et al. 2004), 
and ultimately high-redshift surveys need 
to cover sufficient area (helped by covering independent lines of sight) 
to offer a representative picture of the galaxy population at each epoch.

The comoving cosmological volumes sampled by various example LBG and LAE surveys at $z \simeq 6$ and
$z \simeq 7$ are given for convenient comparison in Table 1.

\begin{table}
\begin{center}
\caption{Example comoving cosmological volumes sampled by different types and scales of 
high-redshift galaxy surveys at $z \simeq 6$ and $z \simeq 7$}.
\label{tab:1}       % Give a unique label

\begin{tabular}{llcrl}

Survey Type & Redshift range & Area & Volume/Mpc$^3$ & Example Reference\\
\hline
LBG WFCAM/VISTA & $z = 5.5 - 6.5$ & 1\,deg$^2$ & 10,000,000\phantom{00}&{\it McLure et al. (2009)}\\
LAE Suprime-Cam(x4)& $z = 5.7 \pm 0.05$ & 1\,deg$^2$ & 1,000,000\phantom{00}&{\it Ouchi et al. (2008)}\\
\\
LBG Suprime-Cam & $z = 6.5 - 7.1$ & 0.25\,deg$^2$ & 1,000,000\phantom{00}&{\it Ouchi et al. (2009)}\\
LAE Suprime-Cam & $z = 6.6 \pm 0.05$ & 0.25\,deg$^2$ & 200,000\phantom{00}&{\it Kashikawa et al. (2011)}\\
\\
LBG HUDF/WFC3  & $z = 6.5 - 7.5$ & 4.5\,arcmin$^2$ & 10,000\phantom{00}&{\it Oesch et al. (2010a)}\\
LBG CANDELS/WFC3\phantom{00} & $z = 6.5 - 7.5$ & 0.2\,deg$^2$& 1,500,000\phantom{00}&{\it Grogin et al. (2011)}\\
\hline
\end{tabular}
\end{center}
\end{table}

\subsection{High-redshift evolution of the LBG luminosity function}

The last $\simeq 5$ years have seen a rolling series of papers on the 
LBG UV luminosity function at $z > 5$, based purely on the ever-improving {\it HST} ACS, 
NICMOS, and now WFC3/IR deep imaging datasets (Bouwens et al. 2006, 2007, 2008, 2011b; Oesch
et al. 2007, 2010a; Trenti et al. 2010)

In a complementary effort, McLure et al. (2009) focussed on determining the bright-end 
of the LBG luminosity function at $z \simeq 5$ and $z \simeq 6$ from ground-based 
data, before extending this work to $z \simeq 7$ and $z \simeq 8$ with WFC3/IR (McLure 
et al. 2010). In addition, the ground-based determination of the LBG luminosity function  
has recently been pushed out to $z \simeq 7$ by Ouchi et al. (2009) and Castellano et al.
(2010a,b).

In general, the results of these various studies are in very good agreement. Specifically,  
McLure et al. (2009) combined their ground-based data 
on bright LBGs with the Bouwens et al. (2007) data on fainter {\it HST}-selected LBGs to 
determine the form of the UV luminosity function at $z \simeq 5$ and $z \simeq 6$, and 
derived Schechter-function parameter values in excellent agreement with Bouwens et al. 
(2007). The form and evolution of the LBG luminosity function deduced from this work is shown 
in the left-hand panel of Fig. 10, including the McLure et al. (2010) extension to $z \simeq 7$ and 
$z \simeq 8$. The simplest way to summarize these results is that the available data are 
consistent with $\alpha = -1.7$ over the full redshift range $z \simeq 5 - 7$,
and that the characteristic luminosity declines by a factor of two 
from $z \simeq 5$ ($M^* \simeq -20.7$) to $z \simeq 6$  ($M^* \simeq -20.0$) (as
always one must caution this does not necessarily imply pure luminosity evolution 
of individual objects; see, for example, Stark et al. 2009).

\begin{figure}

\includegraphics[scale=0.455, angle=0]{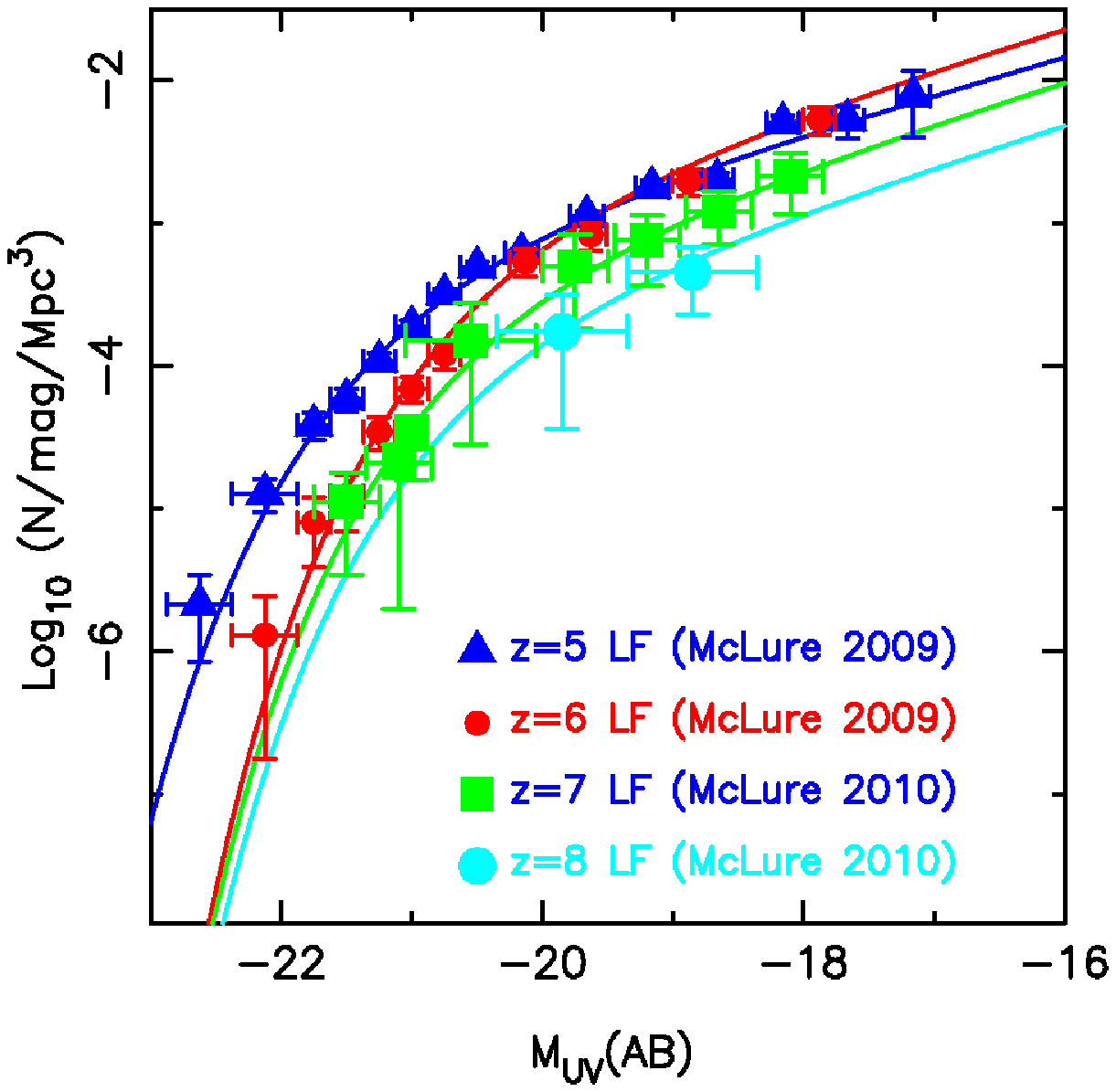}
\hspace*{0.5cm}
\includegraphics[scale=0.44, angle=0]{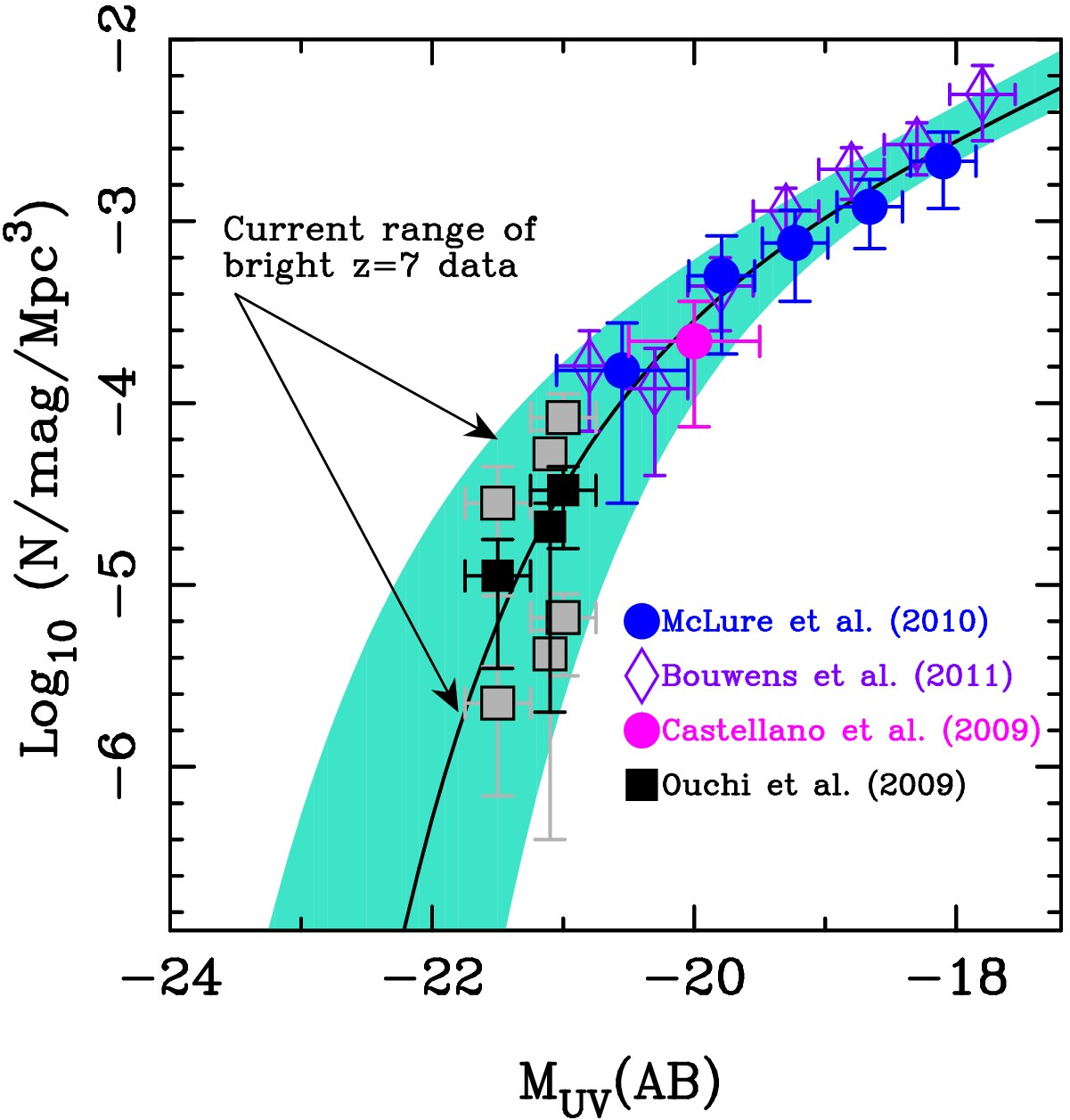}
\caption{The UV continuum LF of LBGs, and its high-redshift evolution.
The left-hand panel shows the $z \simeq 5$ and $z \simeq 6$ LFs determined
by McLure et al. (2009), along with the LFs at $z \simeq 7$ and $z \simeq 8$ 
determined by  McLure et al. (2010) from the recent {\it HST} WFC3 imaging
(the results obtained from a similar analysis by Bouwens et al. 2011b 
are summarized in Fig. 11). The right-hand panel demonstrates the extent of 
general agreement over the form and normalization of the UV LF at $z \simeq 7$
derived by different groups using both ground-based (Subaru \& VLT) and 
{\it HST} WFC3/IR data. While the overall 
level of agreement is impressive, this plot also shows current tension 
over the true value of the faint-end slope, and the lack of information at the very bright end of the LF (courtesy R. McLure).}
\label{fig:10}       % Give a unique label
\end{figure}

From $z \simeq 6$ to $z \simeq 8$ there is good agreement that the number density 
of LBGs continues to decline but uncertainties and degeneracies in the fitted Schechter-function 
parameters mean that it is currently hard to establish whether this evolution is better 
described as density or luminosity evolution. For example, McLure et al. (2010) concluded 
that the $z \simeq 7$ and $z \simeq 8$ luminosity functions are consistent with having the 
same overall shape at at $z \simeq 6$, but with $\phi^*$ a factor of $\simeq 2.5$ and 
5 lower, respectively. Ouchi et al. (2009) also concluded in favour 
of a drop in $\phi^*$ between $z \simeq 6$ and $z \simeq 7$. 
Meanwhile, as shown in Fig. 11, the results of 
Bouwens et al. (2011b) appear to favour some level of continued luminosity evolution
(perhaps also combined with a decline in $\phi^*$ beyond $z \simeq 6$), but their best-fitting values for $\phi^*$, $M^*$ and $\alpha$ 
as a function of redshift are still consistent with the 
results of McLure et al. (2009, 2010) within current uncertainties
(note that at $z \simeq 8$ current data do not really allow a meaningful Schechter-function 
fit).

There are, however, emerging (and potentially important) areas of tension. 
The right-hand panel of Fig. 10 shows the generally good level of agreement 
over the basic form of the LBG LF at $z \simeq 7$ (i.e. 
$\phi^* \simeq 0.8 \times 10^{-3}\,{\rm Mpc^{-3}}$ and $M^* \simeq -20.1$; Ouchi et al. 2009;
McLure et al. 2010; Bouwens et al. 2011b), 
but also reveals issues 
at both the faint and bright ends (issues which we can hope 
will be resolved as the dataset on LBGs at $z \simeq 7 - 8$
continues to improve and grow).

At the faint end there is growing debate over the slope
of the luminosity function. As summarized above, essentially all workers are 
in agreement that the faint-end 
slope, $\alpha$, is steeper by $z \simeq 5$ than in the low-redshift Universe, 
where $\alpha \simeq -1.3$. But recently, Bouwens et al. (2011b), pushing the new WFC3/IR to the limit 
with very small aperture photometry, have provided tentative evidence that the faint-end
slope $\alpha$ may have steepened to $\alpha \simeq -2.0$ by $z \simeq 7$. This ``result'' is 
illustrated in Fig. 10, which shows the confidence intervals on the Schechter parameter values
deduced by Bouwens et al. (2011b) from $z \simeq 4$ to $z \simeq 7$. Clearly the data 
are still consistent with $\alpha = -1.7$ over this entire redshift range, but given 
the luminosity function has definitely steepened between $z \simeq 0$ and $z \simeq 5$, further 
steepening by $z \simeq 7$ is certainly not implausible, and (as discussed above and in section 6) 
would have important implications for the integrated luminosity density, and hence for 
reionization.
Fig.~11 also nicely illustrates the problems of degeneracies between the Schechter parameters; clearly
it will be hard to pin down $\alpha$ without better constraints on $\phi^*$ and $M^*$ which can 
only be provided by the larger-area surveys such as CANDELS and UltraVISTA (Robertson 2010). 
Another key issue is surface brightness bias. As discussed in detail by Grazian et al. 
(2011), because,
for a given total luminosity, {\it HST} is better able to detect the most compact objects 
(especially in the very small ($\simeq 0.3$-arcsec diameter) apertures adopted by 
Bouwens et al. 2011b), the estimated 
completeness of the WFC3/IR surveys at the faintest flux densities is strongly dependent on 
the assumed size distribution of the galaxy population. Thus, it appears that potentially all of any current 
disagreement over the faint-end slope at $z \simeq 7$ can be traced to different assumptions
over galaxy sizes and hence different completeness corrections. 
Finally, there are of course the usual issues over cosmic variance, with 
the faintest points on the luminosity function being determined from the 
WFC3/IR survey of the HUDF which covers only $\simeq 4$\,arcmin$^2$. However, 
as discussed in detail by Bouwens et al. (2011b), it appears that large-scale structure 
uncertainties do not pose a very big problem for luminosity function determinations in the 
luminosity range $-21 < M^* < -18$).

\begin{figure}

\includegraphics[scale=0.59, angle=0]{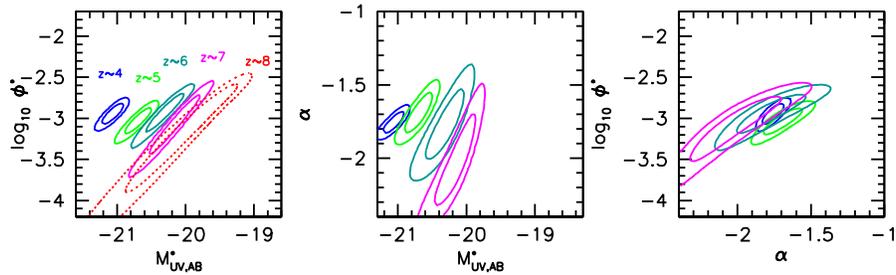}
\caption{68\% and 95\% likelihood contours on the model Schechter-function
  parameters derived by Bouwens et al. (2011b) from their 
  determination of the UV 
  (rest-frame $\sim$1700\AA) continuum LF
  at $z\sim7$ (magenta lines) and $z\sim8$ (dotted
    red lines).  Also shown for comparison are the LF
  determinations at $z\sim4$ (blue lines), $z\sim5$
  (green lines), and $z\sim6$ (cyan lines) from
  Bouwens et al. (2007).  No $z\sim8$ contours are shown in
  the center and right panels given the large uncertainties on the
  $z\sim8$ Schechter parameters.  Fairly uniform 
  evolution in the UV LF (left and middle panels) is seen as a 
  function of redshift, although there remains significant 
  degeneracy between $\phi^*$ and $M^*$.  Most of the evolution in the
  LF appears to be in $M^*$ (particularly from $z\sim7$ to $z\sim4$).
  Within the current uncertainties, there is no evidence for evolution
  in $\phi^*$ or $\alpha$ (rightmost panel)(courtesy R. Bouwens).}
\label{fig:11}       % Give a unique label
\end{figure}

At the bright end, Fig. 10 illustrates that the problem is mainly lack of data, which in turn 
can be traced to a lack of large-area near-infrared surveys of sufficient depth and multi-frequency 
coverage. As discussed above, current ground-based surveys for LBGs at $z \simeq 7$ 
are limited to those undertaken by Ouchi et al. (2009) and Castellano et al. (2010a,b) and suffer 
from somewhat uncertain contamination due to lack of sufficiently deeper longer-wavelength data.
Nevertheless, both Ouchi et al. (2009) and Castellano et al. (2010b) conclude that a decline in the 
number density of brighter LBGs between $z \simeq 6$ and $z \simeq 7$ is now established 
with better than 95\% confidence, even allowing for cosmic variance (the contrary results of Capak et al. 
2011 can be discounted for the reasons discussed in section 3.1.3). 
Significant further improvement
in our knowledge of the bright end of the LBG luminosity function at $z \simeq 7$ and 
$z \simeq 8$ can be expected over the next $\simeq 3$ years, from CANDELS
(Grogin et al. 2011), the WFC3/IR parallel 
programs (Trenti et al. 2011, 2012; Yan et al. 2011) and from UltraVISTA (McCracken et al. 2012; 
Bowler et al. 2012).

\subsection{High-redshift evolution of the Lyman-$\alpha$ luminosity function}

In contrast to the steady decline seen in the LBG ultraviolet continuum luminosity function at high redshift,
there is little sign of any significant 
change in the Lyman-$\alpha$ luminosity function displayed by LAEs from $z \simeq 3$ to $z \simeq 5.5$.
Indeed, as shown in Fig. 12, Ouchi et al. (2008) and Kashikawa et al. (2011)
have presented evidence that the Lyman-$\alpha$ luminosity
function displayed by LAEs selected via narrow-band imaging (and extensive spectroscopic follow-up) 
at $z \simeq 5.7$ is, within the uncertainties,
essentially identical to that seen at $z \simeq 3$. Both studies were unable 
to constrain the faint-end slope of the Lyman-$\alpha$ luminosity function but, assuming 
$\alpha = -1.5$, reported fiducial values for the other Schechter parameters at $z = 5.7$ 
of $\phi^* \simeq 8 \times 10^{-4}\,{\rm Mpc^{-3}}$ and $L^*_{\rm Ly\alpha} = 7 \times 10^{42}\,{\rm erg\,s^{-1}}$.

Why the Lyman-$\alpha$ luminosity function should display different evolution to the LBG continuum luminosity 
function is a subject of considerable current interest. The relationship between LBGs and LAEs is discussed 
at more length in the next subsection (including the evolution of the ultraviolet 
{\it continuum} luminosity function of LAEs), 
but the key point to bear in mind here is that the evolution of the 
Lyman-$\alpha$ luminosity function inevitably reflects not just the evolution in the number density and luminosity
of star-forming galaxies, but also cosmic evolution in the escape fraction of Lyman-$\alpha$ emission. This latter 
could, for example, be expected to increase with increasing redshift due to a decrease in average dust content, 
and/or at some point decrease with increasing redshift due to an increasingly neutral IGM. 

\vspace*{-1cm}

\begin{figure}

\includegraphics[scale=0.325, angle=0]{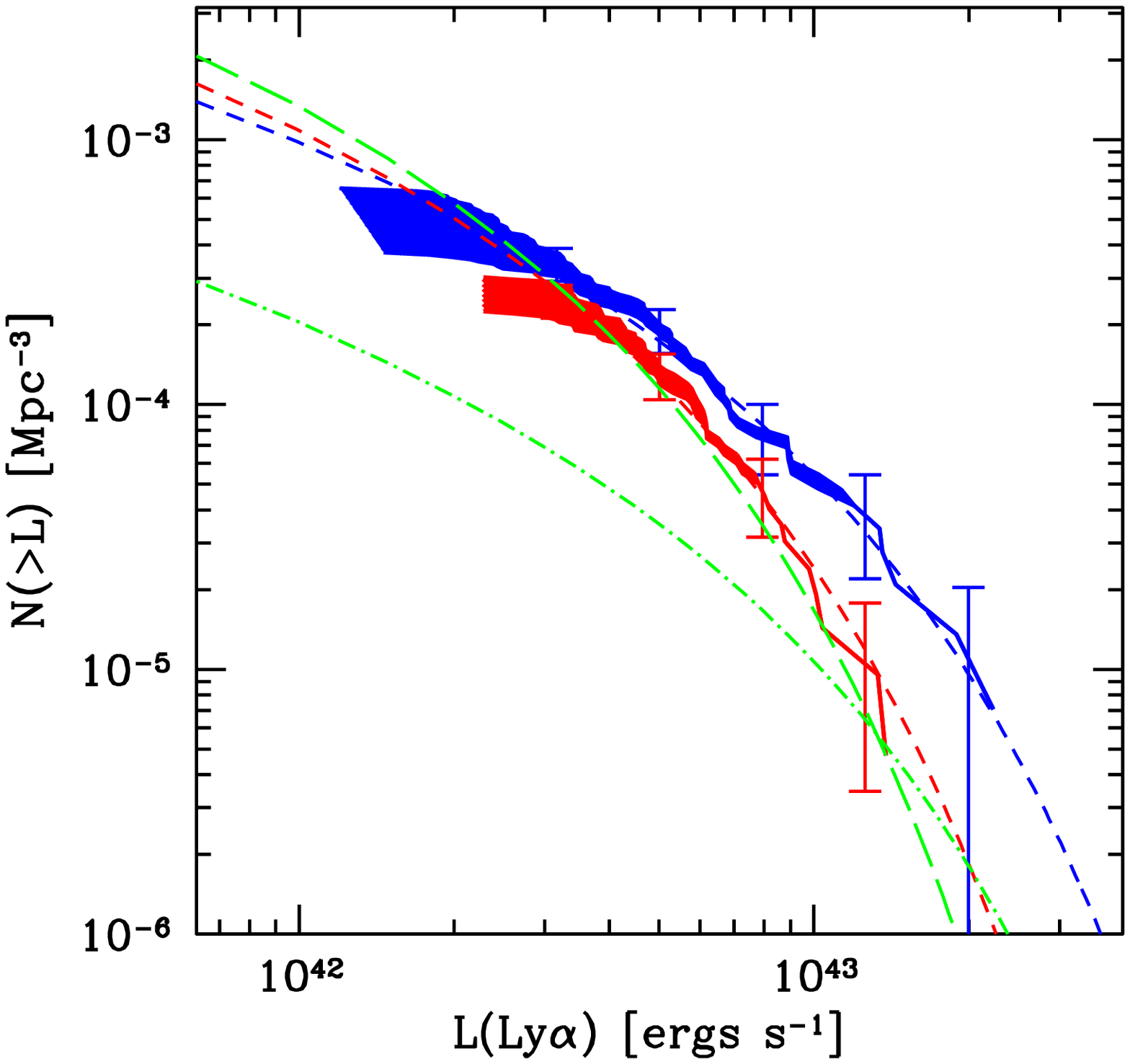}
\includegraphics[scale=0.66, angle=0]{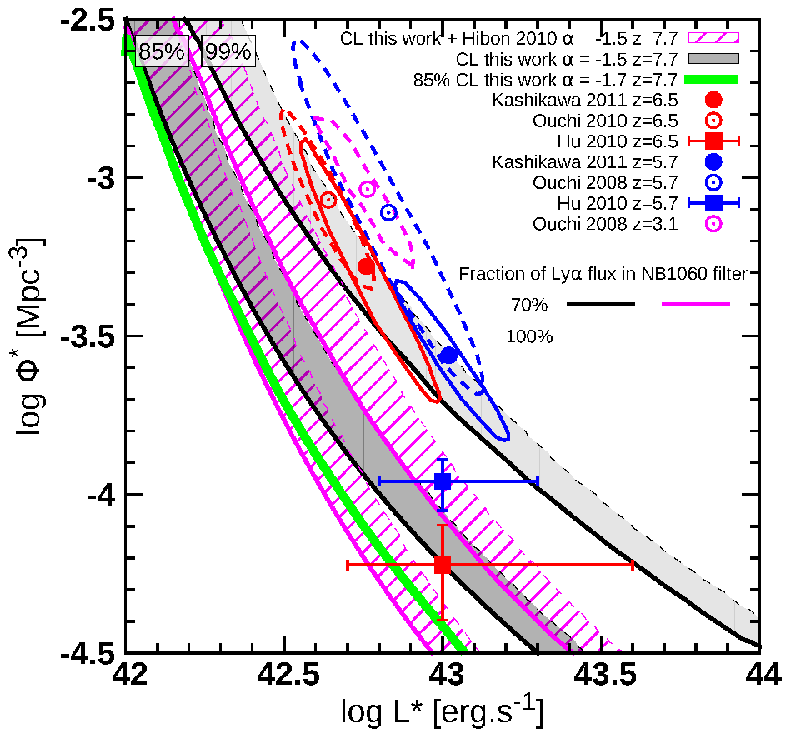}
\caption{Current constraints on the Lyman-$\alpha$ LF at high redshift. The left-hand plot,
taken from Kashikawa et al. (2011), shows a comparison of the {\it cumulative} 
Lyman-$\alpha$ LFs of LAEs at $z=5.7$ (blue-shaded region)
and at $z=6.5$ (red-shaded region).
The upper-edge of each shaded region is based on the assumption that all 
photometrically-selected candidates in the two SDF samples are indeed LAEs, while 
the lower-edge is derived purely on the spectroscopically-confirmed sample at each redshift.
The short-dashed lines (red for $z=6.5$ and blue for $z=5.7$) show the fitted 
Schechter LFs assuming $\alpha=-1.5$. For comparison, the green long-dashed line shows 
the Lyman-$\alpha$ LF at $z=6.5$ determined from the larger area SXDS survey by Ouchi 
et al. (2010), and the green dot-dashed line shows the $z = 6.5$ Lyman-$\alpha$ 
LF determined by Hu et al. (2010) (courtesy N. Kashikawa). The right-hand plot, taken from 
Cl\'{e}ment et al. (2011) summarizes our current knowledge (including some controversial 
disagreement) of $\phi^*$ and $L^*$ for Schechter-function fits to the Lyman-$\alpha$ LF at 
$z = 5.7$ and $z = 6.6$ (again assuming $\alpha = -1.5$), 
as well as attempting to set joint limits on these two parameters at $z = 7.7$ (see text in 
section 4.2 for details; courtesy B. Cl\'{e}ment).}
\label{fig:12}     
\end{figure}

\vspace*{1cm}

Staying with the direct observations for now, at still higher redshifts the situation is somewhat controversial. 
Following Kashikawa et al. (2006), Ouchi et al. (2010) have extended the Subaru 
surveys of narrow-band selected LAEs at $z \simeq 6.6$ to the SXDS field. They conclude that there is a 
modest ($\simeq 20-30$\%) decline in the Lyman-$\alpha$ LF over the redshift interval 
$z \simeq 5.7 - 6.6$ (also shown in the left-hand panel of Fig. 12), and that this decline is best 
described as luminosity evolution, with $L^*_{\rm Ly\alpha}$ falling from  $\simeq 7$ to 
$\simeq 4.5 \times 10^{42}\,{\rm erg\,s^{-1}}$, while $\phi^*$ remains essentially unchanged at 
$\simeq 8 \times 10^{-4}\,{\rm Mpc^{-3}}$ (see right-hand panel of Fig. 12, 
again assuming $\alpha = -1.5$).

The results from Ouchi et al. (2008; 2010) are based on large LAE samples but with only moderate 
levels of spectroscopic confirmation. These results have recently been 
contested by Hu et al. (2010). As summarized in the right-hand panel of Fig. 12, Hu et al. (2010) 
report a comparable value of $L^*$ at $z \simeq 5.7$, but $\phi^*$ an order of 
magnitude lower. They then also report a modest drop in LAE number density by $z \simeq 6.6$, 
but conclude this is better described as density evolution (with $\phi^*$ dropping by a factor of $\simeq 2$).

Also included in Fig. 12 are the latest results from Kashikawa et al. (2011), who used further spectroscopic 
follow-up to increase the percentage of spectroscopically confirmed LAEs in the SDF 
narrow-band selected samples of Taniguchi et al.
(2005) and Shimasaku et al. (2006) to 
70\% at $z \simeq 5.7$ and 81\% at $z \simeq 6.6$. The outcome of the resulting luminosity 
function reanalysis appears, at least at $z \simeq 5.7$,  
to offer some hope of resolving the situation, with Kashikawa et al. (2011) 
reporting a value for $\phi^*$ somewhat lower than (but consistent with) 
the value derived by Ouchi et al. (2008), and at least closer to the $\phi^*$ value 
reported by Hu et al. (2010). 
But at $z \simeq 6.6$ the results from Kashikawa et al. (2011) remain  
at odds with Hu et al. (2010), with $\phi^*$ still an order of magnitude higher,
and modest luminosity evolution since $z \simeq 5.7$ (if anything
offset by slight positive evolution of $\phi^*$, resulting in any significant decline
in number density being confined to the more luminous LAEs).
Given the high spectroscopic confirmation rates in the new Kashikawa et al. (2011) samples,
the claim advanced by Hu et al. (2010) that the previous Ouchi et al. (2008, 2010) and 
Kashikawa et al. (2006) studies were severely affected by high contamination rates in the narrow-band
selected samples now seems untenable. Rather, it apears much more probable that the 
Hu et al. (2010) samples are either 
affected by incompleteness (and hence they have
seriously under-estimated $\phi^*$ for LAEs at high redshift), or that 
our knowledge of the Lyman-$\alpha$ luminosity function at $z \simeq 6.6$ is still
severely confused by the affects of cosmic variance and/or patchy reionization
(see, for example, Nakamura et al. 2011), an issue which is discussed further in section 5.5.

The recent work of Cassata et al. (2011), based on a pure spectroscopic sample of (mostly) 
serendipitous Lyman-$\alpha$ emitters found in deep VIMOS spectroscopic 
surveys with the VLT, also yields results consistent 
with an unchanging Lyman-$\alpha$ luminosity function from $z \simeq 2$ to $z \simeq 6$. 
In addition, their estimated values of $\phi^*$ and $L^*_{\rm Ly\alpha}$ at $z = 5-6$ 
are in excellent agreement with those reported by Ouchi et al. (2008) at $z \simeq 5.7$. Interestingly, 
because the VIMOS spectroscopic surveys can probe to somewhat deeper Lyman-$\alpha$ 
luminosities than the narrow-band imaging surveys, this work has also provided 
useful constraints on the 
evolution of the faint-end slope, $\alpha$, at least at moderate redshifts. 
Specifically, Cassata et al. (2011) conclude that $\alpha$ steepens
from $\simeq -1.6$ at $z \simeq 2.5$ to $\alpha = -1.8$ at $z \simeq 4$. Direct constraints at the highest redshifts 
remain somewhat unclear, but the clear implication is that, as for the LBG luminosity function, the faint-end 
slope is significantly steeper than $\alpha = -1.5$ at $z > 5$ (and hence it is probably more appropriate
to consider $\phi^*$ and $L^*_{\rm Ly\alpha}$ values reported by authors assuming $\alpha = -1.7$ or even $\alpha \simeq -2$).

Finally, also shown in Fig. 12 are limits on the luminosity function parameter values at $z \simeq 7.7$
(albeit assuming $\alpha = -1.5$), imposed 
by the failure of Cl\'{e}ment et al. (2011) to detect any LAEs from deep HAWK-I VLT 1.06\,${\rm \mu}$m narrow-band 
imaging of three $7.5 \times 7.5$\,arcmin fields (probing a volume $\sim 2.5 \times 10^4\,{\rm Mpc^3}$). 
The ability of Cl\'{e}ment et al. (2011) to draw crisp conclusions from this work is hampered by the confusion 
at $z \simeq 6.6$, with the above-mentioned different LFs of Ouchi et al. (2010), 
Hu et al. (2010) and Kashikawa et al. (2011) predicting 11.6, 2.5 and 13.7 objects respectively in the Hawk-I 
imaging (if the luminosity function remains unchanged at higher redshifts). Cl\'{e}ment et al. (2011) 
conclude that an unchanged Lyman-$\alpha$ luminosity function can be excluded at $\simeq 85$\% confidence, but 
that this confidence-level could rise towards $\sim99$\% if one factors in significant quenching of
IGM Lyman-$\alpha$ transmission due to a strong increase in the neutral Hydrogen fraction as we enter the 
epoch of reionization (an issue discussed further in the next subsection). However, 
the issue of whether or not the Lyman-$\alpha$ luminosity function really declines beyond
$z \simeq 7$ undoubtedly remains controversial (e.g. Ota et al. 2010a; 
Tilvi et al. 2010; Hibon et al., 2010, 2011, 2012; Krug et al. 2012) 
and further planned surveys for LAEs at $z \geq 7$ are needed to address this question (e.g. Nilsson et al. 2007).

\subsection{The LBG-LAE connection}

The recent research literature in this field 
is littered with extensive and sometimes confusing discussions
over the differences and similarities between the properties of 
LBGs and LAEs. In the end, however, the LAE population must be a subset
of the LBG population, and the reported differences must be due  
to the biases (sometimes helpful) which are introduced by
the different selection techniques. One key area of much current 
interest is to establish whether/how the fraction of LBGs 
which emit observable Lyman-$\alpha$ varies with cosmic epoch, because this 
has the potential to provide key information on the evolution
of dust and gas in galaxies, and on the neutral hydrogen fraction in the IGM.
There are a number of lines of attack being vigorously pursued, 
and I start by considering how we might 
reconcile the apparently very different high-redshift evolution of the 
Lyman-$\alpha$ and LBG ultraviolet luminosity functions (as summarized 
in the previous two subsections).

\begin{figure}

\includegraphics[scale=0.80, angle=0]{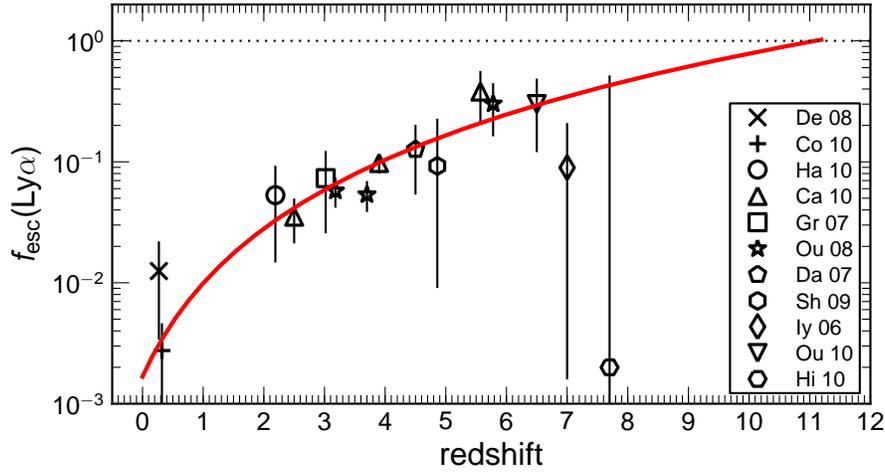}
\caption{The redshift evolution of volume-averaged Lyman-$\alpha$ escape fraction, $f_{\rm esc}^{\rm Ly\alpha}$, 
as deduced by Hayes et al. (2011), normalized to $\simeq 5$\% at $z \simeq 2$ via comparison of the 
Lyman-$\alpha$ and H-$\alpha$ integrated luminosity functions, and deduced at higher redshifts by 
comparison of the Lyman-$\alpha$ and UV continuum luminosity functions discussed in sections 4.1 and 4.2.
The solid red line shows the best-fitting power-law to points 
between redshift 0 and 6, which takes the form $(1+z)^{2.6}$, and appears to be a
good representation of the observed points over this redshift range. It
intersects with the $f_{\rm esc}^{\rm Ly\alpha} =1$ line (dotted) at redshift $z = 11.1$ 
(courtesy M. Hayes).}
\label{fig:13}       % Give a unique label
\end{figure}

\subsubsection{Comparison of the LBG UV and LAE Lyman-$\alpha$ luminosity functions}

First, given the steady negative evolution of the LBG luminosity function from 
$z \simeq 3$ to $z \simeq 6$, and the apparently unchanging form and normalization of the Lyman-$\alpha$
luminosity function over this period, it seems reasonable to deduce that {\it on average} 
the fraction of Lyman-$\alpha$ photons emerging from star-forming galaxies 
{\it relative to the observed continuum emission} increases 
with increasing redshift out to at least $z \simeq 6$. Indeed Hayes et al. (2011) have used this comparison 
to deduce that the {\it volume averaged} Lyman-$\alpha$ escape fraction, $f_{\rm esc}^{\rm Ly\alpha}$, grows 
according to $f_{\rm esc}^{\rm Ly\alpha} \propto (1 + z)^{2.5}$ (normalized at $\simeq 5$\% at $z \simeq 2$ 
through a comparison of the Lyman-$\alpha$ and H-$\alpha$ luminosity functions, currently 
feasible only at $z \simeq 2$). This result is shown in Fig. 13, which also shows that extrapolation of the fit 
to higher redshifts would imply $f_{\rm esc}^{\rm Ly\alpha} = 1$ at $z \simeq 11$ in the absence of any new source 
of Lyman-$\alpha$ opacity, providing clear motivation for continuing the comparison of the 
Lyman-$\alpha$ and LBG continuum luminosity functions to higher redshift if at all possible (see below).

\subsubsection{The LAE UV continuum luminosity function}
The {\it italics} in the preceding paragraph have been chosen with care, because we must proceed carefully.
This is because the situation is confused by the fact that, {\it for those 
galaxies selected as LAEs} (via, for example, narrow-band imaging as discussed in detail above), the ratio of 
average Lyman-$\alpha$ emission to ultraviolet continuum emission apparently 
stays unchanged or even {\it decreases} with increasing redshift. 
We know this  from studies of the ultraviolet continuum luminosity function of LAEs, which I have deliberately 
avoided discussing until now because there are complications in interpreting the UV continuum luminosity function
of objects which have been selected primarily on the basis of the contrast between Lyman-$\alpha$ emission 
and UV continuum emission. Nevertheless, Ouchi et al. (2008) have convincingly shown that, while the Lyman-$\alpha$
luminosity function of LAEs holds steady between $z \simeq 3$ and $z \simeq 5.7$, the UV continuum luminosity 
function of the same objects actually grows with redshift, more than bucking the negative trend displayed by 
LBGs. Then, from $z = 5.7$ to $z \simeq 6.6$, as the Lyman-$\alpha$ luminosity function shows the first signs 
of gentle decline, Kashikawa et al. (2011) find that the UV continuum luminosity function of the LAEs stops
increasing, but seems to hold steady. Another way of saying this is that the average equivalent width 
of Lyman-$\alpha$ emission in LAEs is constant or if anything 
slightly falling with increasing redshift.
Indeed, Kashikawa et al. (2011) report that the median value of Lyman-$\alpha$ equivalent width falls from 
$EW_{rest}  \simeq 90$\,\AA\  at $z \simeq 5.7$ to $EW_{rest}  \simeq 75$\,\AA\  at $z \simeq 6.6$ although, 
interestingly, there is a more pronounced extreme $EW_{rest}$ tail in their 
highest-redshift sample.

\begin{figure}

\includegraphics[scale=0.58, angle=0]{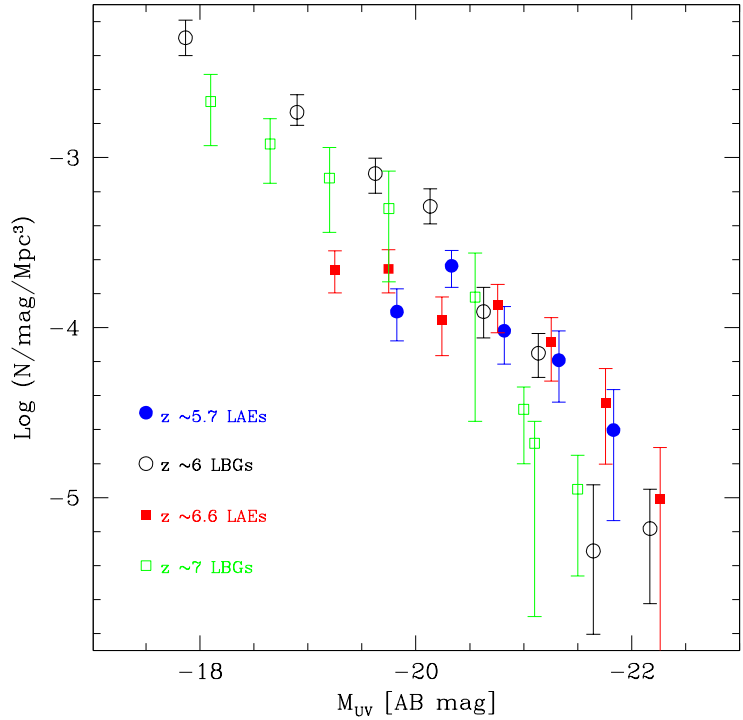}
\caption{A comparison of the high-redshift UV 
continuum LFs of galaxies selected as LBGs and galaxies selected as LAEs.
Shown here are the UV continuum LFs for LBGs at $z \simeq 6$ as determined 
by Bouwens et al. (2007), for LBGs at $z \simeq 7$ as determined by McLure 
et al. (2010), for LAEs at $z \simeq 5.7$ as determined by 
Shimasaku et al. (2006), and for LAEs at $z \simeq 6.6$ as 
as determined by Shimasaku et al. (2006). The LAE UV LFs become 
incomplete at $M_{UV} > -21$ because of the limited depth of the ground-based
broad-band imaging in the large Subaru survey fields (compared to the deeper 
{\it HST} data used to derive the LBG LFs). However, at brighter 
magnitudes the agreement between the $z \simeq 6$ LBG LF and the 
two LAE-derived LFs at $z \simeq 5.7$ and $6.6$ is very good
(courtesy P. Dayal).}
\label{fig:14}       
\end{figure}

Possible physical reasons for why this happens are discussed further below but, whatever the explanation, it is 
clear that the UV continuum luminosity function of LAEs cannot keep rising indefinitely, or it will at some point 
exceed the UV luminosity function of LBGs, which is impossible. Indeed, the two luminosity functions appear to virtually match
at $z \simeq 6$. Specifically, as shown in Fig. 14, and as first demonstrated by Shimasaku et al. (2006), 
by $z \simeq 6$, LAE selection down to $EW_{rest} \simeq 20$\,\AA\ recovers essentially all LBGs with $M_{1500} 
< -20$, to within a factor $\simeq 2$. It is thus no surprise that the UV luminosity function of LAEs must freeze or commence negative 
evolution somewhere between $z \simeq 6$ and $z \simeq 7$, as by then it must start to track (or fall faster than) 
the evolution of the (parent) LBG population.

\subsubsection{The prevalence of Lyman-$\alpha$ emission from LBGs}

If this is true, then it must also follow that the fraction of LBGs which display Lyman-$\alpha$ emission
with $EW_{rest} > 20$\,\AA\ in follow-up spectroscopy must also rise from lower redshifts to near unity at 
$z \simeq 6$. There has been some controversy over this issue, 
but recent observations 
appear to confirm that this is indeed the case. First, 
Stark et al. (2011) have reported that, with increasing 
redshift, an increasing fraction of LBGs display strong Lyman-$\alpha$ emission such that, by $z \simeq 6$, over $50$\% of 
faint LBGs display Lyman-$\alpha$ with $EW_{rest} > 25$\,\AA. Similarly high Lyman-$\alpha$ ``success rates'' 
have now been reported for more luminous ($\simeq 2 L^*$) LBGs at $z \simeq 6$ by Curtis-Lake et al. (2012) (Fig. 15), and 
by Jiang et al. (2011).

\begin{figure}

\includegraphics[scale=0.42, angle=0]{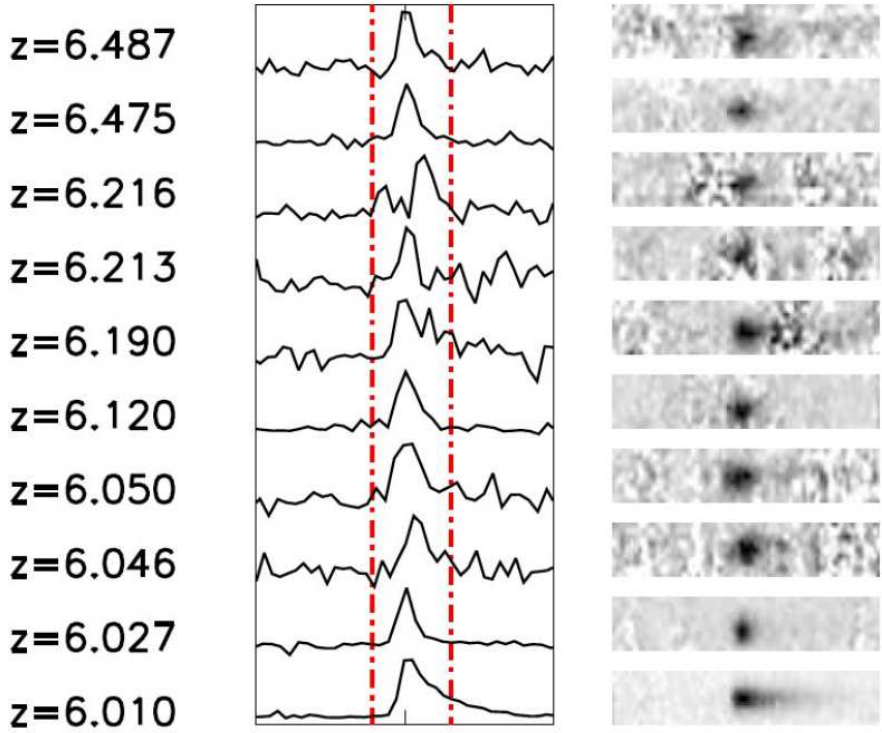}

\caption{A high success rate in the detection of Lyman-$\alpha$ emission
from bright LBGs at $z~=~6~-~6.5$. VLT FORS2 spectra of 10
$z>6$ LBGs selected from the UDS/SXDS field are shown; this represents 70\%
of the targetted high-redshift sample (Curtis-Lake et al. 2012). 
The extracted one-dimensional 
(1D) spectra are shown on the left, with 
the corresponding two-dimensional (2D) spectra on 
the right; Lyman-$\alpha$ emission (often obviously asymmetric) 
is clearly detected from all of these objects (courtesy E. Curtis-Lake).}
\label{fig:15}      
\end{figure}

\subsubsection{Reconciliation to ${\bf z \simeq 6}$}

It thus appears that the average volumetric increase in Lyman-$\alpha$ emission relative to 
ultraviolet continuum emission as summarized by Hayes et al. (2011) to produce Fig. 13 is due to 
an increase with redshift in the {\it fraction} of star-forming galaxies which emit 
at least some detectable Lyman-$\alpha$ emission rather than a systematic increase 
in the Lyman-$\alpha$ to continuum ratio of objects which are selected as LAEs at all epochs.

As already discussed above, one thing which is clear, and perhaps surprising, 
is that by $z \simeq 6$, narrow-band selection of LAEs from the wide-area Subaru surveys 
seems to be an excellent way of determining the bright end of the complete LBG UV luminosity 
function, indicating that the fraction of LBGs which display $EW_{rest} > 10$\,\AA\ 
is approaching unity by this redshift. We stress that this is not the same as 
seeing {\it all} of the Lyman-$\alpha$ emission from the star-forming population; 
it is perfectly possible for virtually all the LBGs to emit enough 
Lyman-$\alpha$ to be detected in deep LAE surveys, while still having some 
way to go before the overall volume-average Lyman-$\alpha$ escape fraction
could be regarded as approaching unity. Put another way, it is not
unreasonable to conclude that the detection
rate of bright LBGs in LAE surveys reaches 100\% at $z \simeq 6$, 
while volume-averaged $f_{\rm esc}^{\rm Ly\alpha}$ as plotted in Fig. 13 
has only reached $\simeq 40-50$\%. As discussed below, there may be 
good astrophysical reasons why the volume-averaged
$f_{\rm esc}^{\rm Ly\alpha}$ never reaches 100\%.

Given that, at least for $M_{1500} < -20$, current LAE and LBG surveys 
appear to be seeing basically the same objects at $z \simeq 6$, it makes 
sense to consider what, if anything, can be deduced about the fainter end 
of the LBG UV luminosity function from those LAEs which are {\it not} detected 
in the continuum. As can be seen from Fig. 14, the ground-based imaging from which the 
LAE samples are selected, not unexpectedly runs out of steam at flux densities which correspond 
to $M_{1500} \simeq -21$ at $z \simeq 6.6$. But there still of course remain many 
(in fact the vast majority of) LAEs which have no significant continuum detections, 
and these objects are not only useful for the Lyman-$\alpha$ luminosity function, but 
potentially also carry information on the faint end of the LBG UV luminosity function. 
The question is of course how to extract this information. One could follow-up 
all the LAEs which are undetected in the ground-based broad imaging with {\it HST} 
WFC3/IR to determine their UV luminosities (i.e. $M_{1500}$); this would 
undoubtedly yield many more detections allowing further extension of the 
UV continuum luminosity function of LAEs to fainter luminosities. However, this 
would still not overcome another incompleteness problem which is that, 
as narrow-band searches are limited not just by equivalent width, but also by basic 
Lyman-$\alpha$ luminosity, the subset of LBGs detectable in the LAE surveys 
becomes confined to those objects with increasingly extreme values of $EW_{rest}$ 
as we sample down to increasingly faint UV luminosities. Then, even with deep WFC3/IR 
follow-up of detectable LAEs, and even assuming all LBGs emit some Lyman-$\alpha$, 
we would still be forced to infer the total number of faint LBGs by extrapolating 
from the observable extreme equivalent-width tail of the LAE/LBG population assuming an 
equivalent-width distribution appropriate for the luminosity and redshift in question. 

This is difficult, and presents an especially severe problem at high-redshift, 
where our knowledge of Lyman-$\alpha$ equivalent-width distributions is confined 
to the highest luminosities. Nevertheless, Kashikawa et al. (2011) have attempted it, 
and discuss in detail how they tried to arrive at an appropriate equivalent-width 
distribution as a function of UV luminosity at $z \simeq 6.6$. A key issue is that 
it is difficult, if not impossible, to determine the UV continuum luminosity 
dependence of Lyman-$\alpha$ $EW_{rest}$ in the underlying LBG population from the 
equivalent-width distribution displayed by the narrow-band selected LAEs themselves, 
as this is in general completely dominated by the joint selection effects of 
Lyman-$\alpha$ luminosity and equivalent width. There has indeed been 
much controversy over this issue, with Nilsson et al. (2009) claiming that, 
at $z \simeq 2-3$ where LAE surveys display good dynamic range, there is 
no evidence for any UV luminosity dependence of the $EW_{rest}$ distribution, contradicting 
previous claims that there was a significant anti-correlation between $EW_{rest}$ 
and UV luminosity. Of course what is really required is complete spectroscopic 
follow-up of LBGs over a wide UV luminosity range, to determine the distribution 
of $EW_{rest}$ as a function of $M_{1500}$, free from the biases introduced by LAE selection. 
This has been attempted by Stark et al. (2010, 2011), and it is the results of this work 
that Kashikawa et al. (2011) have employed to try to estimate the faint end of the 
LBG UV luminosity function at $z \simeq 6$ from the number counts of faint 
(but still extreme equivalent width) LAEs extracted from the narrow-band surveys. 
The problem with this is that even the state-of-the-art work of Stark et al. (2011) 
only really provides $EW_{rest}$ distributions in two luminosity bins at $z \simeq 6$, 
and the apparent luminosity dependence inferred from this work is called into 
some question by the success in Lyman-$\alpha$ detection in bright LBGs at 
$z \simeq 6$ by Curtis-Lake et al. (2012) and Jiang et al. (2011). Thus, at present, 
our understanding of the luminosity dependence of the Lyman-$\alpha$ $EW_{rest}$ 
distribution displayed by LBGs remains poor, and is virtually non-existent for 
LBGs with $M_{1500} > -19$ at $z > 6$.

Nevertheless, the experiment is of interest, and the resulting 
UV LF derived from the Lyman-$\alpha$ LF by Kashikawa et al. (2011) is more like a power-law 
than a Schechter function. Moreover, the implied faint-end slope 
is extremely steep; if a Schechter-function fit is enforced, $\alpha = -2.4$ 
results, even though the UV LF 
was inferred from a Lyman-$\alpha$ Schechter function with $\alpha = -1.5$, which is probably too flat.
It is not yet clear what to make of this result, but it would appear that either the 
equivalent-width distribution of Lyman-$\alpha$ from faint LBGs is biased to even higher values than assumed
(so that even extreme equivalent-width LAEs sample a larger fraction 
of the LBG population at faint $M_{1500}$ than anticipated by Kashikawa et al. 2011), or the
incompleteness in the faint LBG surveys has been under-estimated. This latter explanation seems unlikely
given the already substantial incompleteness corrections made by Bouwens et al. (2011), but is not entirely impossible
if LAEs pick up not just the compact LBGs seen in the HST surveys, but also a more extended
population which is missed with HST but is uncovered by ground-based imaging 
(which is less prone to surface-brightness bias). 
This, however, also seems unlikely; while recent work has certainly demonstrated that the 
Lyman-$\alpha$ emission from high-redshift galaxies is often quite extended (e.g. Finkelstein et al. 2011; Steidel et al. 2011), consistent
with theoretcial predictions (e.g. Zheng et al. 2011), all evidence suggests that the {\it UV continuum emission} from 
these same objects is at least as compact as LBGs at comparable redshifts (i.e. typically 
$r_{h} \leq 1$\,kpc at $z > 5$; Cowie et al. 2011; Malhotra et al. 2012; Gronwall et al. 2011). 
The fact that, due to the complex radiative transfer of Lyman-$\alpha$ photons, 
the 
Lyman-$\alpha$ morphologies of young galaxies are expected to be complex and in general more extended than 
their continuum morphologies is supported by new observational studies of low-redshift Lyman-$\alpha$ emitting galaxies
as illustrated in Fig. 16 (from the Lyman Alpha Reference Sample -- LARS; {\it HST} Program GO12310). However, from the point of view
of luminosity-function comparison, the key point is that while extended low-surface brightness 
Lyman-$\alpha$ emission might be hard to detect with {\it HST}, such LAEs will still not be missed 
by deep {\it HST} broad-band LBG surveys, if virtually all of them display compact continuum emission.
The quest to better constrain the true form of the faint-end slope of the UV LF 
will continue, not least because it is of crucial importance for considering 
whether and when these young galaxies could have reionized the Universe (see section 6.2).
Deeper and more extensive {\it HST} WFC3/IR imaging over the next few years has the potential 
to clarify this still currently controversial issue.

\begin{figure}

\includegraphics[scale=0.335, angle=0]{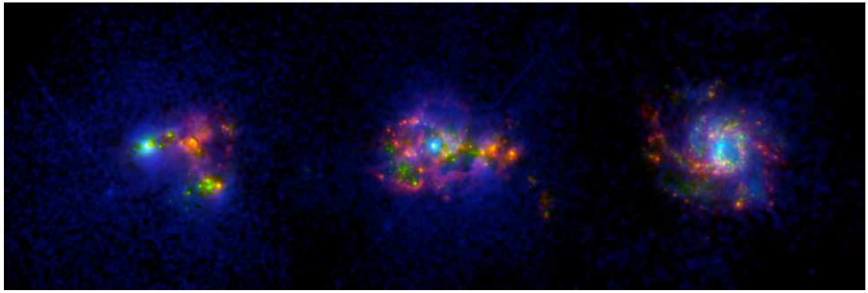}

\caption{Three nearby star-forming galaxies imaged as part of the {\it HST} Lyman-$\alpha$ 
imaging program LARS. Green shows the UV continuum and traces the massive stars, with the 
ionized nebulae they produce shown in Red (tracing $H\alpha$). The Lyman-$\alpha$ photons 
must also be produced in these nebulae, but the Lyman-$\alpha$ image (shown in Blue) reveals 
all these galaxies to be morphologically very different in $H\alpha$ and Lyman-$\alpha$ 
due to the resonant scattering of the Lyman-$\alpha$ photons.
This is at least qualitatively similar to what is found for high-redshift LAEs, in which the Lyman-$\alpha$ 
emission is in general more extended and diffuse than the UV continuum light
(courtesy of Matt Hayes).}

\label{fig:16}       
\end{figure}

\subsubsection{The nature of LAEs}
It is easy to become confused by the (extensive) literature on the properties of LAEs. In part this is 
because different authors adopt a different definition of what is meant by the term LAE. For some, an LAE is any 
galaxy which displays detectable Lyman-$\alpha$ emission, including objects originally selected 
as LBGs and then followed up with spectroscopy. All spectroscopically-confirmed LBGs 
at $z > 5$ must of course be emitters of Lyman-$\alpha$ radiation, and so in an astrophysical sense they 
are indeed LAEs. However, in practice most LAE studies are really confined to objects which have been 
{\it selected} on the basis of Lyman-$\alpha$ emission. Furthermore, many of these studies then proceed
to deliberately confine attention to those LAEs which could {\it not} also have been selected as LBGs from 
the data in hand. Of course there are often good reasons for doing this. For example, Ono et al. (2010)
in their study of the typical UV properties of LAEs at $z \simeq 5.7$ and $z \simeq 6.6$ first excised 
39 of the 295 LAEs from their sample because they were individually detected at IRAC wavelengths, before 
proceeding to stack the data for the remaining LAEs to explore their average continuum colours. This makes 
sense given the objective of this work was to explore the properties of those objects which were 
not detected with IRAC, 
but such deliberate focus on the extreme equivalent-width subset 
of the LAE population does sometimes run the danger of exaggerating the differences between the LBG and LAE 
populations.

To put it another way, in many respects the properties of LAEs, selected on the basis of large $EW_{rest}$,
are largely as would be anticipated from the extreme Lyman-$\alpha$ equivalent-width tail of the LBG 
population. I have already argued above that the observational evidence on luminosity 
functions suggests LAEs are just a subset of LBGs, and that by $z \simeq 6$ the increased 
escape of Lyman-$\alpha$ means that the two populations are one and the same. At least some existing 
{\it HST}-based comparisons of LAEs and LBGs support this viewpoint (e.g. Yuma et al. 2010), as  do 
at least some theoretical predictions (e.g. Dayal \& Ferrara 2012).
It is then simply to be expected that the subset of galaxies selected on the basis of extreme 
Lyman-$\alpha$ $EW_{rest}$ (and hence also typically
faint UV continuum emission) will, on average, have lower stellar masses, younger ages, and lower-metallicities 
than typical LBGs discoverable by current continuum surveys (e.g. Ono et al. 2012).
 
At present, therefore, there is really no convincing evidence that LAEs are 
anything other than a subset of LBGs. 
This is not meant to denigrate the importance of LAE studies; faint narrow-band selection 
provides access to a special {\it subset} of the UV-faint galaxy population over much larger areas/volumes than current 
deep {\it HST} continuum surveys. But the really interesting questions are whether this extreme equivalent-width 
subset represents an increasingly important fraction 
of LBGs with decreasing UV continuum luminosity and, conversely, 
whether some subset of this 
extreme equivalent-width population {\it cannot} be detected in current deep {\it HST} continuum imaging. 
To answer the first question really requires ultra-deep spectroscopic follow-up at $z \simeq 6-7$ of objects 
{\it selected as LBGs} spanning 
a wide range of continuum luminosity $M_{1500}$. To answer the second question, following Cowie 
et al. (2011), further deep {\it HST}
WFC3/IR imaging of objects {\it selected as LAEs} is desirable to establish what subset (if any) of 
the LAE population lacks sufficiently compact UV continuum emission to be selected as a faint LBG given the 
surface brightness biases inherent in the high-resolution deep {\it HST} imaging.

\subsubsection{Beyond ${\bf z \simeq 6.5}$; a decline in Lyman-${\bf \alpha}$?}

Both the follow-up spectroscopy of LBGs, and the discovery of LAEs via narrow-band imaging become 
increasingly more difficult as we approach $z \simeq 7$, due to the declining sensitivity of silicon-based
detectors at $\lambda \simeq 1$\,$\mu$m, the increasing brightness of night sky emission and, of course,
the reduced number density of potential targets (as indicated by the evolution of the LBG luminosity function
discussed above). Nevertheless, even allowing for these difficulties, there is now growing (albeit still tentative) 
evidence that Lyman-$\alpha$ emission from galaxies at $z \simeq 7$ is signicantly less prevalent than at 
$z \simeq 6$. Specifically, while spectroscopic follow-up of LBGs with $z_{phot} > 6.5$  
has indeed yielded several Lyman-$\alpha$ emission-line redshifts up to $z \simeq 7$ (see section 3.1), 
these same studies all report a lower success rate ($\simeq~15-25$\%) than encountered 
at $z \simeq 6$ (Pentericci et al. 2011; Schenker et al. 2012; Ono et al. 2012). In addition, 
such Lyman-$\alpha$ emission as is detected seems typically not very intense, with an especially significant lack
of intermediate Lyman-$\alpha$ equivalent widths, $EW_{rest}  \simeq 20-55$\,\AA. 
The significance of the inferred reduction in detectable Lyman-$\alpha$ obviously becomes enhanced 
if judged against extrapolation of the rising trend in Lyman-$\alpha$ emission out to 
$z \simeq 6$, as discussed above, and plotted in Fig. 13 
(Stark et al. 2011; Hayes et al. 2011; Curtis-Lake et al. 2012).

These results may be viewed as confirming a trend perhaps already hinted at by 
the reported modest decline in the Lyman-$\alpha$ luminosity function between 
$z \simeq 5.7$ and $z \simeq 6.6$ (as outlined above in section 4.2), and the 
tentative (albeit controversial) indications of further decline at $z \geq 7$. In summary, 
there is growing evidence
of a relatively sudden reduction in the transmission of Lyman-$\alpha$ photons between $z \simeq 6$ and $z \simeq 7$.
Given that the galaxy population itself appears, on average, to become increasingly better at 
releasing Lyman-$\alpha$ photons to the observer out to $z \simeq 6$ (perhaps due to a global decline in 
average dust content), the most natural and popular interpretation 
of this decline at $z \simeq 7$ is a significant and fairly 
rapid increase in the neutral hydrogen fraction in the IGM.

This has several implications. First, it suggests that further comparison of LAEs and LBGs over the redshift 
range $z \simeq 6 - 7$ may well have something interesting to tell us about reionization (at least its  
final stages; see section 6.2). Second, it implies that spectroscopic 
redshift determination/confirmation of LBGs at $z > 7$ is likely to be extremely difficult, 
and that, for astrophysical reasons, we may be forced to rely on photometric
redshifts at least until the advent of genuinely deep near-to-mid infrared spectroscopy with {\it JWST} 
(capable of detecting longer-wavelength emission lines including H$\alpha$). Third it suggests that 
the spectacular success of LAE selection via narrow-band imaging out to $z \simeq 6.6$ could be hard to replicate at higher
redshifts, and hence that the future study of galaxies and their evolution at $z \simeq 7-10$ may well be driven  
almost entirely by Lyman-break selection.

\section{Galaxy Properties}

\subsection{Stellar masses}

Stellar mass is one of the most important and useful quantities that can be estimated for a 
high-redshift galaxy. There are two reasons for this. First, since it represents the time-integral 
of past star-formation activity, it can be compared directly with observed 
star-formation rates in even higher-redshift galaxies to set model-independent constraints 
on plausible modes of galaxy evolution (e.g. Stark et al. 2009). 
Second, it enables fairly direct and unambiguous 
comparison with the predictions of different theoretical/computational models of galaxy formation, most 
of which deliver stellar mass functions as one of their basic outputs 
(e.g. Bower et al. 2006; De Lucia \& Blaizot 2007; Choi \& Nagamine 2011; Finlator et al. 2011).

Unfortunately, however, accurate galaxy stellar masses are very hard to derive from data which 
only sample the rest-frame UV continuum. There are two well-known reasons for
this. First, for any reasonable 
stellar initial mass function (IMF) the UV continuum in a galaxy is dominated by light from 
a relatively small number of short-lived massive stars, and thus depends critically on 
recent star-formation activity. Second, the UV continuum is much more strongly affected 
by dust extinction than is light at longer wavelengths, with 1 mag. of extinction in the 
rest-frame $V$-band producing $\simeq 3-4$ mag. of extinction at 
$\lambda_{rest} \simeq 1500$\,\AA\ (e.g. Calzetti et al. 2000).

Ideally, then, stellar masses should be estimated from the rest-frame near-infrared emission, at 
$\lambda_{rest} \simeq 1.6\, {\rm \mu m}$. But, beyond $z \simeq 5$, this is redshifted to $\lambda_{obs} \ge 
10\,{\rm \mu m}$, and so this is not really practical until {\it JWST}. Nevertheless, photometry 
at any wavelength longer that $\lambda_{rest} \simeq 4000$\,\AA\ is enormously helpful in reducing the 
uncertainty in stellar masses, and so the now-proven ability of {\it Spitzer} to 
detect LBGs and LAEs at $z \simeq 5 - 7$ in the two shortest-wavelength IRAC bands (at $\lambda_{obs} = 
3.6\,{\rm \mu m}$ and $\lambda_{obs} = 4.5\,{\rm \mu m}$) has been crucial in enabling 
meaningful estimates of their stellar masses (e.g. Yan et al. 2006; 
Stark et al. 2007a; Eyles et al. 2007; 
Labbe et al. 2006, 2010a,b; Ouchi et al. 2009a;  Gonz\'{a}lez et al. 2010, 2011; McLure et al. 2011).

However, because even these {\it Spitzer} IRAC bands still sample rest-frame {\it optical} emission 
at $z > 5$ (see Fig. 1), 
the derived stellar masses remain significantly affected by star-formation history, which thus 
needs to either be assumed (often a constant star-formation rate is simply 
adopted for a galaxy's entire history -- e.g. Gonz\'{a}lez et al. 2010) 
or inferred from full SED fitting to as much multi-waveband photometry as is available 
(e.g. Labb\'{e} et al. 2010b). 
This is, of course, a natural bi-product of the SED-fitting approach to deriving photometric 
redshifts, but it presents a number of challenges. First, it requires the 
careful 
combination of {\it Spitzer} and {\it HST} data which differ 
by an order-of-magnitude in angular resolution. Second, there are areas of disagreement between different 
evolutionary-synthesis models of galaxy evolution (e.g. Jimenez et al. 2000;
Bruzual \& Charlot 2003; Maraston 2005; Conroy \& Gunn 2010), 
although in practice these are not very serious
when it comes to modelling the rest-frame ultraviolet-to-optical SEDS of galaxies which must be less than 1 billion
years old (e.g. uncertainty and controversy over the strength of the asymptotic red giant branch is 
not really an issue when modelling the UV-to-optical SEDS of young galaxies; Maraston 2005; Conroy \& Gunn 2010; 
Labb\'{e} et al. 2010a). Third, 
and probably most serious, there are often significant degeneracies between age, dust-extinction, and 
metallicity, which can be hard or impossible to remove given only moderate signal-to-noise photometry 
in only a few wavebands. Fourth, as has recently become more apparent
(e.g. Ono et al. 2010; Labb\'{e} et al. 2010b; McLure et al. 2011; 
Gonz\'{a}lez et al. 2012), very different stellar masses can 
be produced depending on what is assumed about the strength of the nebular emission-lines and continuum from a galaxy's
inter-stellar medium relatively to the continuum emission from its stars 
(an issue discussed further in section 6.1).
Fifth, one cannot escape the need 
to assume a stellar IMF to deduce a total stellar mass, as most of the mass 
is locked up in low-mass stars which are not detected!

It would be a mistake to over-emphasize the issue of the IMF, because it is an assumption which can also 
be changed in the model predictions (e.g. Dav\'{e} 2008). Thus, for example, the systematic 
factor of $\simeq 1.8$ 
reduction in inferred stellar mass that results from changing 
from the Salpeter (1955) IMF to that of Chabrier (2003) (see also Weidner et al. 2011) 
need not necessarily prevent useful comparison with theory. 
In addition, another key quantity which follows on from the derivation
of stellar mass is relatively immune to the assumed IMF. This is the specific star-formation rate ({\it sSFR}),
defined here as the ratio of star-formation rate to stellar mass already in place. The extrapolation to smaller 
masses invoked by the IMF assumption applies to both the numerator and denominator when calculating this quantity,
making it reasonably robust (albeit still highly vulnerable to any 
uncertainties in dust extinction). 
This, combined with the attraction that {\it sSFR} encapsulates a basic measure
of ``current'' to past star-formation activity, has made {\it sSFR} a key focus of many recent 
studies of high-redshift galaxies. 
Indeed, one of the most interesting, and controversial results to emerge 
from this work in recent years is that star-forming 
galaxies lie on a ``main sequence'' which can be characterised by 
a single value of {\it sSFR} which is a function of epoch. Moreover, as shown in Fig. 17, 
the current observational evidence suggests that this characteristic {\it sSFR}, after rising by a factor 
of $\simeq 40$ from $z = 0$ to $z \simeq 2$ (Noeske et al. 2007; Daddi et al. 2007)
plateaus at $2 -3$\,Gyr$^{-1}$ at all higher redshifts 
(e.g. Stark et al. 2009; Gonz\'{a}lez et al. 2010). 

\begin{figure}

\includegraphics[scale=0.60, angle=0]{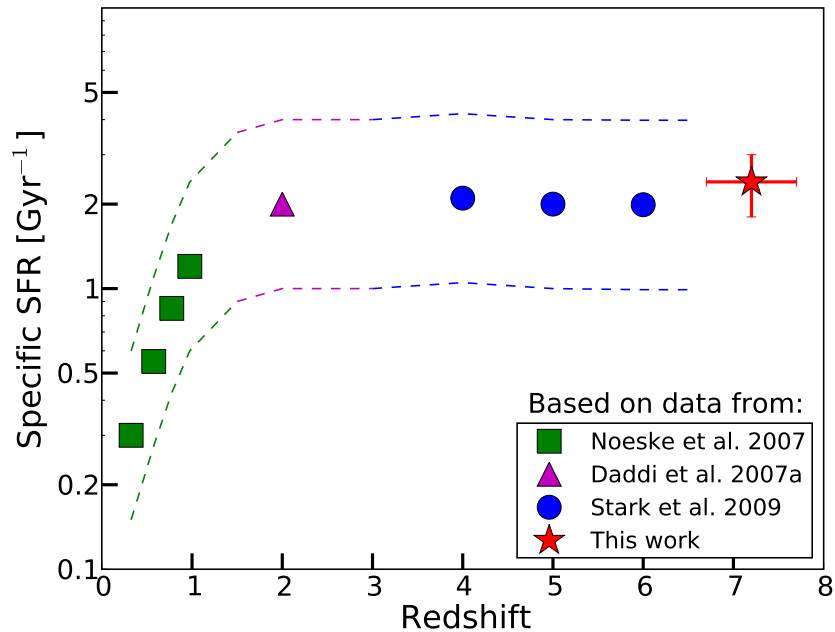}

\caption{Average $\langle sSFR \rangle$ determined at $z\sim7$ by 
Gonz\'{a}lez et al. (2010) for a median stellar mass of 
5$\times10^9\,{\rm M_{\odot}}$ compared to the average values determined 
by other authors at lower redshifts, but comparable stellar masses
(Noeske et al. 2007; Daddi et al. 2007; Stark et al. 2009).
This plot implies that $\langle sSFR \rangle$ stays remarkably constant,
at $\simeq 2\,{\rm Gyr^{-1}}$ over the redshift range $2 < z < 7$, 
suggesting that the star-formation--mass relation does not 
evolve strongly during the first $\simeq 3$ billion
years of galaxy evolution (courtesy V. Gonz\'{a}lez).}
\label{fig:17}       % Give a unique label
\end{figure}

Our current best estimate of the evolving 
stellar mass function of LBGs at $z = 4$, 5, 6 \& 7 is shown in Fig. 18. 
This summarizes the work of 
Gonz\'{a}lez et al. (2011), who used the WFC3/IR ERS data in tandem 
with the GOODS-South IRAC and {\it HST } ACS data to establish 
a mass to luminosity relation $M_{star} \propto L^{1.7}_{1500}$ at $z \sim 4$,
and then applied this to convert the UV LFs of Bouwens et al. (2007, 2010) 
into stellar mass functions at $z \simeq 4$, 5, 6 \& 7.

This plot also usefully illustrates both the level of agreement 
{\it and} current tension with current theoretical models, several of which 
predict extremely steep faint-end mass-function slopes, $\alpha_{mass} 
\simeq -2$ to $-3$ (e.g. Choi \& Nagamine 
2010; Finlator et al. 2011). The observationally-inferred mass-functions 
in Fig. 17 appear to display significantly flatter low-mass slopes than this, 
with $\alpha_{mass} \simeq -1.4$ to $-1.6$.  
However, it must be emphasized that Fig. 18 is based 
on the assumption that all galaxies at these redshifts follow 
the same mass-to-light relation as inferred at $z \simeq 4$
(and even this relation includes many objects without individual
IRAC detections, and once again 
involves the simplifying assumption of constant star-formation rate).
Thus the fact that
the low-mass end of the mass function has a flatter slope than the 
faint-end slope of the UV LF is a simple consequence of applying
$M_{star} \propto L^{1.7}_{1500}$.
In detail this is clearly wrong, but the question is how wrong? 

\begin{figure}

\includegraphics[scale=0.42, angle=0]{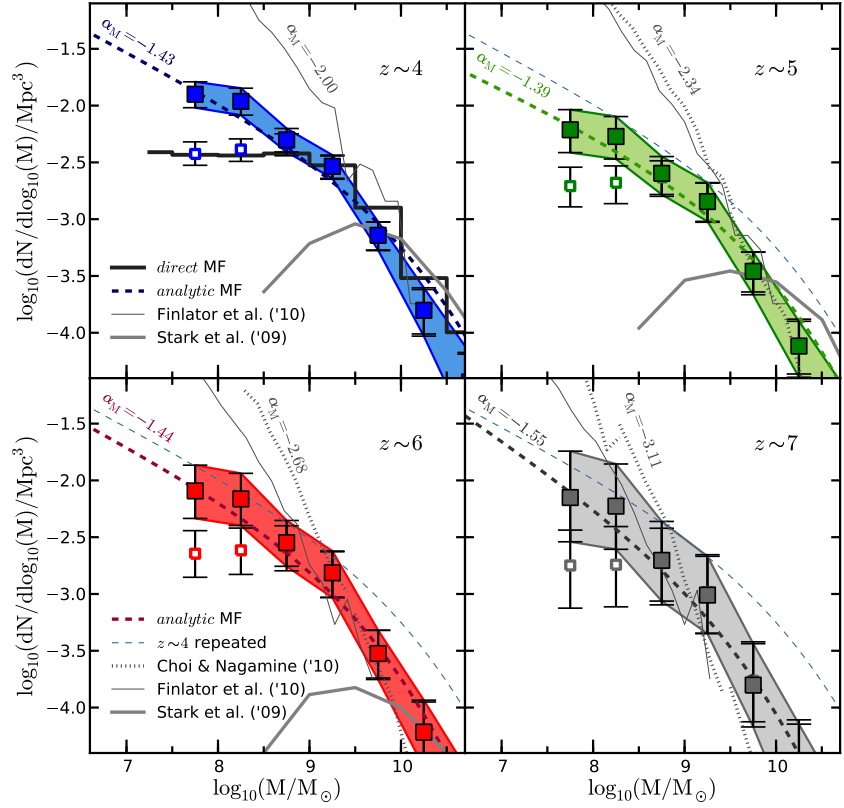}

\caption{The stellar mass functions of LBGs 
at $z\sim4,~5,~6,$ and 7 as produced by Gonz\'{a}lez et al. (2011) 
by applying the $M_{star}-L_{1500}$ 
distribution found for $z\sim4~B$-dropouts (within the WFC3/IR ERS field)
to the UV LFs of Bouwens et al. (2007, 2010) at $z\sim4-7$. 
For masses $M_{star}~>~10^{9.5}\,{\rm M_{\odot}}$, the 
$z<7$ mass functions are in reasonable  agreement with the earlier 
determinations by Stark et al. (2009) and McLure et al. (2009).
The thick dashed curve in each panel represents the 
\emph{analytic} mass functions derived from an idealized $M_{star}-L_{1500}$ 
relation which, given the adopted form of $M_{star} \propto L^{1.7}_{1500}$,
inevitably display somewhat flatter low-mass slopes 
$\alpha_{mass} \simeq -1.4$ to $-1.6$,
than the faint-end slopes in the 
LBG UV LFs ($\alpha=-1.7$ to$-2.0$; see section 4.1). 
The $z \simeq 4$ \emph{analytical} mass function  is repeated in
the other panels for comparison (thin dashed curve).
The dotted and thin solid lines show the simulated mass functions
from Choi \& Nagamine (2010) and Finlator et al. (2011) 
(courtesy V. Gonz\'{a}lez).}
\label{fig:18}      
\end{figure}

This issue has recently been explored by McLure et al. (2011) who, for 21 
galaxies at $z > 6.5$ which {\it do} have IRAC detections, explored the extent 
to which $M_{star}$ for individual galaxies changes if the assumption
of a universal $M_{star}-L_{1500}$ relation is relaxed, and the full parameter space of
age, star-formation history, dust-extinction and metallicity is explored
in search of the best model fit. The results of this analysis are shown in 
Fig. 19. For individual objects, $M_{star}$ can vary considerably depending
on the adopted model, and it appears that stellar mass can span a factor 
of up to $\simeq 50$ at a given UV luminosity. However, for 
most objects it is found that the size of the mass 
uncertainty is generally limited
to typically a factor of $\simeq 2 -3$, partly
by the lack of available cosmological time
(see also Labb\'{e} et al. 2010b); this is one (the only?)
benefit of working at $z > 5$, namely that the impact on 
$M_{star}$ of a plausible range of star-formation histories is damped somewhat 
by the fact that less than 1 billion years is available.  Another 
interesting outcome of this analysis is that, despite the increase in 
scatter in $M_{star}$, the {\it average} value was indeed still 
found to be consistent with 
$\langle sSFR \rangle \simeq 2-3\,{\rm Gyr^{-1}}$, and McLure et al.
(2011) also confirmed that the mass-luminosity relation at $z \simeq 7$ 
is at least consistent with the $z \simeq 4$ relation adopted
by Gonz\'{a}lez et al. (2011).

\begin{figure}
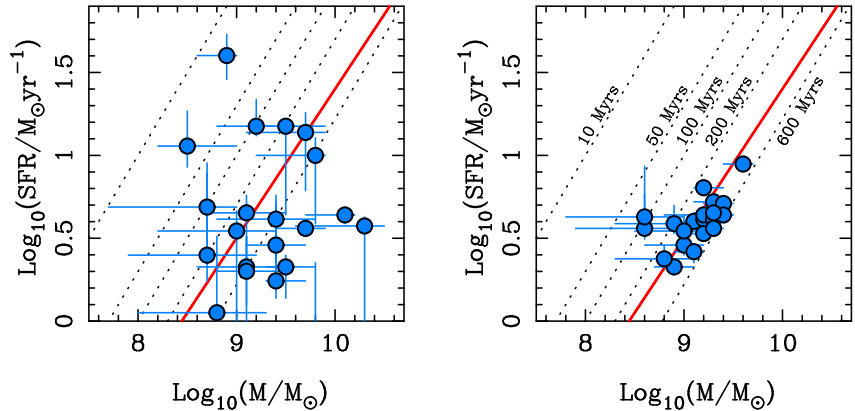


\includegraphics[scale=0.55, angle=270]{fig19a.eps}
\hspace*{0.5cm}
\includegraphics[scale=0.55, angle=270]{fig19b.eps}

\caption{Star-formation rate versus stellar mass for the twenty-one objects in 
the final robust $z > 6.5$ LBG sample of McLure et al. (2011) 
with IRAC detections at either 3.6$\mu$m or
3.6+4.5$\mu$m. In the left-hand panel the star-formation rates
and stellar masses have been measured from the best-fitting SED
template drawn from a wide range of star-formation histories,
metallicities and reddening. In the
right-hand panel the star-formation rates and stellar masses have been
estimated from the best-fitting constant star-formation 
model. 
The $1\sigma$ errors on both parameters have been calculated by determining 
the $\Delta \chi^{2}=1$ interval, after marginalising over all other
free parameters. The solid line in both panels is the SFR$-M_{star}$ relation derived by Daddi et al. (2007)
for star-forming galaxies at $z\simeq 2$ and corresponds to a
sSFR of $\simeq 2.5$ Gyr$^{-1}$. The 
dotted lines illustrate how the SFR$-M_{star}$ relation for a galaxy 
with a constant star-formation rate and zero reddening varies as a function of stellar population age, as marked in the right-hand panel
(courtesy R. McLure).}
\label{fig:19}       
\end{figure}

From these plots it can be seen that the typical stellar mass of an $L^*$ LBG 
detected at $z \simeq 7$ in the current deep WFC3/IR imaging surveys is
$M_{star} \simeq 2 \times 10^9\,{\rm M_{\odot}}$, and the faintest LBGs 
uncovered at these redshifts have masses as low as 
$M_{star} \simeq 1\times 10^8\,{\rm M_{\odot}}$ (see also Finkelstein et al. 
2010). 
This is impressive, as is the effort to establish the typical 
masses of faint LAEs from stacking of the available 
(somewhat shallower) IRAC imaging over the wider-area narrow-band 
Subaru surveys. This work indicates even lower typical stellar masses for LAEs 
selected at $z \simeq 5.7$ and $z \simeq 6.6$, with 
$M_{star} \simeq 1 - 10\times 10^7\,{\rm M_{\odot}}$ (Ono et al. 2010).   

This is not to say that a few significantly more massive LAEs have not been 
uncovered. For example, Ouchi et al. (2009a) discovered ``Himiko'', a giant LAE 
at $z = 6.595$, whose relatively straightforward IRAC detection implies a stellar mass
in the range $M_{star} = 0.9 - 5 \times 10^{11}\,{\rm M_{\odot}}$.
The example of Himiko shows that reasonably massive galaxies can be uncovered
at $z \simeq 7$ given sufficiently large survey areas (in this case 
the parent $z \simeq 6.6$ narrow-band survey covered $\simeq 1\,{\rm deg^2}$,
sampling a comoving volume of $\simeq $800,000\,Mpc$^3$; Ouchi et al. 2010). Such 
discoveries are ``expected'', and 
are entirely consistent with the mass functions shown in Fig. 18.
However, one should obviously be sceptical about claims of enormously massive 
(e.g. $M_{star} > 5 \times 10^{11}\,{\rm M_{\odot}}$) galaxies 
at $z > 5$, especially at very high redshift and/or if discovered 
from very small surveys (e.g. Mobasher et al. 2005). In fact, 
Dunlop et al. (2007) 
found no convincing evidence of any galaxies with 
$M_{star} > 3 \times 10^{11}\,{\rm M_{\odot}}$ at $ z > 4$ in  
125\,arcmin$^2$ of the GOODS-South field, a result which is again consistent 
with the high-mass end of the mass functions shown in Fig. 18.

\subsection{Star-formation histories}

Given the growing evidence that $\langle sSFR \rangle$ is approximately constant 
at early times, it is tempting to conclude that the star-formation rates of 
high-redshift galaxies are {\it exponentially increasing} with time (rather 
than staying constant or exponentially decaying, as generally previously assumed -- 
e.g. Eyles et al. 2007; Stark et al. 2009). 
While the analysis of McLure et al. (2011) indicates that it may be a mistake 
to assume all individual galaxies grow in this self-similar manner, 
and simple exponential growth is not really supported by the 
Gonz\'{a}lez et al. (2011) mass-luminosity relationship, on   
average something at least close to this scenario does indeed appear to be broadly consistent with 
much of the available data from $z \simeq 8$ 
down to $z \simeq 3$ (Gonz\'{a}lez et al. 2010; Papovich et al. 2011).
Interestingly, the latest results from hydrodynamical simulations
strongly predict that the star-formation rates of high-redshift galaxies should be 
increasing approximately exponentially (e.g. Finlator et al. 2011). However, the same simulations 
make the generic prediction that $\langle sSFR \rangle$ should continue to rise 
with increasing redshift beyond $z \simeq 2$, with $\langle sSFR \rangle \propto (1+z)^{2.5}$, tracking
the halo mass accretion rate (e.g. Dekel et al. 2009). 
Thus, if the improving data continue to support an unevolving 
value for $\langle sSFR \rangle$ at high redshift, 
it will likely be necessary to invoke an additional 
feedback mechanism to supress star-formation in the theoretical models at high redshift.
 
It is probably premature to attempt to say anything much more detailed about 
the evolution of the stellar populations in LBGs and LAEs at $z \simeq 5-8$. Without 
high-quality spectroscopy, the determination of detailed star-formation 
histories is inevitably confused by the complications and degeneracies
arising from uncertain dust extinction,
nebular emission, and metallicity, 
issues which are discussed a bit further below 
in the context of UV slopes and ionizing photon escape fractions. Certainly,
the best-fitting star-formation histories for $z \simeq 7$ LBGs deduced by McLure
et al. (2011) range from 10\,Myr-old ``Bursts'' to models involving constant 
star-formation over 700\,Myr, and for each individual object a wide range of 
alternative star-formation histories is generally statistically permitted by the broad-band 
photometric data and the uncertainty in metallicity. Meanwhile, for LAEs, ages as young as 
1--3\,Myr have been inferred for the fainter, bluer objects (e.g. Ono et al. 2010).

Of course, as argued by Stark et al. (2009), in reality the 
star formation in these young galaxies may be highly intermittent or
episodic. Interestingly, it is possible to try to estimate the ``duty cycles'' of LBGs 
and LAEs by reconciling their clustering properties with their number density. As discussed 
by Ouchi et al. (2010), the clustering of LAEs at $z \simeq 6.6$ can be used to infer a typical
halo mass of $M_{halo} \simeq 10^8 - 10^9\,{\rm M_{\odot}}$, 
and then comparison of the predicted number-density of such halos 
with the observed number density of LAEs implies that these galaxies/halos
are observable as LAEs 
for $1-10$\% of the time. These estimates are clearly 
still highly uncertain, not least because the clustering properties of LAEs and LBGs are still not 
very well determined (see section 5.5 below), and indeed, 
based on the luminosity function comparison discussed above, 
it could be argued that the duty cycles of LAE and 
LBG activity must be virtually the same at $z \simeq 6 - 7$.
Nevertheless, such calculations have the potential 
to provide genuinely useful constraints on duty cycles, 
as future surveys for $z \simeq 5-8$ LAEs and LBGs increase in both 
area and depth.

\subsection{Ultraviolet slopes}

The first galaxies, by definition, are expected to contain very young stellar 
populations of very low metallicity. However, the possibility of detecting 
unambiguous observable signatures of such primordial 
stellar populations with current 
or indeed planned future instrumentation is currently 
a matter of considerable debate.

For example, one long-sought distinctive 
spectral signature of the first generation 
of galaxies is relatively strong HeII emission at $\lambda_{rest} = 1640$\AA\ 
(e.g. Shapley et al. 2003; Nagao et al. 2008; di Serego Alighieri
et al. 2008). However, near-infrared spectroscopy of the sensitivity required 
to detect this line at $z > 7$ will certainly not be available until the 
{\it JWST}, and even then some theoretical
predictions indicate that it is unlikely to be found in detectable objects
(Salvaterra, Ferrara \& Dayal 2011, but see also
Pawlik, Milosavljevic \& Bromm 2011). 

By necessity, therefore, recent attention has focussed on whether 
the broad-band near-infrared photometry which has now been 
successfully used to discover galaxies at $z \simeq 6.5 - 8.5$ 
can actually be used 
to establish the rest-frame continuum slopes of the highest redshift galaxies.
Specifically, very young, metal-poor stellar populations
are arguably expected to result in substantially bluer continuum slopes around 
$\lambda_{rest} \simeq 1500$\AA\ than have been detected to date in galaxies 
discovered at any lower 
redshift $z < 6.5$ (e.g. Steidel et al. 1999; Meurer et al. 1999; 
Adelberger \& Steidel 2000; Ouchi et al. 2004; Stanway et al. 2005; Bouwens et al. 2006; Hathi et al. 2008; 
Bouwens et al. 2009b; Erb et al. 2010).

It has become the normal convention 
to parameterise the ultra-violet continuum slopes of 
galaxies in terms of a power-law index, $\beta$, where $f_{\lambda} \propto
\lambda^{\beta}$ (e.g. Meurer et al. 1999; 
thus, $\beta = -2$ corresponds to a source 
which has a flat spectrum in terms of $f_{\nu}$, and hence 
has zero colour in the AB magnitude system). As discussed by several authors,
while the bluest galaxies observed at $z \simeq 3 - 4$ have $\beta \simeq -2$, 
values as low (i.e. blue) as $\beta = -3$ can in principle be produced 
by a young, low-metallicity stellar population (e.g. 
Bouwens et al. 2010b; Schaerer 2002). However, as illustrated in Fig. 20, for 
this idealized prediction to actually be realized in practice, several 
conditions have to be satisfied simultaneously, 
namely i) the stellar population 
has to be very young 
(e.g. $t < 30$\,Myr for metallicity $Z \simeq 10^{-3}\,{\rm Z_{\odot}}$, or 
$t < 3$\,Myr for $Z \simeq 10^{-2}\,{\rm Z_{\odot}}$), 
ii) the starlight 
must obviously be completely free from any significant 
dust extinction,
and iii) the starlight must also {\it not} be significantly contaminated by 
(redder) nebular continuum (a condition which has important implications 
for UV photon escape fraction, and hence reionization --  see, for example,
Robertson et al. 2010).

\begin{figure}

\includegraphics[scale=0.45, angle=0]{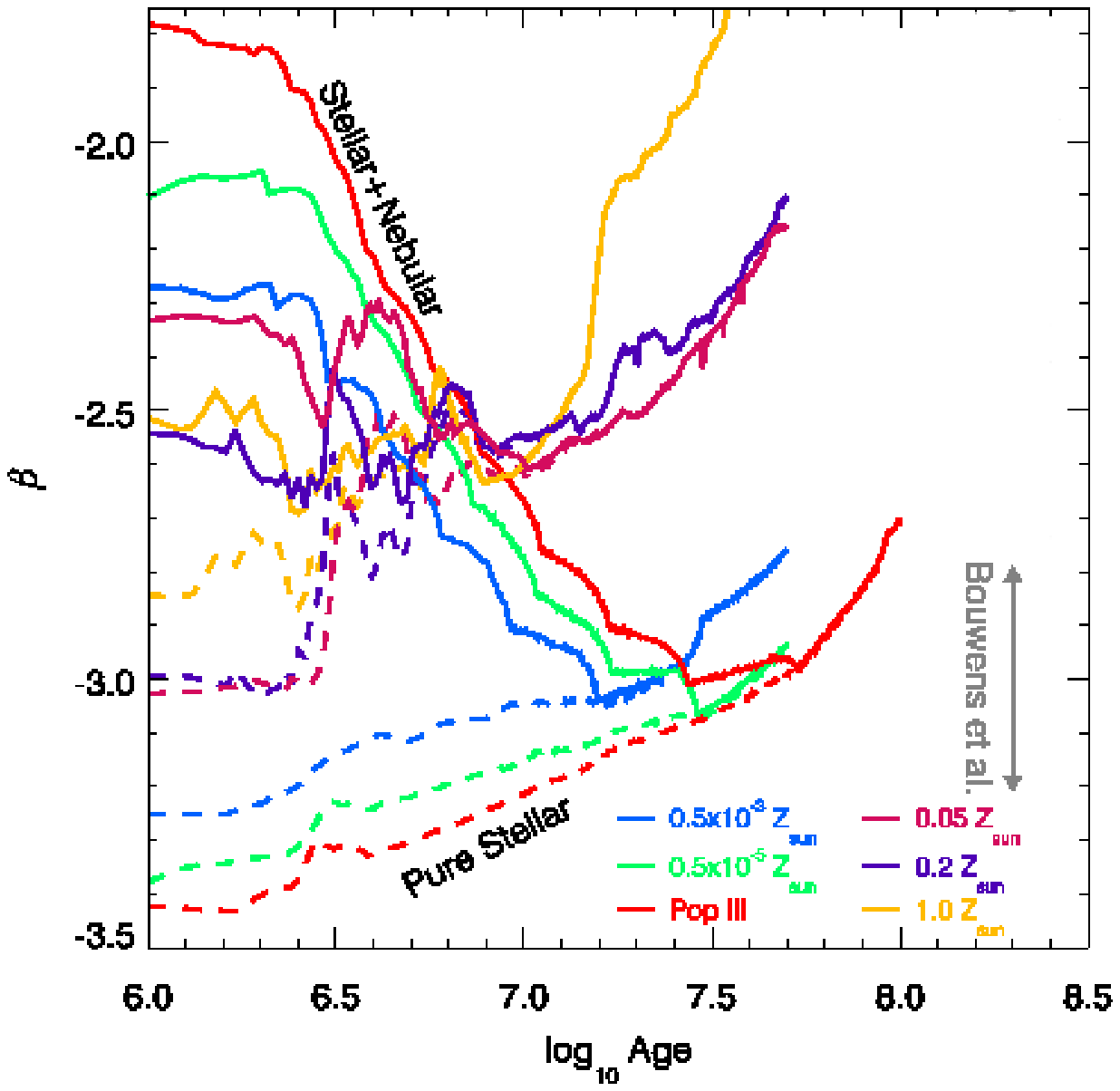}
\includegraphics[scale=0.45, angle=0]{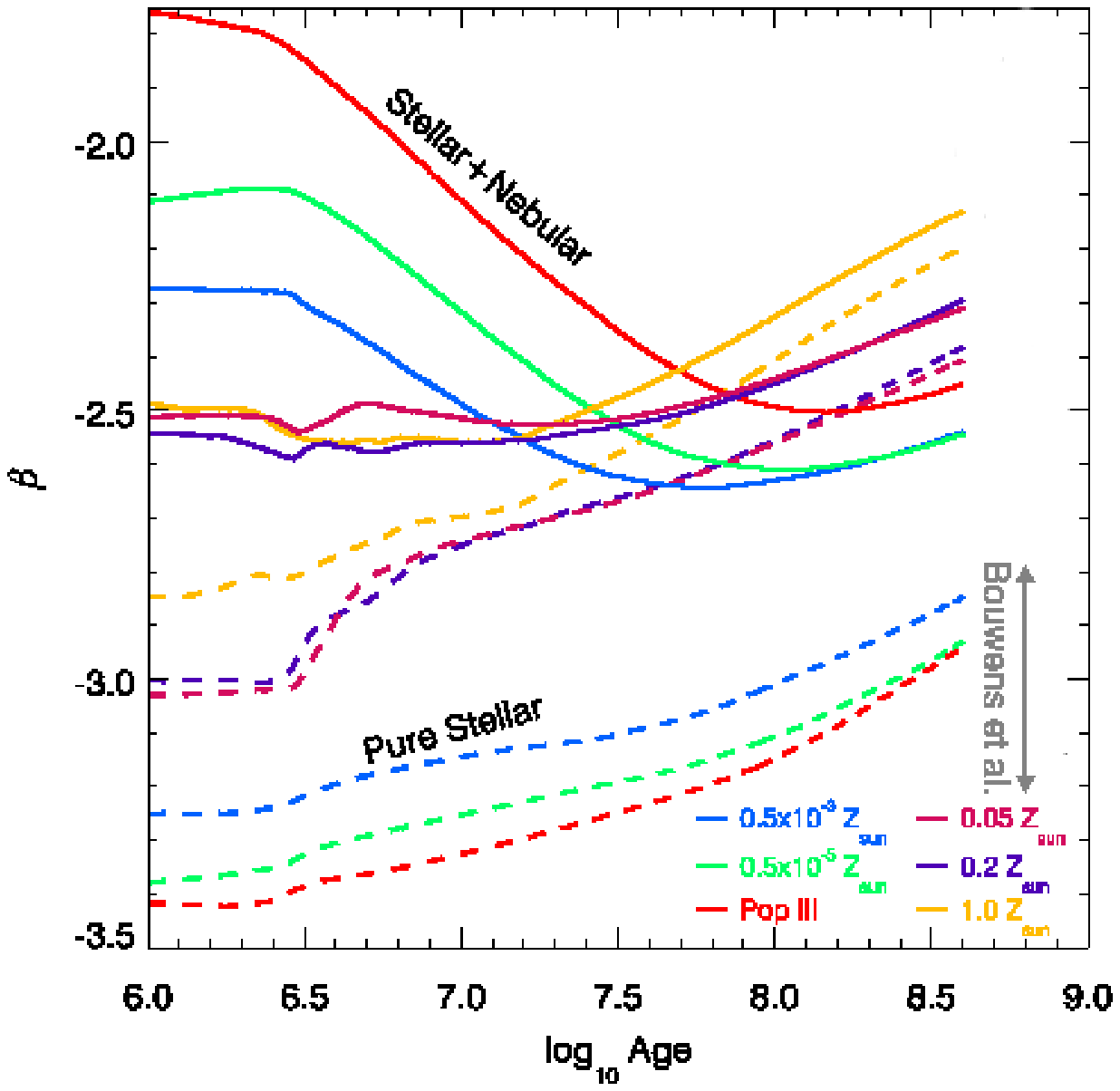}

\caption{Theoretical predictions of galaxy UV-slope $\beta$ showing the values that are expected for 
stellar populations of different age and metallicity, and the reason for 
the interest generated by intial claims that faint galaxies at $z \simeq 7$ display $\langle \beta \rangle \simeq -3$
(Bouwens et al. 2010a). 
The left-hand panel shows the predicted evolution of $\beta$ 
for instantaneous starbursts of differing metallicity, with and without predicted nebular continuum (Schaerer 2003). The right-hand panel shows equivalent information for the (arguably more realistic) case of continuous star-formation. More recent measurements
of $\langle \beta \rangle$ have failed to find evidence for values $\langle \beta \rangle < -2.5$ at $z \simeq 7$, and have converged
on $\langle \beta \rangle = -2.1 \pm 0.1$ for brighter ($L^*$) galaxies at $z \simeq 7$. These plots show that
while extreme values of $\langle \beta \rangle < -2.8$ would imply 
 both very low metallicity, {\it and} a high UV-photon escape fraction, the interpretation of the more moderate values 
actually observed (i.e. $\langle \beta \rangle = -2.0 \rightarrow -2.5$), is much less 
straightforward.}
\label{fig:20}      
\end{figure}

For this reason, the report by Bouwens et al. (2010a) (supported to
some extent by Finkelstein et al. 2010) that the 
faintest galaxies detected at $z > 6.5$ apparently display an average 
value of $\langle \beta \rangle = -3.0 \pm 0.2$ was both exciting and arguably 
surprising, and was immediately subjected to detailed theoretical 
interpretation (e.g. Taniguchi et al. 2010).

However, a series of further observational studies of UV slopes over the last year have 
failed to confirm this result, revealing that the original measurement was biased towards 
excessively blue slopes in a subtle but significant way (e.g. Dunlop et al. 2012; 
Wilkins et al. 2011b; Finkelstein et al. 2012; Bouwens et al. 2012). It now 
seems likely that, for the faintest galaxies uncovered so far at $z \simeq 7$, the 
average UV power-law index lies somewhere in the range $\langle \beta \rangle = -2.5 \rightarrow - 2.0$. 
As can be seen from Fig. 20, the correct interpretation of such slopes is unclear, as 
they can be produced by different mixes of age, metallicity, and nebular contributions;
it is only in the extreme case of  $\langle \beta \rangle \simeq -3.0$ that the interpretation 
in favour of exotic stellar populations and large escape fractions becomes relatively clean.

Although these more moderate UV slopes are arguably in better accord with theoretical 
expectations (e.g. Dayal \& Ferrara 2012) it may still be the case 
that some individual galaxies at $z \simeq 7$ with $\beta = -3$ have already been discovered
among the faint LBG or LAE samples (e.g. Ono et al. 2010). 
Unfortunately, however, this is impossible to check with current data 
because $\beta$ is such a sensitive 
function of observed colour. At $z \simeq 7$, the estimate of $\beta$ 
for an {\it HST}-selected LBG has currently to be based on a single colour, with 

\begin{equation}
\beta = 4.43(J_{125}-H_{160})-2.
\end{equation}

Thus, a perfectly ``reasonable'' photometric uncertainty of 
$\simeq 15$\% in $J_{125}$ and $H_{160}$ translates to a $\simeq 20$\% uncertainty
in colour and hence to an 1-$\sigma$ uncertainty of $\pm 0.9$ in $\beta$. Better 
photometric accuracy, ideally combined with additional near-infrared wavebands to allow
improved power-law or SED fitting (as is already possible at lower redshifts; 
Finkelstein et al. 2012; Castellano et al. 2012) is required to enable
a proper investigation of the UV slopes of the faintest galaxies at $z \simeq 7 - 8$.

In contrast to this confusion at the faintest luminosities, 
there does at least seem to be general agreement that the brighter galaxies found at $z \simeq 7$ 
(with $M_{1500} \simeq -21$) have 
 $\langle \beta \rangle = -2.1 \pm 0.1$. This is basically as expected for a few hundred Myr-old
star-forming galaxy, with solar metallicity and virtually no dust extinction, although 
other interpretations are possible (see Fig. 20). 
Certainly, for galaxies at these brighter luminosities, any evolution
in $\langle \beta \rangle$ with redshift has generally been interpreted as arising 
predominantly from a change in the level of dust obscuration, as already discussed above in the 
context of Lyman-$\alpha$ emission. In particular, Bouwens et al. (2009b) have reported that  
the average value of $\beta$ exhibited by brighter LBGs declines from $\langle \beta \rangle \simeq -1.5$ at 
$z \simeq 4$ to $\langle \beta \rangle \simeq -2$ at $z \simeq 6$, and have interpreted 
this as reflecting a progressive reduction
in average extinction with increasing redshift.
This is an important result, because it leads to the conclusion that the 
correction to be applied to observed UV luminosity density 
to account for dust-obscured star-formation steadily decreases with increasing 
redshift. This has obvious implications for the inferred 
history of cosmic star-formation density as discussed further below in section 6.1. 
However, the precise redshift dependence of average extinction, and indeed whether
there exists a clear relationship between UV luminosity and spectral slope 
at high redshift is 
still the subject of some controversy (Dunlop et al. 2012;
Bouwens et al. 2012; Finkelstein et al. 2012; Castellano et al. 2012).

\subsection{Galaxy sizes and morphologies}

Only the most basic information is known about the morphologies of 
LBGs and LAEs at $z > 5$, for the simple reason that the objects are faint, 
and are thus generally detected at only modest signal:noise ratios. Oesch et al. (2010b) 
investigated the first WFC3/IR imaging of LBGs at $z \simeq 7 - 8$, and 
were able to show that almost all of these galaxies are marginally resolved, with an average intrinsic size 
of $\simeq 0.7 \pm 0.3$\,kpc. Thus, known extreme-redshift LBGs are clearly very compact
(certainly too compact to be resolved with typical ground-based imaging),
and the detection of extended features is, to date, rare. Comparison with 
the sizes of LBGs at lower redshift implies that average size decreases gently
from $z \simeq 4$ to $z \simeq 7$, following approximately a relationship of the 
form $(1 + z)^{-1}$ (Fig. 21). 
A slow decrease in average size at a fixed luminosity 
with increasing look-back time 
is anticipated from semi-analytic models of galaxy 
formation (e.g. Mo et al. 1998, 1999; Somerville et al. 2008; Firmani et al. 2009), 
and is consistent with earlier observations of lower-redshift
LBGs (e.g. Ferguson et al. 2004; Bouwens et al. 2004b) and disc galaxies (e.g. Buitrago et al. 2008). 
However, it is still unclear whether the apparently observed relationship at high redshift is, at least in part, a consequence of the fact 
that galaxy detection with {\it HST} is 
biased towards objects which have high surface brightness 
(e.g. Grazian et al. 2011). Oesch et al. (2010b) claim that this is not a significant 
problem, but {\it HST} imaging covering a greater dynamic range, and providing larger samples for stacking,
should certainly help to clarify this issue in the near future.

\begin{figure}

\includegraphics[scale=0.75, angle=0]{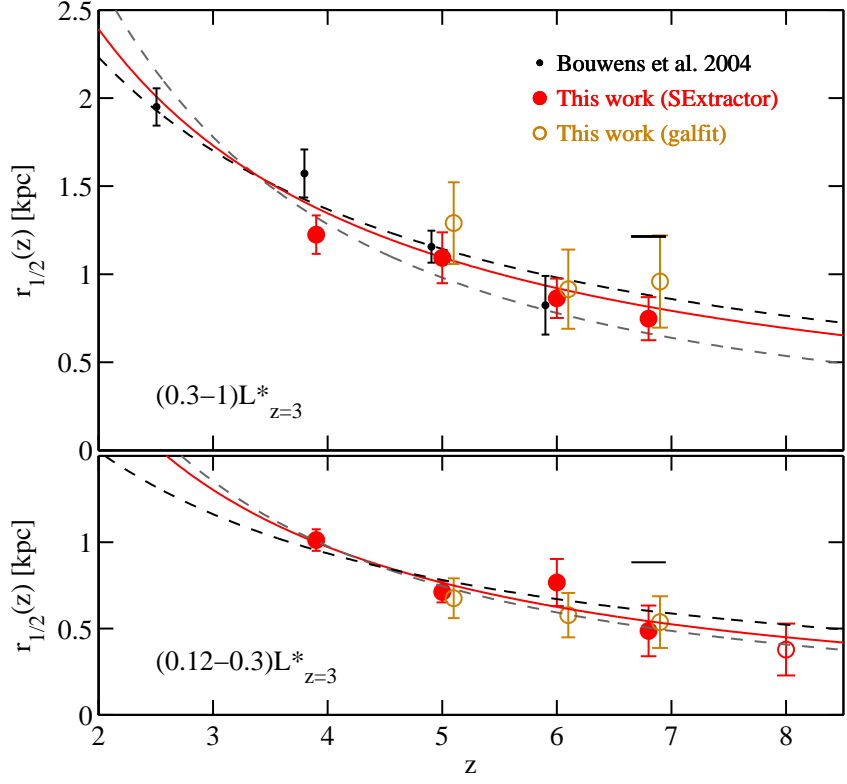}

\caption{The observed evolution of the mean size of Lyman-break 
galaxies reported by Oesch et al. (2010b), over 
the redshift range $z\sim2-8$ in two 
different luminosity ranges (0.3-1)$L^*_{z=3}$ (top) 
and (0.12-0.3)$L^*_{z=3}$ (bottom) (where $L^*_{z=3}$ is the characteristic 
luminosity of a LBG at $z \simeq 3$). Different symbol styles correspond to 
different ways of analysing the data to extract size estimates.
The dashed lines indicate the scaling expected for a fixed dark matter
halo mass 
($\propto (1+z)^{-1} \equiv \propto H(z)^{-2/3}$; black) 
or at fixed halo circular velocity ($\propto (1+z)^{-3/2} \propto H(z)^{-1}$; 
gray). The solid red lines indicate the best fit to the observed evolution,
which is described as proportional to $(1+z)^{-m}$, 
with $m=1.12\pm0.17$ for the brighter luminosity bin, and $m=1.32\pm0.52$ at fainter luminosities (but both are formally identical, and 
consistent with $m \simeq 1$). The extent to which this apparent
relationship is influenced by the surface-brightness bias 
inherent in deep {\it HST} imaging is still a matter of some 
debate (courtesy P. Oesch)}
\label{fig:21}       
\end{figure}

At somewhat lower redshifts, Taniguchi et al. (2009) used 
the {\it HST} ACS single-orbit $I_{814}$ imaging in the COSMOS field to attempt to investigate
the morphologies of 85 LAEs at $z \simeq 5.7$, selected via Subaru narrow-band imaging.
The results of this study are somewhat inconclusive, with only 47 LAEs being detected, and 
approximately half of these being apparently unresolved. Nevertheless, the result is a typical half-light
radius of $\simeq 0.8$\,kpc, clearly not inconsistent with that displayed by LBGs at comparable
redshifts. Taniguchi et al. (2009) also report that fits to the light profile of their stacked
images favour a S\'{e}rsic index $n \sim 0.7$, more consistent with disc-like or irregular 
galaxies than with a de Vaucouleurs spheroid.

\subsection{Clustering}

A measurement of the clustering of high-redshift galaxies is of interest primarily for 
estimating the characteristic mass of the dark matter halos in which they reside. If this can 
be established with meaningful accuracy, then a comparison of the observed galaxy 
number density with that of the relevant halos (as predicted within the concordance 
${\rm \Lambda}$-CDM cosmological model) can yield an estimate of the halo occupation fraction,
or, equivalently, the duty cycle of a given class of high-redshift galaxy.

The measurement of galaxy clustering at $z > 5$ is still in its infancy, due primarily
to the small sample sizes delivered by current facilities. To date it has been most 
profitably pursued using the samples of several hundred LAEs selected over degree-scale 
fields via the narrow-band Subaru imaging targetted at $z \simeq 5.7$ (Ouchi et al. 2005)
and $z \simeq 6.6$ (Ouchi et al. 2010). Fig. 22 shows the distribution
on the sky of the $\simeq 200$ LAEs at $z \simeq 6.6$ uncovered in the SXDS field by Ouchi et al. (2010).

To quantify the significance and strength of any clustering present in such images, the standard 
technique is to calculate the Angular Correlation Function, $\omega(\theta)$, which 
represents the excess (or deficit) of objects at a given angular radius from a galaxy 
relative to that expected from a purely random distribution of galaxies with the observed
number counts. This is usually calculated following the prescription of Landy \& Szalay 
(1993), which gives

\begin{equation}
\omega_{\rm obs}(\theta)
  = [DD(\theta)-2DR(\theta)+RR(\theta)]/RR(\theta),
\label{eq:landyszalay}
\end{equation}

\noindent
where $DD(\theta)$, $DR(\theta)$, and $RR(\theta)$ are the numbers of
galaxy-galaxy, galaxy-random, and random-random pairs normalized by
the total number of pairs in each of the three samples.

Considerable care and simulation work is required to calculate $\omega_{\rm obs}(\theta)$ 
especially when, as shown by the grey regions in Fig. 22, several areas of the image have to 
be masked out due to bright stars or image artefacts. As described in detail by Ouchi 
et al. (2005, 2010), $\omega_{\rm obs}(\theta)$ is converted to a best estimate of 
$\omega(\theta)$, then used to derive a clustering amplitude $A_{\omega}$ assuming a 
power-law correlation function $\omega(\theta) = A_\omega \theta^{-\beta}$, in which 
$\beta$ has generally to be fixed rather than fitted given the limited sample size
(usually $\beta = 0.8$ is adopted on the basis of clustering analyses at lower redshift).
Finally $A_\omega$ is coverted into a physical correlation length $r_0$ using knowledge 
of the redshift distribution (which is relatively straightforward to establish 
for narrow-band selected LAEs).
\begin{figure}

\includegraphics[scale=0.6, angle=0]{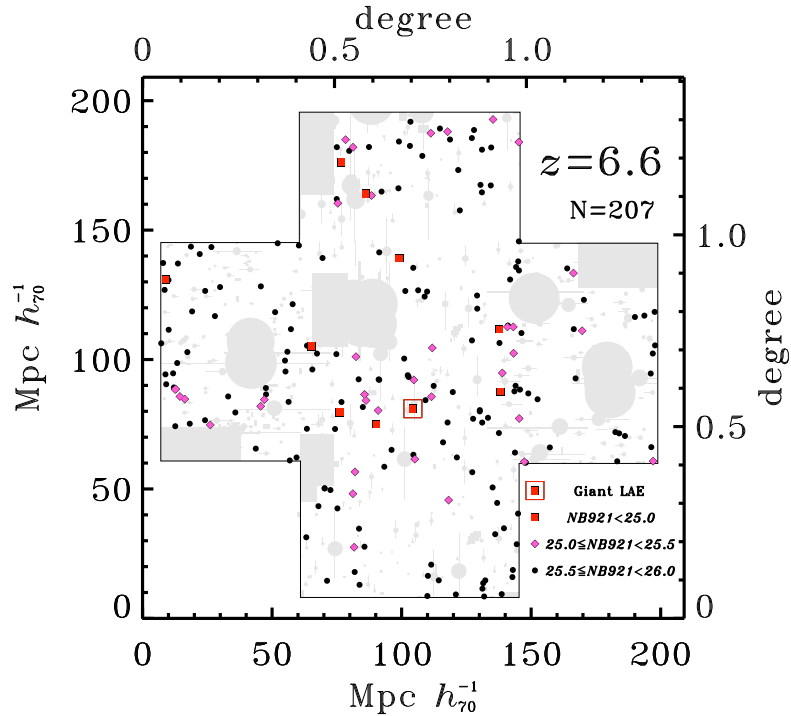}
\caption{The distribution on the sky of the 207 LAEs at $z = 6.565 \pm 0.054$ 
detected by Ouchi et al. (2010) via narrow-band imaging 
with Subaru in the SXDS field. 
Red squares, magenta diamonds, and black circles
indicate the positions of bright ($NB921<25.0$),
medium-bright ($25.0\le NB921\le 25.5$),
and faint ($25.5\le NB921\le 26.0$) LAEs,
respectively.
The red square surrounded by  
a red open square indicates 
the giant LAE, {\it Himiko}, at $z = 6.595$
reported by Ouchi et al. (2009a), and already briefly discussed  
in section 5.1 (courtesy M. Ouchi).}
\label{fig:22}     
\end{figure}

The result of these studies is that significant clustering has now been detected 
in the LAE population at $z > 5$, and that the best estimate of the correlation length
for LAEs at $z \simeq 5-7$ is $r_0 = 3 - 7$\,Mpc (for $H_0 = 70\,{\rm km\,s^{-1}\,Mpc^{-1}}$).
This in turn can be used to infer an average mass for the dark-matter halos hosting these 
LAEs of $10^{10}-10^{11}\,{\rm M_{\odot}}$. A similar analysis for LBGs at $5 < z < 6$ 
has been performed by McLure et al. (2009), who report a correlation length 
of $r_0 = 8^{+2}_{-1.5}\,{\rm Mpc}$ (for $H_0 = 70\,{\rm km\,s^{-1}\,Mpc^{-1}}$) and 
a resulting characteristic dark-matter halo mass of $10^{11.5}-10^{12}\,{\rm M_{\odot}}$.
This significantly larger halo mass for the LBGs is not unexpected, given that they 
are considerably rarer, more massive objects (than typical narrow-band selected LAEs), having  
been selected from a substantially larger cosmological volume, down to much brighter continuum 
flux limits. These results are consistent 
with those derived by Overzier et al. (2006) for LBGs at a mean redshift of $z \simeq 5.9$.
Interestingly, the $z > 5$ LBGs studied by McLure et al. (2009) are bright enough
to allow a reasonable estimate of average stellar mass, $\log_{10}(M/M_{\odot}) = 10.0^{+0.2}_{-0.4}$,
which is consistent with the results of the clustering analysis for plausible values of 
the ratio of stellar to dark matter.

Unfortunately the large uncertainty in current estimates of characteristic halo mass,
combined with the steepness of the halo mass function, means that such clustering 
measurements cannot yet be used to yield meaningfully-accurate duty cycles
for LAEs and LBGs (e.g. Ouchi et al. (2010) report an inferred duty cycle
of $\simeq 1$\% for $z \simeq 6.6$ LAEs, but acknowledge this is extremely uncertain).
Nonetheless, these pioneering studies provide 
hope for meaningful measurements with the much larger LAE and LBG samples anticipated 
from Hyper-Suprime Cam on Subaru (Takada 2010) 
and from the {\it EUCLID} Deep survey over the next decade (Laurejis et al. 2011).

As discussed in section 6.2, such future large-scale surveys also have the potential 
to search for one of the long anticipated signatures of reionization, namely 
an enhancement in the clustering of LAEs relative to LBGs due to patchy reionization. At present,
Ouchi et al. (2010) report no evidence for such a clustering amplitude boost at $z \simeq 6.6$.

\section{Global perspective}

\subsection{A consistent picture of galaxy evolution?}

\subsubsection{Cosmic star-formation history}

To gain a broader view of the time evolution of action in the
Universe the evolving galaxy UV luminosity function discussed in 
section 4.1 can be integrated (over luminosity) to yield the evolving {\it comoving UV luminosity density}. 
This might be of academic interest were it not for the fact that this quantity can, in principle, be converted into {\it star-formation density},
to gain a global view of the evolution of
star-formation activity per unit comoving volume over cosmic time.

This calculation has been performed by many authors over a wide range of redshifts since being first promoted by Lilly et al. (1996) and Madau et al. (1996). 
It is, however, a calculation fraught with danger as it involves (and can depend critically upon) several extrapolations; the galaxy luminosity function has to
be correctly extrapolated to the lowest
luminosities, the stellar mass function has to be correctly extrapolated
to the lowest masses, and any mass and/or time dependence of the obscuring effects of cosmic dust has to be correctly accounted for and removed. In addition, care has to be taken to account for highly-obscured populations which may be entirely missed in UV-selected galaxy samples.

A full review of the many, and continually-improving studies of cosmic star-formation history is obviously beyond the scope of this $z > 5$ review.
However, to place the high-redshift results in context, it is fair to
say that there is now broad agreement that star-formation density rises by an order of magnitude as we look back from $z = 0$ to $z = 1$, increases 
by a further factor of 2 or 3 by $z \simeq 2$, and then appears to plateau out to $z \simeq 3$ before declining at still higher redshifts (see e.g. Hopkins \& Beacom 2006;
Dunlop 2011). This evolution is shown in Fig. 23 (taken from Bouwens 
et al. 2011b), but this particular figure has been deliberately designed 
to focus on the apparently smooth and steady decline of star-formation 
density from $z \simeq 3$ to $z > 8$.

\begin{figure}

\includegraphics[scale=0.6, angle=0]{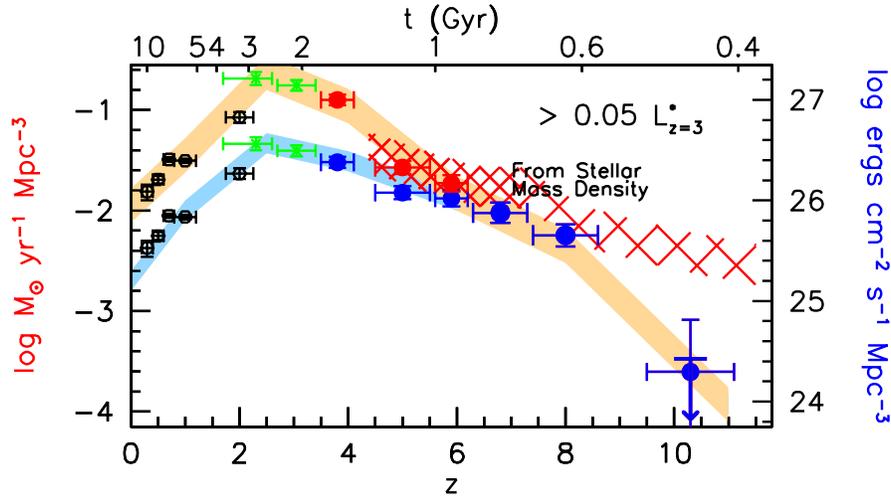}

\caption{A compilation of UV luminosity density measurements, and hence 
inferred evolution of cosmic star-formation rate density as presented 
by Bouwens et al. (2011b). The data-points at $z = 4 - 8$ are basically based 
on integrating the luminosity functions described in Fig. 11 down 
$M_{UV} \simeq -18$ AB mag, while data points at lower redshift from 
Reddy \& Steidel (2009), Bouwens et al. (2007) and Schiminovich 
et al. (2005) are provided for context. The blue data points and lower shaded 
regions indicate UV luminosity density prior to any correction for dust 
obscuration. The upper locus indicates the effects of correcting for 
redshift dependent dust onscuration which, as apparent, is assumed here 
to decline to zero by $z \simeq 7$ following Bouwens et al. (2009b). The 
red hatched region is intended to indicate the SFR density derived from 
differentiating the growth in stellar mass density delineated below in 
Fig. 24. Results at $z > 8$ are probably not meaningful, as the 
stellar mass density is based on an assumed extrapolation, while the $z \simeq 10$ datapoint is based on the single $z \simeq 10$ galaxy in the HUDF claimed by 
Bouwens et al. (2011a) and additional 
tentative upper limits based on non detections (Oesch et al. 2012). (courtesy R. Bouwens)}
\label{fig:23}
\end{figure}

Crucial to the precise form of this plot is the assumed strength and 
redshift dependence of typical dust obscuration as a function of redshift,
which in this case is assumed to decline from a factor of $\simeq 7$
at $z \simeq 2$ to essentially zero at $z \simeq 7$. The true redshift 
dependence of dust extinction in LBGs of course remains a matter of debate,
but several independent pieces of evidence 
point towards a high-redshift decline (e.g. Fig. 13), and even with 
dust corrections it is hard to escape the conclusion that 
SFR density is significantly lower at $z \simeq 7$ than at the peak 
epoch corresponding to $z \simeq 2 -3$.

Finally, it is worth noting that while the datapoint at $z \simeq 10$ should 
probably be taken with a pinch of salt (but see also Oesch et al. 2012), the 
decline in SFR density from $z = 4$ to $z = 8$ is in fact fairly precipitous 
when viewed in terms of the $\simeq 1 $\,Gyr of elapsed cosmic time.
Current observations of the high-redshift UV LF thus support the view that we 
are witnessing the rapid emergence of the star-forming galaxy population.

\subsubsection{The growth of cosmic stellar mass density}                       

\begin{figure}

\includegraphics[scale=0.9, angle=0]{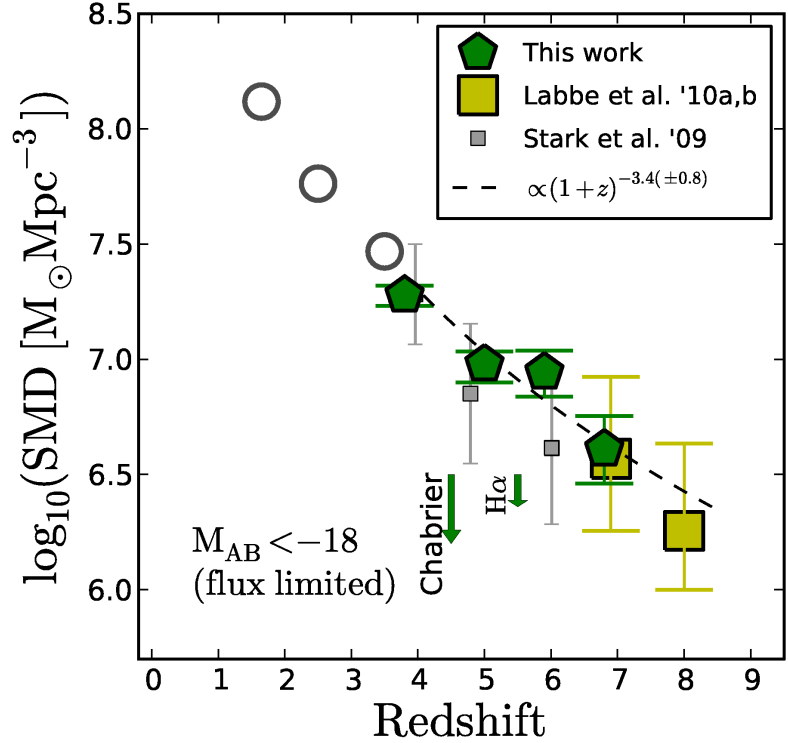}

\caption{Stellar Mass Density as a function of
    redshift for sources brighter than
    $M_{1500}=-18$ as derived by Gonz\'{a}lez et al. (2011). These 
    stellar-mass density values were produced by integrating the 
    mass functions shown in Fig. 18 to 
    $M_{1500}=-18$ at $z=4,~5,~6,~\rm{and}~7$.  Shown for comparison 
    are the stellar-mass density determinations from 
    from Stark et al. (2009) corrected to the same $M_{1500}=-18$ limit.
    The low-redshift open circles were derived by integrating
    the Marchesini et al. (2009) mass functions between
    $\rm{8.3<log_{10}M_{star}/M_\odot)<13}$ and multiplying by 1.6
    to match the Salpeter IMF. A constant SFH and 0.2$\,$Z$_\odot$
    metallicity was assumed to derive all stellar masses at 
    $z \ge 4$. The
    effect of a possible 20\% correction due to contamination by
    H$\alpha$ is shown, as is the effect of using a different IMF.
    The
    integrated mass growth shown here is well described by
    $\rm{log_{10}(SMD)\propto(1+z)^{-3.4\pm0.8}}$ (courtesy V. Gonz\'{a}lez).}
\label{fig:24}       
\end{figure}

An alternative route to determining the cosmic history of star-formation
is provided by integrating the evolving galaxy stellar mass functions (over stellar mass) to map out the build-up of {\it comoving stellar mass density} with cosmic time. In principle this can be used as a check on the validity of the assumptions (concerning, for example, dust) used to estimate cosmic
star-formation history as described above, because the stellar mass density in place at any epoch should (modulo some stellar mass loss) equate to the time integral of all preceding star-formation activity. At $z > 5$ these 
calculations are arguably still premature. However, given data of sufficient 
quality, they may in fact be more straightforward and less uncertain 
than at more modest redshifts, principally because serious
dust obscuration may be less of a problem in the young, relatively metal-poor Universe.

Fig. 24 shows a compilation of recent determinations of 
stellar mass density at high redshift taken from Gonz\'{a}lez et al. (2011), 
with the dark-green pentagons in effect being based on
integration of the stellar mass functions shown in Fig. 18 (again down to 
$M_{UV} \simeq -18$, for ease of comparison with Fig. 23). 
Despite the well-documented uncertainties in 
current measurements of the stellar mass function at early times, this
figure indicates that there is now reasonably good evidence for 
a smooth, monotonic rise in the integrated stellar mass density with decreasing redshift.
Moreover, while there have in the past been some problems in reconciling 
the growth in stellar mass density with the directly observed SFR
density (e.g. Wilkins, Thretham \& Hopkins 2008), 
this situation has improved (e.g. Reddy \& Steidel 2009), 
and the reasonable agreement seen here between the time differential of 
Fig. 24 with the SFR density plotted in Fig. 23 (as indicated by the red
hatched region) arguably provides some confidence that neither measurement is 
too far off. It also suggests that fears the IRAC fluxes from many 
$z \simeq 7$ galaxies are dominated by extreme nebular emission lines 
(rather than starlight) may have been somewhat  
exaggerated (Schaerer \& Barros 2010; see also McLure et al. 2011).

It is however, probably premature to conclude that the agreement is 
sufficiently good to support the assumption of zero dust obscuration 
at high redshift, especially given the current limitations 
in constraining the low-mass end of the stellar mass function at $z > 4$.
Deeper {\it HST} data will undoubtedly help further progress in this area,
as of course will the higher resolution mid-infrared imaging to be 
delivered by {\it JWST} (see section 7).

\subsection{Cosmic reionization}

\subsubsection{Current constraints on reionization}

A second arena in which the integrated UV luminosity density of the evolving high-redshift galaxy population is of interest is in the study of cosmic reionization. 
The reionization of the hydrogen gas that permeates the Universe was a landmark event in cosmic history. It marked the end of the so-called cosmic
``dark ages'', when the first stars and galaxies 
formed, and when the intergalactic gas was heated to tens of thousands 
of degrees Kelvin from much smaller temperatures. 
This global transition had far-reaching effects on the formation of the early 
cosmological structures and left deep impressions on subsequent galaxy 
and star formation, some of which almost certainly persist to the present day. 

The study of this epoch is thus arguably the key
frontier in completing our understanding of cosmic history, and
naturally the focus of much current research. Nevertheless, 
despite the considerable recent progress in both theory and observations 
(for recent reviews see Robertson et al. (2010), McQuinn et al. (2010)) 
all that is really established  about this 
crucial era is that it was completed by redshift $z \simeq 6$ 
(as evidenced by the Gunn-Peterson troughs in the spectra of the most distant quasars; Fan et al. 2006) 
and probably commenced around 
$z \sim 15$ (as suggested by the latest WMAP microwave polarisation 
measurements, which favour a mean redshift of reionization 
of $10.4 \pm 1.4$; Komatsu et al. 2009). However, as discussed by Dunkley et al. (2009), within these bounds the reionization history is essentially unknown,
and with current data we cannot even distinguish whether it was ``sharp'' or 
extended.

Unsurprisingly, therefore, understanding reionization is one of the key science goals for
a number of current and near-future large observational projects. In 
particular, it is a key science driver for the new generation of major
low-frequency radio projects (e.g. LOFAR, MWA and SKA) which aim to map out
the cosmic evolution of the {\it neutral atomic} Hydrogen via 21-cm emission
and absorption. However, radio observations at these high redshifts are overwhelmingly difficult, 
due to the faintness of the emission and the very strong foregrounds, and in any case
such radio surveys cannot tell us 
about the sources of the ionizing flux.

A key and interesting question, then, is {\it whether and when} the apparently rapidly evolving UV-selected galaxy population is capable of delivering enough ionizing photons per unit time per unit volume which can escape from their host galaxies to reionize the inter galactic medium.

\subsubsection{The galaxy population at ${\bf z \simeq 7}$, 
and the supply of reionizing photons}

Clearly the complete ionization of hydrogen in the intergalactic medium requires sustained sources of Lyman continuum 
photons with wavelengths $\lambda < 912$\,\AA\ 
(corresponding to the ionization energy of ground-state hydrogen, $E > 13.6$\,eV).
If the emerging population of young faint galaxies revealed in the HST surveys is responsible 
for reionizing the Universe then, as discussed by many authors, the process of reionization should, at least in broad 
terms, follow their time-dependent density (e.g. Robertson et al. 2010; Trenti et al. 2010). However, it is not straightforward to establish 
the number density of ionizing photons delivered by galaxies at $z \simeq 6 - 10$, because they are 
essentially unobservable due to the fact they are absorbed by neutral hydrogen (as they must be 
if they are doing their job of reionizing the hydrogen gas). We are therefore forced to infer the number density 
of ionizing photons from the observable evolving UV luminosity density at $\lambda_{rest} \simeq 1500$\,\AA, coupled 
with estimates of the rate of ionizing photons produced per unit solar mass of star-formation activity, and an estimate 
of what fraction of the ionizing Lyman-continuum photons produced by young stars 
can actually escape their host galaxies to help with reionization of the surrounding
inter-galactic medium ($f_{esc}$).

Finally, we also require some knowledge of the ``clumpiness'' ($C$) of the inter-galactic medium in the young 
Universe. For the IGM to be ionized simply requires that recombinations 
are balanced by ionizations. The recombination rate depends on the IGM temperature and the physical hydrogen density
which declines with time according to the universal expansion factor $R(t)^{-3} \propto (1+z)^3$. 
However, it is enhanced in locally overdense regions by the clumping factor $C = \langle n_H^2 \rangle / \langle n_H \rangle^2$
(i.e. $C=1$ corresponds to a uniform IGM). Early cosmological simulations indicated that the IGM 
clumping factor at $z \simeq 6$ could be as high as $C \simeq 30$ (e.g. Gnedin \& Ostriker 1997) 
making reionization difficult due to self-shielding within dense clumps. However, more recent simulations suggest 
that the IGM clumping factor lies in the range $1 < C < 6$, making reionization easier to achieve 
(e.g. Bolton \& Haehnelt 2007; Pawlik, Schaye \& van Scherpenzeel 2009) 

\begin{figure}

\includegraphics[scale=0.93, angle=0]{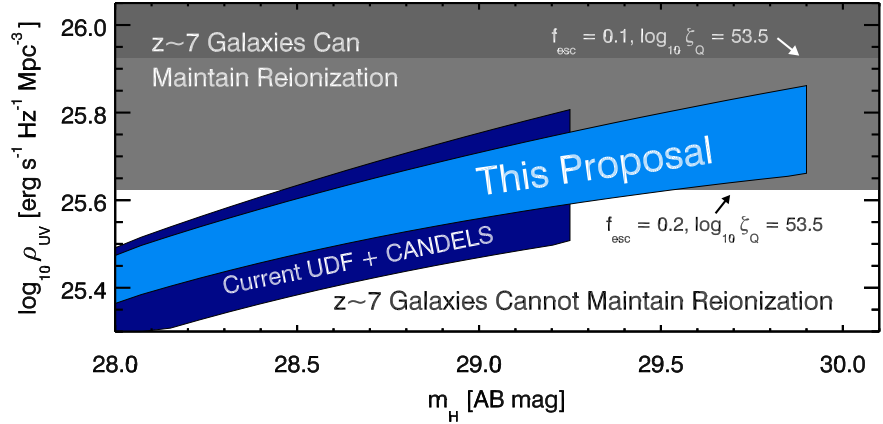}

\caption{
Did Galaxies Reionize the Universe? Expected constraints on the rate of 
ionizing photons $dn_{ion}/dt$, from the combination of the HUDF and on-going CANDELS {\it HST} surveys. 
The shaded regions show a 3$\sigma$ uncertainty in the UV luminosity density $\rho_{UV}$ and the improvement 
made possible via the planned deeper 29.9(AB) (stacked $J_{140}$ + $H_{160}$) UDF12 imaging program. 
The light and dark grey regions show the $\rho_{UV}$ ranges necessary to maintain reionization for 
escape fractions $f_{esc}=0.2$ and 0.1 respectively (for other assumptions see Robertson et al. 2010) (courtesy B. Robertson).}
\label{fig:25}      
\end{figure}

An illustration of current constraints on the ability of galaxies to 
reionize the Universe at $z \simeq 7$ is shown in Fig. 25 (adapted from Robertson et al. 2010).
This is based on our current knowledge of the 
galaxy LF at this epoch, an assumed IGM clumping factor $C=2$, a production rate of ionizing photons 
per unit star formation of $10^{53.5}\,{\rm s^{-1}M_{\odot}^{-1} yr}$, and alternative escape fraction assumptions
of $f_{esc} = 0.1$ and $f_{esc} = 0.2$
(values for $f_{esc}$ which at least have some tentative support
from observations at $z \simeq 3$; Shapley et al. 2006, Iwata et al. 2009). 
All of these numbers remain highly uncertain, but it can be seen 
that for these, arguably not unreasonable assumptions, confirmation that the $z \simeq 7$ LF of McLure et al. 
(2010) continues to rise steeply down to $H \simeq 30$ AB mag may be sufficient to prove that 
the emerging population of young galaxies could have reionized the Universe by $z \simeq 7$. Of course 
this conclusion would become even more secure if the even steeper faint-end slope in the LF favoured  
by Bouwens et al. (2011) (see section 4.1) is confirmed, and especially if more accurate 
determinations of the UV-slope parameter $\beta$ favour high escape fractions (see section 5.3). This provides 
strong motivation for the even deeper HST WFC3/IR observations in the HUDF planned in the UDF12 project in 
summer 2012 ({\it HST} programme GO-12498). 

In addition, Fig. 25 conservatively assumes that we only count up the photons from the {\it observable} 
galaxies which, for the $H_{160} \simeq 30$ limit of the planned HST imaging, corresponds to an absolute 
magnitude limit of $M_{1500} \simeq -17$. Extrapolation to still fainter luminosities will provide 
yet more ionizing photons, especially if the LF remains steep, and even more so if escape fraction 
rises with decreasing luminosity. However, it is currently unclear how far down in 
luminosity one can safely integrate. At low redshifts, {\it GALEX} imaging
in the Coma and Virgo clusters suggests a turnover in the LF  
around $M_{UV} \simeq -14$ (Hammer et al. 2012; Boselli et al. 2011)
corresponding to a deficit of dwarfs below masses of $10^8\,{\rm M_{\odot}}$, while the field UV LFs 
appear to keep rising to fainter magnitudes ($\simeq -11$; e.g. Treyer et al. 2005). However, the relevance 
of these low-redshift results is unclear, given that Schechter function fits at low-redshift yield 
a faint-end slope of only $\alpha = -1.4$.

If the new HST imaging continues to strengthen the argument that galaxies could have reionized the Universe 
by $z \simeq 7$, then attention will shift to the issue of how to reconcile such relatively late 
reionization with the WMAP results.

\subsubsection{Lyman-$\alpha$ emission}

Additional information on the progress of reionization can be gleaned from observations
of Lyman $\alpha$ emission from high-redshift galaxies, which have the potential 
to inform us about the ionization state of the IGM as a function of redshift.

This work complements detailed analyses of Lyman-$\alpha$ emission from high-redshift
quasars, where studies of the size of proximity zone (i.e. the ionized region) around, for example,
the $z = 7.085$ quasar LAS J1120+0641 (whose spectrum was shown in Fig. 2), have been used to argue 
that the IGM is significantly more neutral at $ z \simeq 7$ than at $z \simeq 6.5$ (Mortlock et al. 2011; 
Bolton et al. 2011). The problem with such studies of very bright, but hence very rare objects is that it 
is hard to know whether the sightline is typical, and it is also hard to decide whether a small
proximity zone may simply refect the fact a given quasar has only recently turned on.
Thus, while observations of Lyman-$\alpha$ emission from the much fainter galaxy population are obviously
much more challenging, they offer the prospect of statistically representative results based on multiple sightlines.

As already discussed in detail in section 4.3, and as concisely described by Finlator (2012), a partially 
neutral IGM has the effect of scattering the Lyman-$\alpha$ emission from a galaxy into 
a low surface brightness halo, and this has several measurable consequences which, for convenience,  I summarize again briefly here.

First, the luminosity function of LAEs may evolve. However, as discussed in sections 4.2 and 4.3,  the interpretation
of current measurements of the evolving LAE LF is complicated by the fact any evolution seen reflects
a mix in the underlying evolution of the galaxy mass function, evolution in the intrisic ISM of the galaxies,
and the desired signature of the evolving IGM. This complication, coupled with the very poor constraints 
currently available on the form of the LAE LF at $z \simeq 7$, mean that it appears too early to attempt 
to draw any definitive conclusions on the progress of reionization from this work.

Second, the Lyman-$\alpha$ escape fraction should evolve. As explained in section 4.3.3, 
follow-up spectroscopy of objects selected as LBGs has, in principle, the ability 
to cleanly separate the evolution of the underlying galaxy population from the 
evolution of Lyman-$\alpha$ escape fraction. These observations are being keenly pursued, and 
current indications are that average Lyman-$\alpha$ escape fraction increases out to $z \simeq 6$, but that
this trend shows signs of reversal at $z \simeq 7$ (Pentericci et al. 2011; 
Schenker et al. 2012; Ono et al. 2012). 
In addition, tentative claims have been advanced that this drop in escape fraction is more dramatic for 
faint objects than for bright ones, and that this may indicate reionization proceeds from high- 
to low-density environments, as suggested by an inside-out reionization model (Ono et al. 2012).
However, the spectroscopy at $z \simeq 7$
is challenging, and 
current results are somewhat controversial and based on very small samples.
Moreover, the $z \simeq 6$ baseline against which it can be judged 
is still in the process of being properly pinned down (Curtis-Lake et al. 2012).
Nevertheless, continued work in this area has the potential to yield relatively clear-cut results, and should be 
enormously aided by the advent of the new generation of multi-object near-infrared spectrographs, as 
summarized below in section 7.

Third, the mean shape of the Lyman-$\alpha$ line emission from galaxies should evolve, as 
any increasing neutral; fraction impacts on the blue side of the Lyman-$\alpha$ line more than the red.
However, current contraints on this are necessarily confined to stacking measurements, and 
to date have proved inconclusive (e.g. Ouchi et al. 2010).

\begin{figure}[ht]
\begin{center}
\includegraphics[width=2.0in]{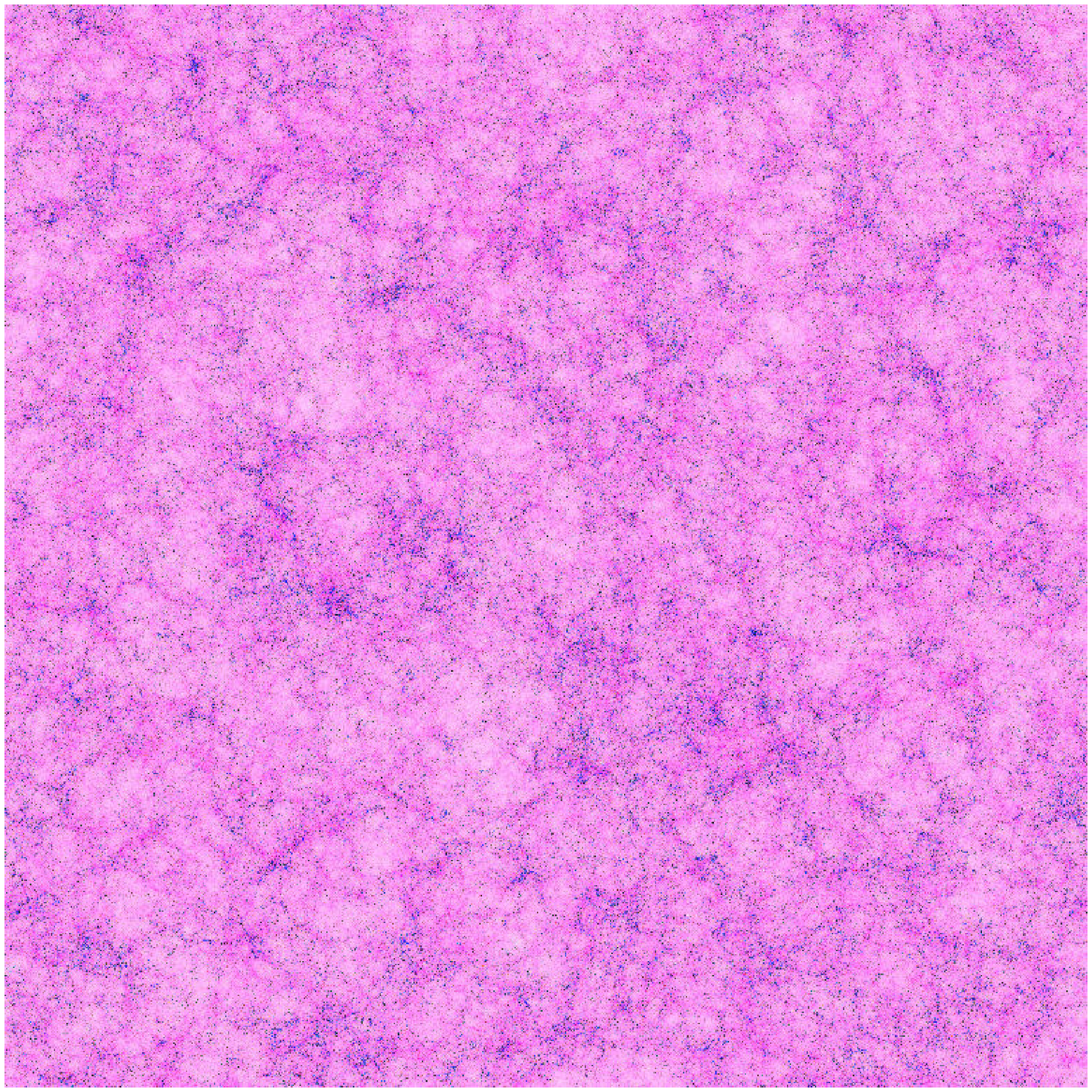}
    \hspace{10mm}
\includegraphics[width=2.0in]{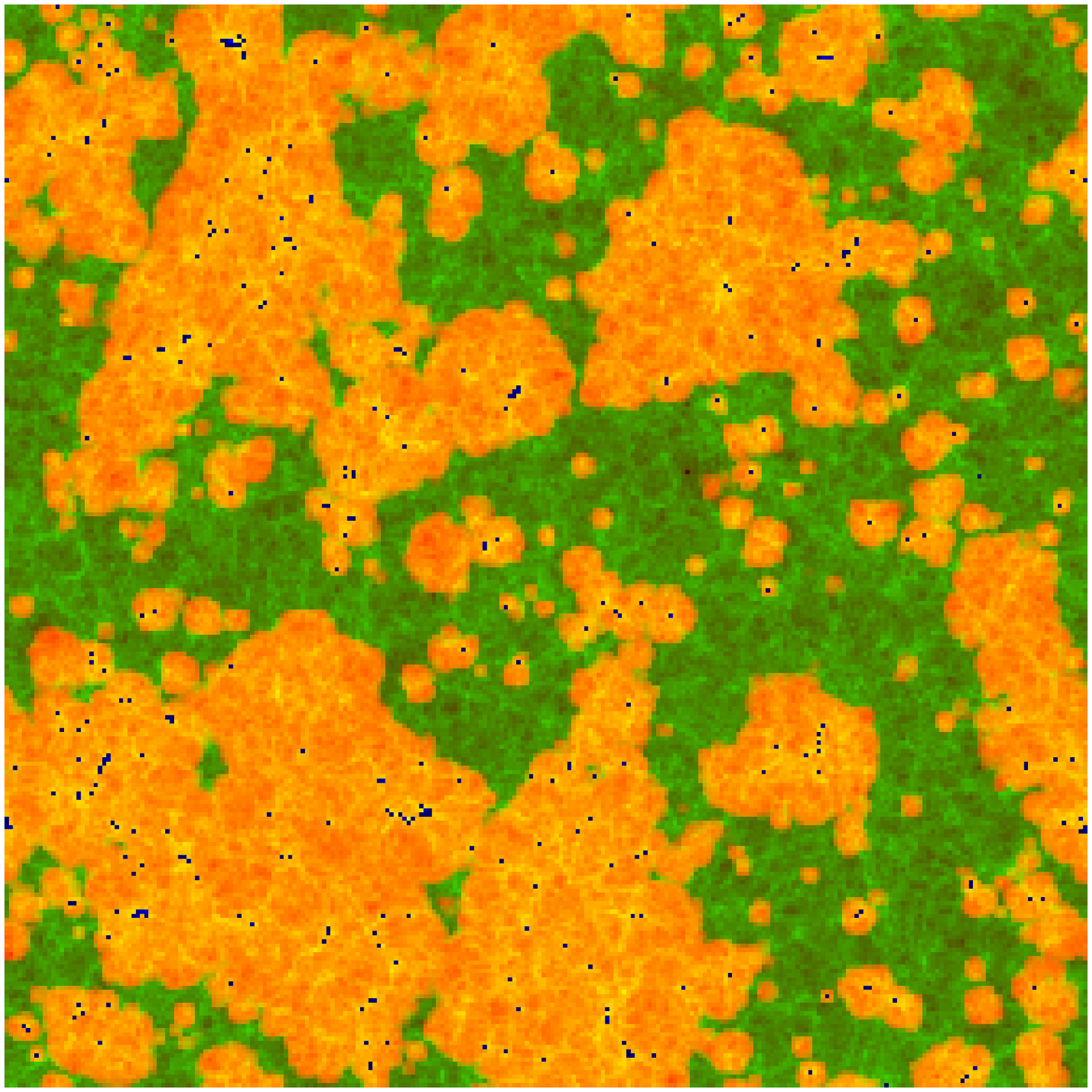}
\caption{\small {\bf Left}  Early structure formation in 
$\Lambda$CDM (at $z=6$) from an N-body simulation with $5488^3$ (165 
billion) particles and $(425\,\rm h^{-1} Mpc)^3$ volume. Shown are the 
dark-matter density (pink) and halos (blue). 
This synthetic image corresponds to 
$3.5 \times 3.5$ degrees on the sky. 
{\bf Right:} The geometry of the epoch of reionization, as illustrated by
a slice through a $(165\,\rm Mpc)^3$
simulation volume at $z=9$. Shown are the density (green/yellow), 
ionized fraction (red/orange), and ionizing sources (dark dots) 
(courtesy I. Iliev).}
\label{fig:26}
\end{center}
\vspace{-0.7cm}
\end{figure}

Finally, the clustering of LAEs should increase with redshift 
as we look back into the epoch of reionization. 
If galaxies produced the photons that reionized the IGM, then 
their clustering should have influenced the history of reionization, and the first galaxies 
are certainly expected to have been highly clustered. Specifically, by the 
time the neutral fraction of hydrogen has dropped to $\simeq 50$\%, the average ionized region is 
expected to have a radius of $\simeq 10$\,Mpc (comoving), created and maintained by many hundreds of
small galaxies working in concert. Because a Lyman-$\alpha$ photon redshifts out of resonance after 
travelling $\simeq 1$\,Mpc, most Lyman-$\alpha$ emission produced by the galaxies 
which together have created this bubble should emerge unscathed (i.e. unscattered) by the IGM. Consequently,
as illustrated by the state-of-the-art simulations shown in Fig. 26, the clumpy nature of reionization
means that LAEs are predicted to appear more clustered than LBGs during the reionization 
epoch. Indeed, under some scenarios the apparent clustering of LAEs can be well in excess 
of the intrinsic clustering of halos in the concordance cosmology. Observing
such enhanced clustering would confirm the prediction that the $HII$ 
regions during reionization are large (McQuinn et al. 2007).
This prediction has arguably gained some tentative observational support from the latest large-area surveys 
for Ly$\alpha$ emitters at $z \simeq 6.5$, where it has been found that,
depending on luminosity, the number density of LAEs varies by a factor of 
$2-10$ between different 
$\simeq 1/4$ deg$^2$ fields (Ouchi et al. 2010; Nakamura et al. 2011). However, 
Ouchi et al. (2010) report no evidence for any significant clustering amplitude boost at $z \simeq 6.6$, 
and it seems clear that the meaningful search for 
this effect must await surveys of large samples of LAEs and LBGs at $z > 7$ covering many square degrees.

\section{Conclusion and future prospects}

Over the last decade we have witnessed a revolution in our knowledge of galaxies in the first billion years of cosmic 
time. Arguably the next 10 years should be even more exciting.

\begin{figure}

\includegraphics[scale=0.18, angle=0]{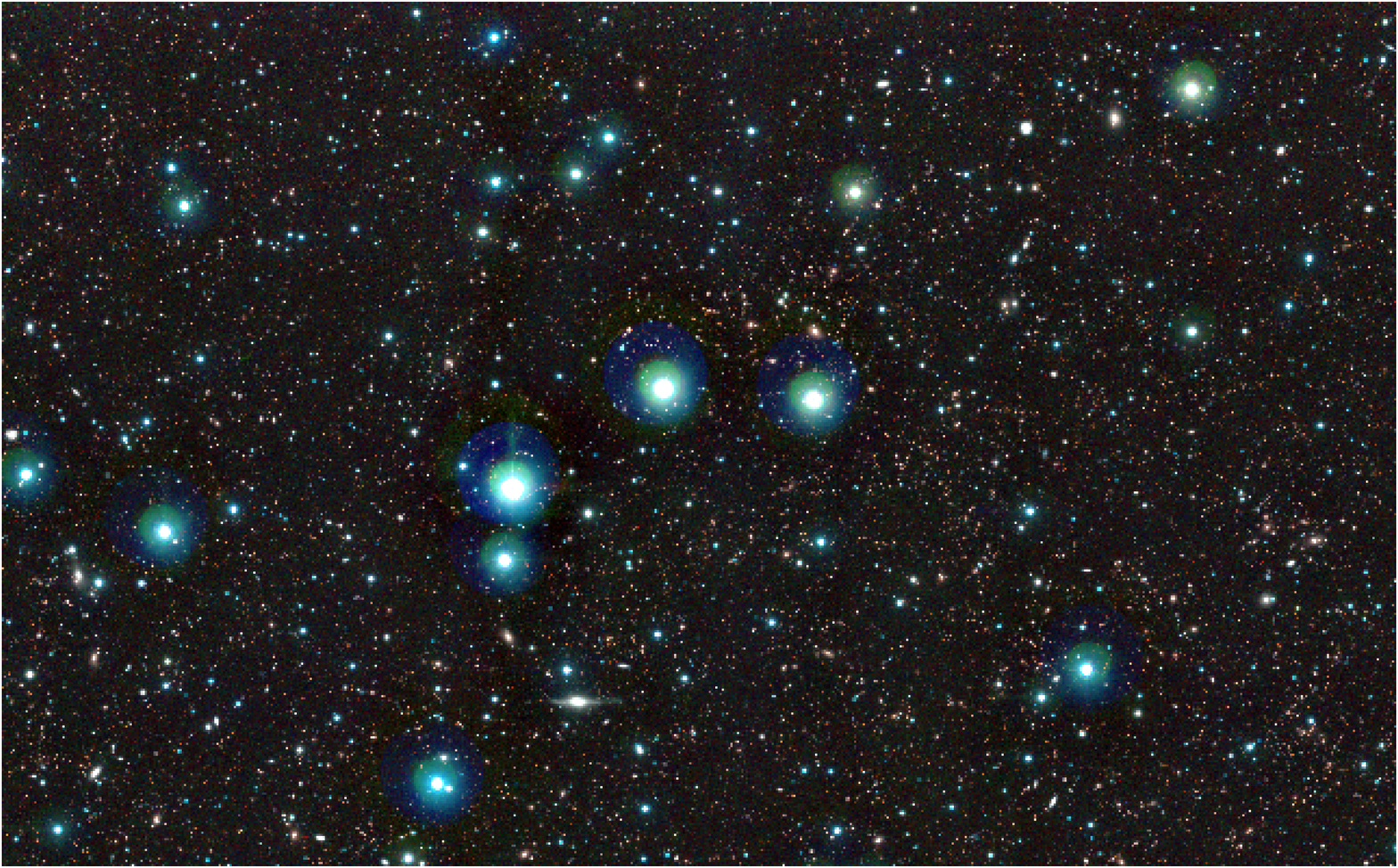}

\caption{A small sub-region of the first-year UltraVISTA near-infrared 
image of the COSMOS field, presented as a $Y+J+K_s$ colour image
(McCracken et al. 2012). This 1.5 deg$^2$ image reaches $Y+J \simeq 25$ AB 
mag, and has already revealed $\simeq 200,00$ galaxies, of which 
$\simeq 5$ appear to be massive galaxies at $z \simeq 7$ (Bowler
et al. 2012). The final UltraVISTA imaging should reach 5-10 times deeper, 
and enable the search for rare massive galaxies out to $z \simeq 10$
(courtesy UltraVISTA/Terapix/CNRS/CASU).}
\label{fig:27}      
\end{figure}

The accurate measurement of the bright end of the evolving galaxy UV luminosity function should soon be improved by 
combining the data over $\simeq 0.2$ deg$^2$ from the {\it HST} CANDELS project (Grogin et al. 2011; Finkelstein et al. 2012; Oesch et al. 2012) 
with the brighter but even larger-area multi-colour imaging being produced by the new generation of ground-based near-infrared surveys, 
such as UltraVISTA (Fig. 27; Bowler et al. 2012; McCracken et al. 2012).
Beyond this, the {\it EUCLID} 
satellite, as part of the ``deep'' component of its mission,
is expected to survey several tens of square degrees down to $J \simeq 26$
mag.
(Laureijs et al. 2011). 
Near-infrared narrow-band imaging surveys can also be expected to continue to expand in scope (e.g. ELVIS, at $z \simeq 8.8$; 
Nilsson et al. 2007). Crucially, 
these wide-area near-infrared imaging surveys will be complemented by well-matched wide-area optical imaging provided, for example, 
by Hyper-Suprime CAM on Subaru (Takada 2010).

At the faint end, attempts will continue to exploit the power of WFC3/IR on {\it HST} to the full, with further 
ultra-deep near-infrared imaging planned over several 
square arcmin of sky, and at the end of the decade NIRCam on the {\it JWST} should extend this work out to $z > 10$.
At the same time the angular resolution limitations of IRAC on {\it Spitzer} will be overcome with MIRI on {\it JWST}, which should deliver 
{\it rest-frame optical} imaging of the highest-redshift galaxies of a quality comparable to that currently achieved with WFC3 on {\it HST}
(Fig. 28). This should enormously improve our knowledge of the 
rest-frame optical morphologies, and the 
stellar masses of the highest-redshift galaxies. It should also enable much more accurate measurement of the UV-optical SEDs 
of faint galaxies at $z \simeq 5 - 10$, including accurate determination of their UV slopes (with resulting implications for 
age, metallicity and escape fraction as discussed above).

\begin{figure}

\includegraphics[scale=0.057, angle=0]{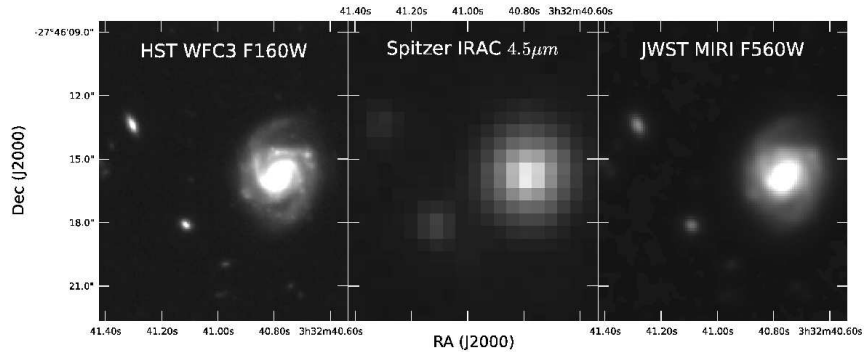}

\caption{{\it HST} WFC3 and {\it Spitzer} IRAC observations of a small sub-region 
of the HUDF field, alongside a simulated image of the same region 
as expected from MIRI on the {\it JWST}. The {\it HST} image is the product 
of 22 hours of on-source integration with WFC3 in the $H_{160}$ band. The
simulated 5.6\,$\mu$m {\it JWST} 
image was produced using the $H$-band morphology and the IRAC 
fluxes as input, and assumes 28 hours of intergation with MIRI. The 
unconfused mid-infrared imaging which will be delivered by MIRI at IRAC 
wavelengths will enable the study of the rest-frame optical morphologies 
out beyond $z \simeq 10$, and will also allow much more robust determinations of the stellar masses of the most distant galaxies 
(courtesy A. Rogers).}
\label{fig:28}       
\end{figure}

Spectroscopic follow-up of the brighter $z > 7$ galaxies (perhaps down to $J \simeq 28$) is set to be transformed by the new generation of 
ground-based multi-object near-infrared spectrographs including FMOS on Subaru (Kimura M. et al., 2010), 
KMOS on the VLT (Sharples et al. 2006), and MOSFIRE on Keck (McLean et al. 2008), before this work should be extended to even fainter magnitudes with NIRSpec 
on {\it JWST} (Birkmann et al. 2011). This should clarify the currently confused picture of Lyman-$\alpha$ emission from LBGs, with 
important implications for our understanding of the progress of reionization. Wide-field near-infrared GRISM spectroscopy with {\it EUCLID} may enable 
the first meaningful study of the clustering of LAEs relative to LBGs at $z > 7$, where an enhancement of the apparent clustering 
of Lyman-$\alpha$ emitters is a prediction of some models for reionization 
(e.g. 
McQuinn et al. 2007). The next generation of giant ground-based near-infrared 
telescopes equipped with sophisticated adaptive-optics systems (TMT, E-ELT, GMT) will also enable detailed near-infrared high-resolution spectroscopy of the most distant galaxies.

Finally, in the rather near future, we can expect a revolution in the search for and study of galaxies at $z > 5$ at sub-millimetre wavelengths. 
We already know that the most distant quasars are detectable in the sub-mm, so we can anticipate that 
a significant population of rare high-mass dusty galaxies should be uncovered by combining existing {\it Herschel} SPIRE imaging with longer-wavelength 
data from the 
SCUBA-2 Cosmology Legacy Survey now underway at the JCMT. Crucially, detailed millimetre spectroscopy of such objects will be relatively 
straightforward with ALMA. Over the next few years ALMA can also be 
exploited to undertake the first sub-mm surveys of sufficient depth and angular-resolution to complement the Ultra Deep Field studies previously 
only possible at shorter wavelengths with HST. 
This work should enormously clarify our understanding 
of the role of dust and molecules 
at the highest redshifts, completing our census of cosmic star-formation history at early times, and transforming our understanding 
of the production-rate of the first metals in the Universe.

\begin{acknowledgement}
James Dunlop gratefully acknowledges the support of the Royal Society through a Wolfson Research Merit award and the support of the European Research 
Council via the award of an Advanced Grant. He also wishes to acknowledge the many substantial contributions of his collaborators in the study 
of high-redshift galaxies, and the (mostly) good-natured and productive rivalry engendered by the various 
competing groups working at this exciting research frontier.

\end{acknowledgement}

\end{document}